\documentclass[aps,twocolumn,prd,10pt,showpacs,showkeys,preprintnumbers,superscriptaddress,nobibnotes,floatfix,longbibliography]{revtex4-1}
\pdfoutput=1
\usepackage{amsmath,amsfonts,amssymb,mathrsfs,graphicx,color}

\newcommand{\ie}{{\it i.e.}}

\newcommand{\eg}{{\it e.g.}}

\newcommand{\fig}{Fig.}

\newcommand{\Ref}{Ref.}
\newcommand{\Refs}{Refs.}

\newcommand{\deltacp}{\delta_\mathrm{CP}}

\newcommand{\figu}[1]{\fig~\ref{fig:#1}}

\newcommand{\bi}{\begin{itemize}}
\newcommand{\ei}{\end{itemize}}

\begin{document}

\title{Theoretically palatable flavor combinations of astrophysical neutrinos}

\author{Mauricio Bustamante}
\affiliation{Center for Cosmology and AstroParticle Physics (CCAPP), Ohio State University,
        Columbus, OH 43210, USA}
\affiliation{Department of Physics, Ohio State University, Columbus, OH 43210, USA}

\author{John F.~Beacom}
\affiliation{Center for Cosmology and AstroParticle Physics (CCAPP), Ohio State University,
        Columbus, OH 43210, USA}
\affiliation{Department of Physics, Ohio State University, Columbus, OH 43210, USA}
\affiliation{Department of Astronomy, Ohio State University, Columbus, OH 43210, USA}

\author{Walter Winter}
\affiliation{DESY, Platanenallee 6, D-15738 Zeuthen, Germany \\
{\tt bustamanteramirez.1@osu.edu, beacom.7@osu.edu, walter.winter@desy.de} \smallskip}

% \date{\today}
\date{September 03, 2015}

\begin{abstract}

The flavor composition of high-energy astrophysical neutrinos can reveal the physics governing their production, propagation, and interaction. The IceCube Collaboration has published the first experimental determination of the ratio of the flux in each flavor to the total. We present, as a theoretical counterpart, new results for the allowed ranges of flavor ratios at Earth for arbitrary flavor ratios in the sources. Our results will allow IceCube to more quickly identify when their data imply standard physics, a general class of new physics with arbitrary (incoherent) combinations of mass eigenstates, or new physics that goes beyond that, \eg, with terms that dominate the Hamiltonian at high energy.

\end{abstract}

% \pacs{13.15.+g, 13.35.Hb, 14.60.Pq, 14.60.St, 95.30.Cq, 95.85.Ry}
% 13.15.+g: Neutrino interactions
% 13.35.Hb: Neutrinos, decays of
% 14.60.Pq: Neutrino oscillations
% 14.60.St: Neutrinos in nonstandard model
% 95.30.Cq: Elementary particles in astrophysics
% 95.85.Ry: Neutrinos in astronomical observations

% PRL does not use keywords
%\keywords{}

\maketitle

{\bf Introduction.---} The discovery of astrophysical neutrinos with energies up to a few PeV by the IceCube Collaboration~\cite{Aartsen:2013bka,Aartsen:2013jdh,Aartsen:2013eka,Aartsen:2014gkd} is tremendously important for multi-messenger astronomy as well as for new tests of neutrino properties. While the origin of these neutrinos is still unclear, there are important clues in the energy spectrum and sky distribution, and a component from cosmic distances ($\sim$ Gpc) is required \cite{Lipari:2007su,Murase:2013ffa,Laha:2013lka,Murase:2013rfa,Winter:2013cla,Esmaili:2013gha,Razzaque:2013uoa,Ahlers:2013xia,Liu:2013wia,Lunardini:2013gva,Kashiyama:2014rza,Chang:2014hua,Aartsen:2014cva,Bhattacharya:2014sta,Bhattacharya:2014yha,Aartsen:2014ivk,Murase:2014tsa,Anchordoqui:2014pca,Aartsen:2014aqy,Ahlers:2015moa}. These are the most extreme energies and distances for detected neutrinos.

The flavor composition is also expected to be important, because the ratio of flux in each flavor to the total cancels the unknown normalization. The ratios depend on the physical conditions at the source, the effects of standard flavor mixing, and on potential new physics \cite{Barenboim:2003jm,Xing:2006uk,Pakvasa:2007dc,Lipari:2007su,Esmaili:2009dz,Choubey:2009jq,Lai:2009ke,Mena:2014sja,Xu:2014via,Fu:2014isa,Palomares-Ruiz:2015mka,Aartsen:2015ivb,Palladino:2015vna}. 

The first IceCube results on flavor composition have been published recently~\cite{Aartsen:2015ivb}, and were followed by results obtained with a combined-likelihood analysis of several data sets with more statistics~\cite{Aartsen:2015ita}. Accordingly, there has been intense interest in deducing flavor ratios from IceCube data \cite{Winter:2013cla,Mena:2014sja,Winter:2014pya,Palomares-Ruiz:2015mka,Palladino:2015zua}. 

In this Letter, we use ternary plots or ``flavor triangles'' to show the flavor composition at Earth. We systematically explore which regions of this plot can be populated from theoretical perspectives ---without or with new physics--- including the uncertainties in source flavor composition and neutrino mixing parameters. We also note prospects for the proposed volume upgrade, IceCube-Gen2~\cite{Aartsen:2014njl}.

We make no distinction between $\nu$ and $\bar{\nu}$, because, except for yet-unobserved high-energy events, IceCube cannot distinguish between them. (In addition, their cross sections agree to better than $\simeq 5\%$ in this energy range~\cite{Gandhi:1995tf,Gandhi:1998ri}.)

All plots shown in the main text are for the normal neutrino mass hierarchy (NH), in which $\nu_1$ is the lightest mass eigenstate. Corresponding plots for the inverted hierarchy (IH), in which $\nu_3$ is lightest, are given in the Supplemental Material; the differences are modest.

%%%%%%%%%%%%%%%%%%%%%%%%%%%%%%%%%%%%%%%%%%%%%%%%%%%%%%%%%%%%%%%%%%%%%%%%%%%%%%%
%%%%%%%%%%%%%%%%%%%%%%%%%%%%%%%%%%%%%%%%%%%%%%%%%%%%%%%%%%%%%%%%%%%%%%%%%%%%%%%

{\bf Flavor identification in IceCube.---} IceCube can discriminate between muon tracks (from $\nu_\mu$, mostly) and cascades (from charged-current interactions of $\nu_e$ and $\nu_\tau$, mainly, and from neutral-current interactions of all flavors). If higher-energy events are observed, it will be possible to isolate $\bar{\nu}_e$ cascades via the Glashow resonance~\cite{Anchordoqui:2004eb,Bhattacharya:2011qu,Barger:2014iua}, and $\nu_\tau$ and $\bar{\nu}_\tau$ via double-bang and lollipop topologies~\cite{Learned:1994wg,Beacom:2003nh,Bugaev:2003sw}. In their absence, there is an experimental degeneracy between the electron and tau neutrino flavor content at Earth \cite{Palomares-Ruiz:2015mka,Aartsen:2015ivb}. In contrast, theoretically predicted flavor ratios, even in models with new physics, have a $\mu$-$\tau$ symmetry due to that mixing angle being near-maximal.

%%%%%%%%%%%%%%%%%%%%%%%%%%%%%%%%%%%%%%%%%%%%%%%%%%%%%%%%%%%%%%%%%%%%%%%%%%%%%%%
%%%%%%%%%%%%%%%%%%%%%%%%%%%%%%%%%%%%%%%%%%%%%%%%%%%%%%%%%%%%%%%%%%%%%%%%%%%%%%%

{\bf Flavor composition at the source.---} The flavor composition at the source could be quite different depending on the physical conditions. For the pion decay chain, which is often used as standard (``pion beam''), one expects a composition $(f_{e,\text{S}}:f_{\mu,\text{S}}:f_{\tau,\text{S}})$=$(\frac{1}{3}:\frac{2}{3}:0)_\text{S}$, with $f_{\alpha,\text{S}}$ the ratio of $\nu_{\alpha}+\bar{\nu}_{\alpha}$ to the total flux, where $f_{e,\text{S}}+f_{\mu,\text{S}}+f_{\tau,\text{S}}=1$. Synchrotron cooling of secondary muons in strong magnetic fields leads to a transition to $(0:1:0)_\text{S}$ (``muon damped'') at higher energies, which depends on the field strength; see, \eg, \Refs~\cite{Kashti:2005qa,Lipari:2007su,Kachelriess:2007tr,Hummer:2010ai,Winter:2014pya}. If these muons pile up at lower energies~\cite{Hummer:2010ai}, or if there are contributions from charmed meson decays \cite{Kachelriess:2006fi,Enberg:2008jm,Choubey:2009jq}, then $(\frac{1}{2}:\frac{1}{2}:0)_\text{S}$ is expected. Neutron decays \cite{Lipari:2007su} lead to $(1:0:0)_\text{S}$. Small deviations, $\lesssim 5\%$ in the $\nu_e$/$\nu_\mu$ ratio, are expected from effects such as the helicity dependence of muon decays~\cite{Lipari:2007su,Hummer:2010vx}. If several of the above processes in the source compete, arbitrary flavor compositions $(f_{e,\text{S}}:1-f_{e,\text{S}}:0)$ can be obtained~\cite{Hummer:2010ai}. If, in addition, $\nu_\tau$ are produced, such as by oscillations in a matter  envelope~\cite{Lunardini:2000fy,Razzaque:2009kq,Sahu:2010ap}, even $(f_{e,\text{S}}:f_{\mu,\text{S}}:1-f_{e,\text{S}}-f_{\mu,\text{S}})$ (with $0 \le f_{\mu,\text{S}} \le 1- f_{e,\text{S}}$) could be possible. Dark matter annihilation or decay could yield any mixture, but $(\frac{1}{3}:\frac{1}{3}:\frac{1}{3})_\text{S}$ is the most natural.

%%%%%%%%%%%%%%%%%%%%%%%%%%%%%%%%%%%%%%%%%%%%%%%%%%%%%%%%%%%%%%%%%%%%%%%%%%%%%%%
%%%%%%%%%%%%%%%%%%%%%%%%%%%%%%%%%%%%%%%%%%%%%%%%%%%%%%%%%%%%%%%%%%%%%%%%%%%%%%%

{\bf Flavor composition at Earth.---} Here we focus on a diffuse flux, which is composed of small contributions from many sources over a wide range of distances, and detected with energy resolution $\gtrsim 10\%$ (and binned more coarsely).  In this case, the neutrinos are, at least effectively, an incoherent mixture of mass eigenstates.  Even for the solar $\Delta m_\odot^2 \approx 8 \cdot 10^{-5}$ eV$^2$ and PeV energies, the vacuum oscillation length is only $\sim 10^{-13}$ Gpc, much smaller than the complete baseline. (Depending on the physics in the production region, there can be also wave packet decoherence
in the source~\cite{Farzan:2008eg,Akhmedov:2012uu,Jones:2014sfa}.) As a consequence, the flavor composition at Earth \cite{Farzan:2008eg} is $f_{\beta,\oplus} = \sum_{i,\alpha} | U_{\beta i} |^2 \, | U_{\alpha i} |^2 \, f_{\alpha,\text{S}}$, with $U$ the PMNS matrix \cite{Agashe:2014kda}, implying $\sum_\beta f_{\beta,\oplus} = 1$. For a pion beam, the flavor composition evolves roughly into flavor equipartition at the detector, $(\frac{1}{3}:\frac{1}{3}:\frac{1}{3})_\oplus$.

\begin{figure}[t!]
 \centering
 \includegraphics[width=0.45\textwidth]{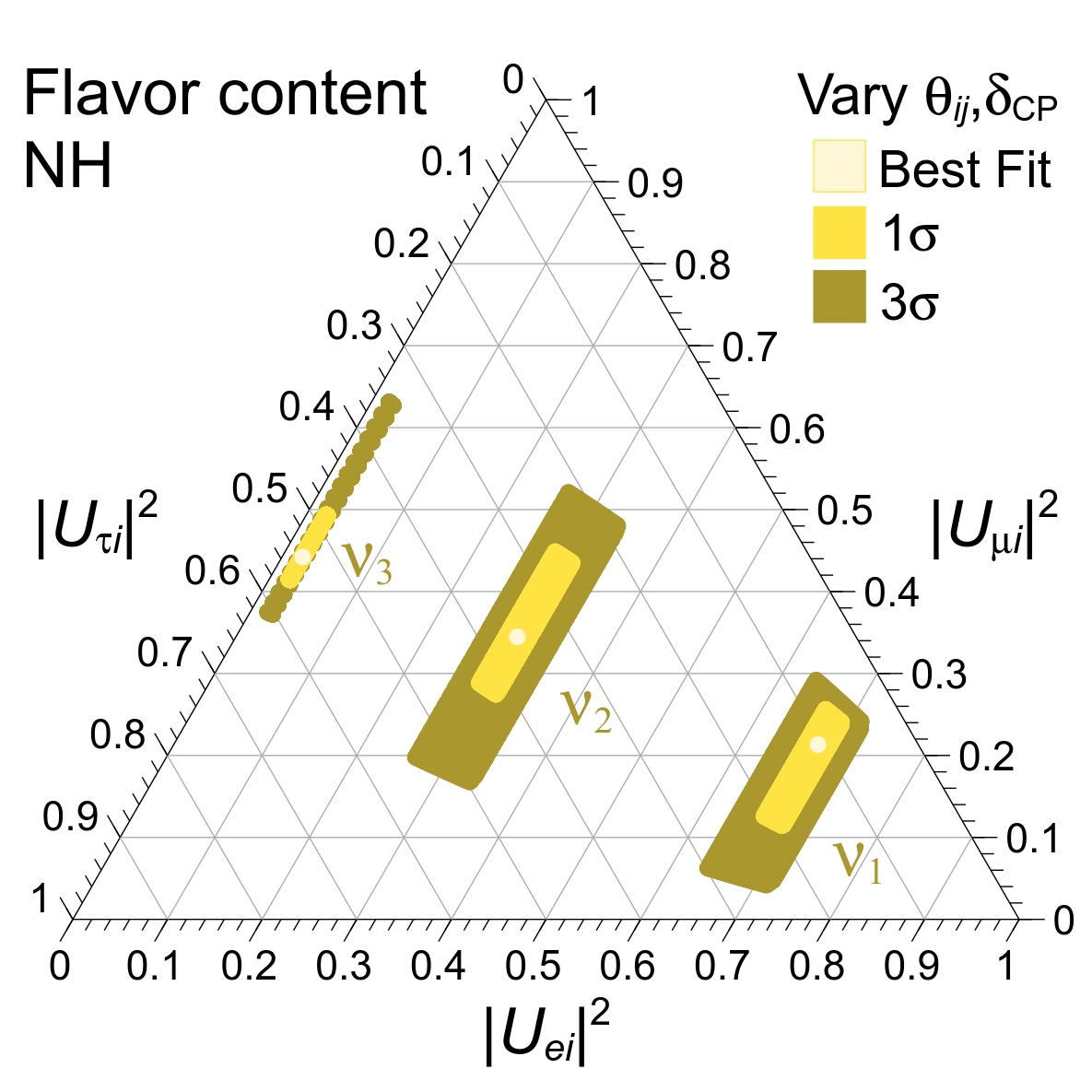}
 \caption{\label{fig:FlavorContentNH}Flavor content of the three active mass eigenstates. The regions are given by the best-fit values of the mixing parameters (light yellow), and their $1\sigma$ (darker) and $3\sigma$ (darkest) uncertainty regions~\cite{Gonzalez-Garcia:2014bfa}, assuming a normal mass hierarchy (NH). The tilt of the tick marks indicates the orientation with which to read the flavor content.}
\end{figure}

\begin{figure}[t!]
 \centering
 \includegraphics[width=0.45\textwidth]{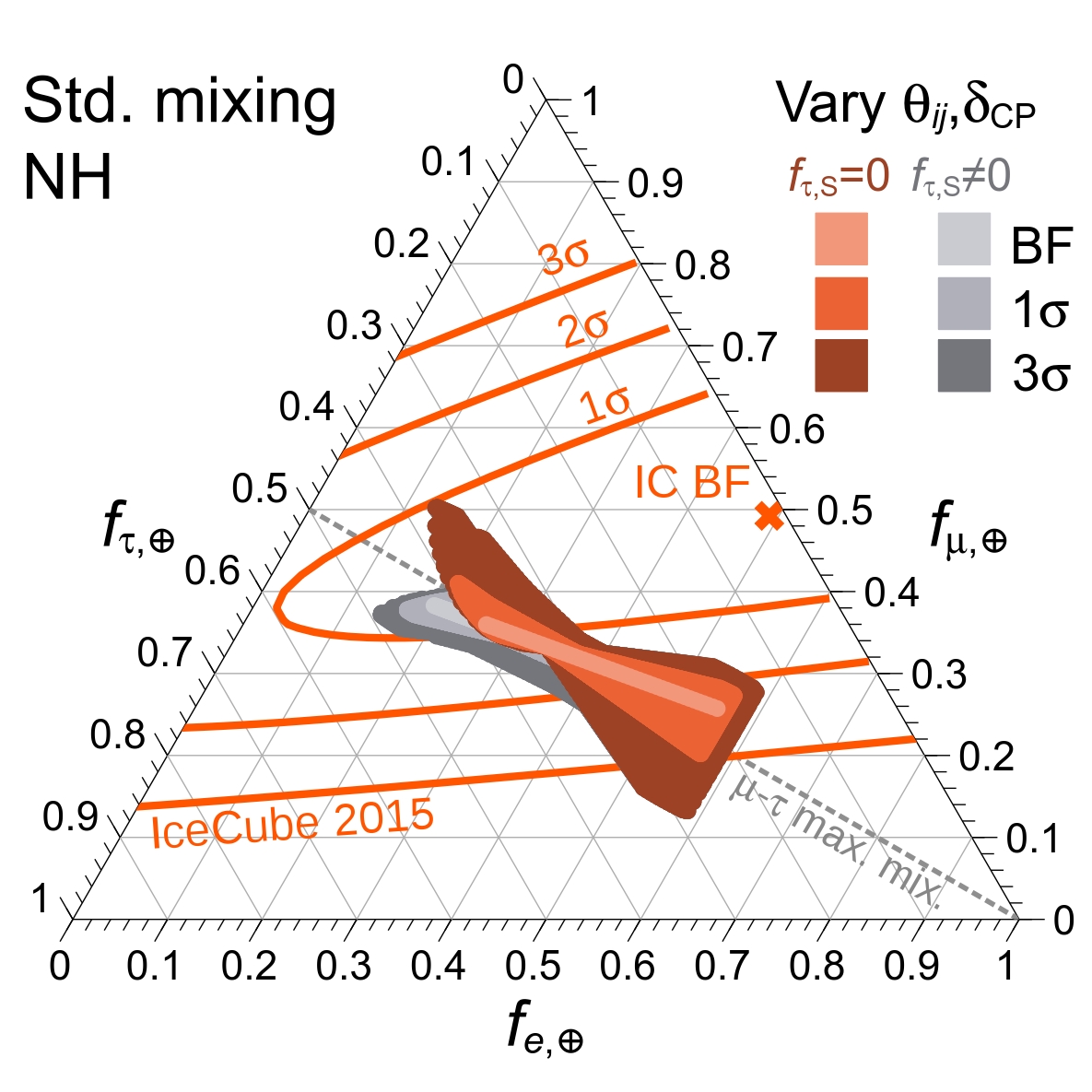}
 \caption{\label{fig:StdMixingFullVarVsNoTauNH}Allowed flavor ratios at Earth with no new physics. The flavor ratios at the source are arbitrary (gray) or contain no tau flavor (red). The IceCube results are from \Ref~\cite{Aartsen:2015ita}.}
\end{figure}

New physics in neutrino propagation might modify the flavor composition. We categorize classes of new-physics models below.

%%%%%%%%%%%%%%%%%%%%%%%%%%%%%%%%%%%%%%%%%%%%%%%%%%%%%%%%%%%%%%%%%%%%%%%%%%%%%%%
%%%%%%%%%%%%%%%%%%%%%%%%%%%%%%%%%%%%%%%%%%%%%%%%%%%%%%%%%%%%%%%%%%%%%%%%%%%%%%%

{\bf Flavor content of the mass eigenstates.---}

Figure \ref{fig:FlavorContentNH} shows the flavor content $\lvert U_{\alpha i} \rvert^2$ of the mass eigenstates, which is the fundamental input that determines flavor ratios at Earth without or with new physics. It also illustrates the underlying three-flavor unitarity of our analysis, \ie, $\lvert U_{\alpha 1} \rvert^2 + \lvert U_{\alpha 2} \rvert^2 + \lvert U_{\alpha 3} \rvert^2 = 1$, which allows the flavor content to be displayed in a ternary plot~\cite{Dalitz:1953cp}. This is appropriate because the mixing angles to sterile neutrinos must be quite small \cite{Kopp:2013vaa,Giunti:2013aea}.

The long axis of each region is set by the uncertainty in $\theta_{23}$ and $\deltacp$, while the short axis is set by the uncertainty in $\theta_{12}$. The effect of the uncertainty in $\theta_{13}$ is tiny. Even if $\theta_{23}$ were to be precisely determined soon, it is less likely that $\deltacp$ will be, and the uncertainty in the latter will still span a large range in $\left\vert U_{\tau1} \right\vert^2$ and $\left\vert U_{\tau2} \right\vert^2$.

%%%%%%%%%%%%%%%%%%%%%%%%%%%%%%%%%%%%%%%%%%%%%%%%%%%%%%%%%%%%%%%%%%%%%%%%%%%%%%%
%%%%%%%%%%%%%%%%%%%%%%%%%%%%%%%%%%%%%%%%%%%%%%%%%%%%%%%%%%%%%%%%%%%%%%%%%%%%%%%

{\bf Standard flavor mixing.---} Figure \ref{fig:StdMixingFullVarVsNoTauNH} shows the allowed region for the flavor composition at Earth assuming arbitrary flavor composition at the source and standard neutrino mixing (including parameter uncertainties). The region is quite small: even at $3\sigma$ it covers only about 10\% of the available space. There is little difference between $f_{\tau,\text{S}}=0$ and $f_{\tau,\text{S}} \neq 0$.

There is a theoretical symmetry along the line $(f_{e,\oplus}:(1-f_{e,\oplus})/2:(1-f_{e,\oplus})/2)$ coming from nearly-maximal mixing. On the other hand, the experimental degeneracy pulls towards $(f_{e,\oplus}:f_{\mu,\oplus}:1-f_{\mu,\oplus}-f_{e,\oplus})$, with $f_{e,\oplus} \leq 1-f_{\mu,\oplus}$, on account of the difficulty of distinguishing between electromagnetic and hadronic cascades. Thus, theory and experiment are complementary, which enhances the discriminating power of flavor ratios.

\begin{figure}[t!]
 \centering
 \includegraphics[width=0.45\textwidth]{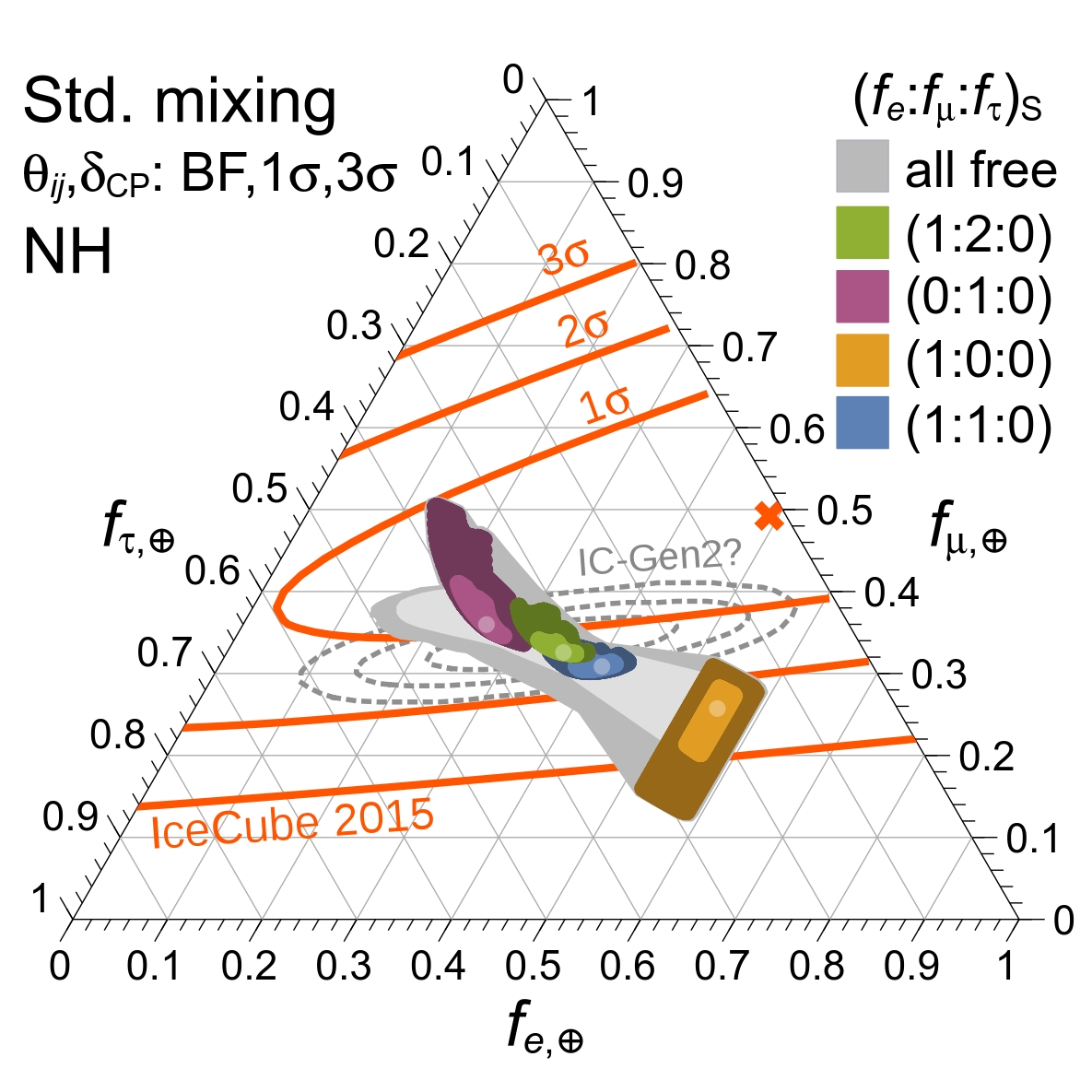}
 \caption{\label{fig:StdMixingSelectedSourceRatiosNH}Allowed flavor ratios at Earth for different choices of source ratios, assuming standard mixing. Projected $1\sigma$, $2\sigma$, and $3\sigma$ exclusion curves from IceCube-Gen2 are included for comparison (gray, dotted); see main text.}
\end{figure}

The region shown includes the possibility of energy-dependent flavor composition at the source; see the Supplemental Material for an example. It also includes the possibility that the diffuse flux has contributions from sources with different flavor compositions, because of the linear mapping between those at the source and those at Earth.

Whereas the first IceCube flavor ratio analysis~\cite{Aartsen:2015ivb} used only three years of contained-vertex events, the updated analysis~\cite{Aartsen:2015ita}, whose exclusion curves are shown in \figu{StdMixingFullVarVsNoTauNH}, combines several different data sets collected over four years, including through-going muons. The exclusion curves of both analyses are compatible.

Figure \ref{fig:StdMixingSelectedSourceRatiosNH} shows that if the flavor composition at the source could be restricted from astrophysical arguments, the allowed regions at Earth could become tiny (and will shrink when the mixing parameters are better known). A source composition of $\left(1:0:0\right)_\text{S}$ is already disfavored at $\gtrsim 2\sigma$. While the current IceCube fit is compatible with the standard $\left(\frac{1}{3}:\frac{1}{3}:\frac{1}{3}\right)_\oplus$ at $1\sigma$, the best-fit point cannot be reached within the Standard Model.

An upgrade of IceCube would have excellent discrimination power, as indicated by the projected sensitivity curves we estimate for IceCube-Gen2 and show in \figu{StdMixingSelectedSourceRatiosNH}.  We reduced the IceCube uncertainties by a factor 5, corresponding to an exposure increased by a factor $\sim 25$ ($\sim 6$ times larger effective area~\cite{Aartsen:2014njl} and twelve years instead of three).  The true sensitivity might be worse (due to sparser instrumentation) or better (due to new techniques or to the discovery of flavor-identifying signals~\cite{Glashow:1960zz,Berezinsky:1977sf,Berezinsky:1981bt,Learned:1994wg,Athar:2000rx,Bugaev:2003sw,Anchordoqui:2004eb,DeYoung:2006fg,Hummer:2010ai,Xing:2011zm,Bhattacharya:2011qu,Bhattacharya:2012fh,Abbasi:2012cu,2013PhRvD..87c7302B}). To be conservative, we assumed the best fit will correspond to the most-frequently considered composition, $(\frac{1}{3}:\frac{1}{3}:\frac{1}{3})_\oplus$, for which it will be most difficult to test for new physics.

\begin{figure}[t!]
 \centering
 \includegraphics[width=0.45\textwidth]{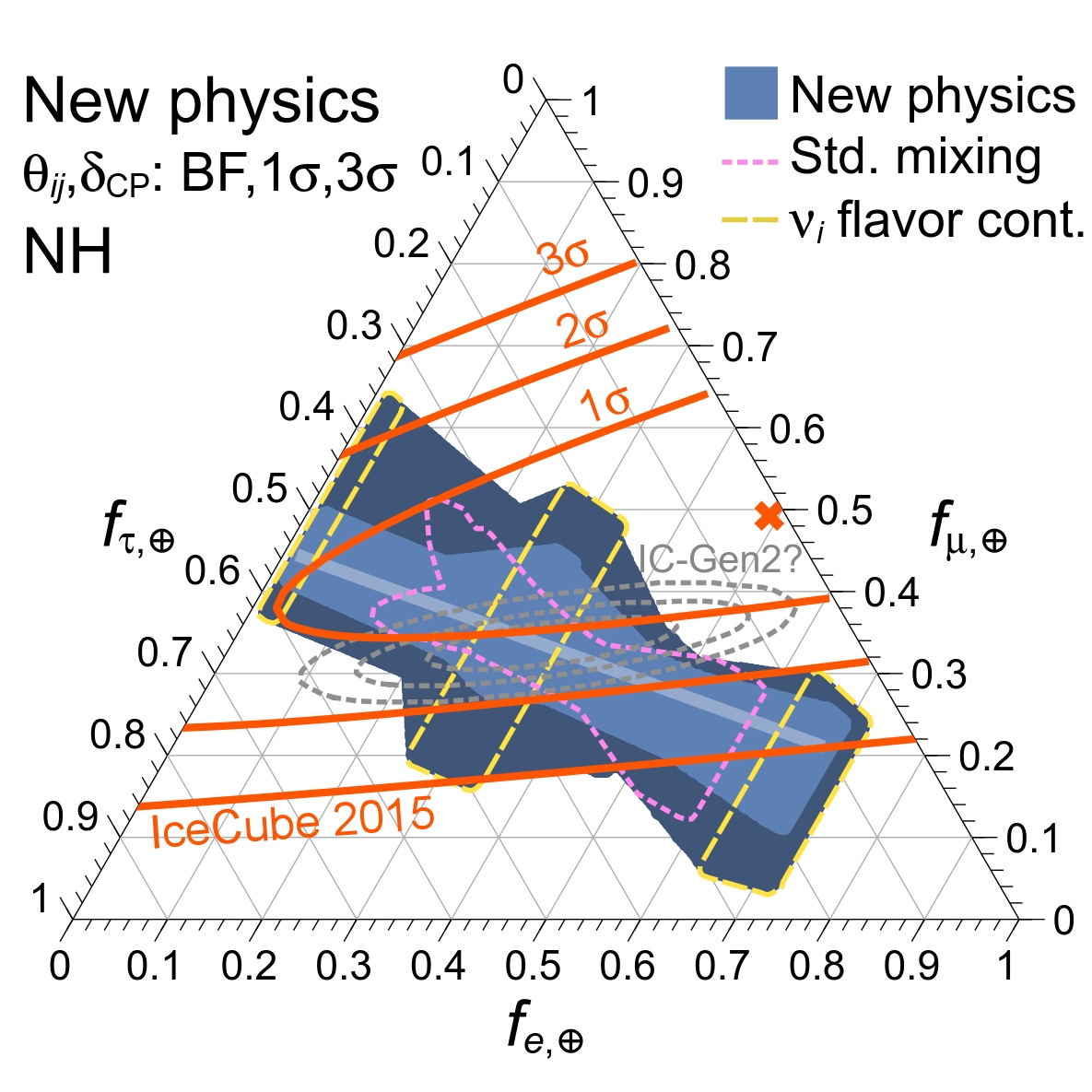}
 \caption{\label{fig:RegionNewPhysicsNH}Allowed flavor ratios at Earth in a general class of new-physics models. These produce linear combinations of the flavor content of $\nu_3$, $\nu_2$, and $\nu_1$, shown as yellow (dashed) curves, from left to right. The standard mixing $3\sigma$ region from \figu{StdMixingFullVarVsNoTauNH} is shown as a magenta (dotted) curve.}
\end{figure}

%%%%%%%%%%%%%%%%%%%%%%%%%%%%%%%%%%%%%%%%%%%%%%%%%%%%%%%%%%%%%%%%%%%%%%%%%%%%%%%
%%%%%%%%%%%%%%%%%%%%%%%%%%%%%%%%%%%%%%%%%%%%%%%%%%%%%%%%%%%%%%%%%%%%%%%%%%%%%%%

{\bf Flavor ratios with new physics.---} New physics can modify the flavor composition at production, during propagation, or in interaction. In the first two cases, it will affect the flavor composition that reaches the detector; this is our focus. In the last case ---which includes, \eg, non-standard interactions~\cite{Blennow:2009rp} and renormalization group running of the mixing parameters \cite{Bustamante:2010bf}--- we assume that new physics, possibly energy-dependent, can be separated by probing the interaction length in Earth via the angular dependence of the neutrino flux~\cite{Jain:2002kz,Hussain:2003vi,Borriello:2007cs,Marfatia:2015hva}.

In extreme scenarios, there could be only one mass eigenstate present at detection, and the flavor composition would correspond to that of one eigenstate. This could happen if all but one mass eigenstate completely decays or if matter-affected mixing at the source singles out a specific one for emission.

Figure \ref{fig:RegionNewPhysicsNH} shows the allowed region if we restrict ourselves to a general class of new-physics models ---those in which arbitrary combinations of incoherent mass eigenstates are allowed (we give examples below of models that can access the area outside this region). The $\alpha$-flavor content of an allowed point is computed as $k_1 \left\vert U_{\alpha 1} \right\vert^2 + k_2 \left\vert U_{\alpha 2} \right\vert^2 + k_3 \left\vert U_{\alpha 3} \right\vert^2$, where the $k_i$ are varied under the constraint $k_1 + k_2 + k_3 = 1$ and the values of the mixing parameters are fixed. To generate the complete region, we repeat the procedure by varying the mixing parameters within their uncertainties.

For a particular new-physics model, the functional forms and values of the $k_i$ are determined by its parameters. The most dramatic examples include all variants of neutrino decay among mass eigenstates, both partial and complete~\cite{Beacom:2002vi,Barenboim:2003jm,Maltoni:2008jr,Baerwald:2012kc,Pagliaroli:2015rca}, and secret neutrino interactions~\cite{Kolb:1987qy,Ioka:2014kca,Ng:2014pca,Blum:2014ewa,Cherry:2014xra,Kamada:2015era,DiFranzo:2015qea}; the $k_i$ in these cases depend on neutrino lifetimes and new coupling constants, respectively.  Other examples are pseudo-Dirac neutrinos~\cite{Beacom:2003eu,Esmaili:2012ac,Joshipura:2013yba} and decoherence on the Planck-scale structure of spacetime~\cite{Ellis:1983jz,Banks:1983by,Benatti:2000ph,Gago:2002na,Morgan:2004vv,Hooper:2005jp,Anchordoqui:2005gj}.

Even with this general class of new-physics models, only about $25\%$ of the flavor triangle can be accessed. The current IceCube best fit cannot be reached even by invoking this class of physics models.  IceCube-Gen2 will be needed to strongly constrain such new-physics models.

Interestingly, there is more than one way in which the standard $\left(\frac{1}{3}:\frac{1}{3}:\frac{1}{3}\right)_\oplus$ composition can be generated, such as through the standard mixing of $\left(\frac{1}{3}:\frac{2}{3}:0\right)_\text{S}$, or through a fortuitous incoherent mix of mass eigenstates due to decay.

Already, complete decay in the most often used neutrino decay scenario (only $\nu_1$ stable) for the NH can be ruled out at $\gtrsim 2\sigma$ (see \Ref~\cite{Pagliaroli:2015rca} for a weaker exclusion at $1\sigma$ based on their own analysis of tracks and cascades), and bounds on the neutrino lifetimes can be set \cite{BBWprep}.

To access the white region in \figu{RegionNewPhysicsNH}, a broader class of new-physics models is required. Possible examples are models with violation of CPT and/or Lorentz invariance (which alter the dispersion relations)~\cite{Colladay:1998fq,Kostelecky:2003xn,Barenboim:2003jm,Hooper:2005jp,Ellis:2011ek,Bustamante:2010nq}, or the equivalence principle~\cite{Gasperini:1989rt,Butler:1993wi,Glashow:1997gx}, and coupling to a torsion field~\cite{DeSabbata:1981ek}.

All these have in common that they either invalidate the concept of decoherence in the astrophysical neutrino flavor composition or they change the values of the mixing parameters.  
\Ref~\cite{Arguelles:2015dca} adopted a generic effective theory approach in which the new-physics terms dominate the propagation Hamiltonian at high energies, and showed that such models are indeed able to populate almost the full triangle.

Another possibility is the existence of extra dimensions, which could lead to matter-like resonant mixing between active and sterile flavors~\cite{Aeikens:2014yga}. Boosted dark matter~\cite{Agashe:2014yua,Bhattacharya:2014yha,Kopp:2015bfa} could generate neutrino-like events, even mimicking pure-flavor signatures.

%%%%%%%%%%%%%%%%%%%%%%%%%%%%%%%%%%%%%%%%%%%%%%%%%%%%%%%%%%%%%%%%%%%%%%%%%%%%%%%
%%%%%%%%%%%%%%%%%%%%%%%%%%%%%%%%%%%%%%%%%%%%%%%%%%%%%%%%%%%%%%%%%%%%%%%%%%%%%%%

{\bf Conclusions.---} We have demonstrated that the allowed region of neutrino flavor composition at Earth under standard mixing is quite small, in spite of the uncertainties in the mixing parameters and flavor composition at the sources. The allowed region remains small even in the presence of a general class of new-physics models whose effect is to change the incoherent mix of mass eigenstates during propagation (\eg, neutrino decay and secret interactions). These results hardly depend on the mass hierarchy, and they hold for energy-dependent flavor compositions at the source or energy-dependent new physics, even when simultaneously present~\cite{Mehta:2011qb}; see the Supplemental Material.

In order to access the larger space of possible flavor combinations, a broader class of new physics during propagation ---flavor-violating or capable of modifying the values of the mixing parameters--- or at detection is required. Interestingly, the current IceCube best-fit composition lies in this region, though the standard $(\frac{1}{3}:\frac{1}{3}:\frac{1}{3})_\oplus$ case is not excluded. 

The power of IceCube to determine the composition is enhanced by the complementarity between its experimental $\nu_e$-$\nu_\tau$ degeneracy and the theoretical $\nu_\mu$-$\nu_\tau$ symmetry coming from nearly-maximal mixing. The current bounds are not only compatible with most source compositions, but also with many potential new physics effects. However, the most favored neutrino decay scenario (only $\nu_1$ stable) can be already ruled out at $\gtrsim 2\sigma$.

The smaller the allowed region with only standard mixing shown in \figu{StdMixingFullVarVsNoTauNH} and \figu{StdMixingSelectedSourceRatiosNH}, the more sensitive IceCube is to new physics.  Likewise, the smaller the new-physics region shown in \figu{RegionNewPhysicsNH}, the more sensitive IceCube is to the broader class of new physics.  The recent successes in measuring neutrino mixing parameters have been essential to making these regions small. Our results provide new perspectives that will sharpen and accelerate tests of flavor ratios.

Ideally, flavor ratios would be determined using a single class of point sources at known distances. No high-energy astrophysical sources have been resolved yet, however. We have shown that, even using a diffuse flux, flavor ratios can reveal information about source conditions and neutrino properties.

Data from a volume upgrade of IceCube in combination with improved measurements of the mixing parameters, including $\deltacp$, have the potential to nail down the flavor composition at the source or to identify new physics in propagation. However, it is not possible to extract the value of $\deltacp$ from astrophysical data alone if the flavor composition at the source is not known; see the Supplemental Material.

To fully exploit the power of neutrino flavors, advances in four directions are needed:
\begin{enumerate}
 \item 
  A volume upgrade of IceCube (IceCube-Gen2~\cite{Aartsen:2014njl}) or a corresponding experiment in seawater (\eg, KM3NeT~\cite{Kappes:2007ci}).
 \item
  Reduction of the uncertainties in the values of the mixing parameters (especially $\theta_{23}$ and $\deltacp$).
 \item
  Improvements in experimental techniques to reconstruct neutrino flavor and energy.
 \item
  More systematic model building to better understand, or constrain, the region of flavor ratios at Earth that could be accessed by new physics.
\end{enumerate}

Given the wealth of information about neutrino production, propagation, and interaction that the flavor composition provides, its precise determination should become a high-priority goal of ongoing and near-future experimental analyses.

%%%%%%%%%%%%%%%%%%%%%%%%%%%%%%%%%%%%%%%%%%%%%%%%%%%%%%%%%%%%%%%%%%%%%%%%%%%%%%%
%%%%%%%%%%%%%%%%%%%%%%%%%%%%%%%%%%%%%%%%%%%%%%%%%%%%%%%%%%%%%%%%%%%%%%%%%%%%%%%

\vspace*{0.3cm}

{\bf Acknowledgements.}~We thank Markus Ackermann, Shunsaku Horiuchi, Kohta Murase, Kenny Ng, Nathan Whitehorn, and Guanying Zhu for useful discussions and comments; we thank  Marek Kowalski, Lars Mohrmann for that and for help reproducing the curves in \Ref~\cite{Aartsen:2015ita}. JFB is supported by NSF Grant PHY-1404311. MB thanks the Institute for Nuclear Theory at the University of Washington for its hospitality during 14 -- 26 June, 2015, and the Department of Energy for partial support during the completion of this work. WW thanks CCAPP for the hospitality during 29 April -- 3 May, 2015, where substantial progress was made on this project, and acknowledges support from the ``Helmholtz Alliance for Astroparticle Physics HAP'' funded by the Initiative and Networking Fund of the Helmholtz Association. This project has received funding from the
European Research Council (ERC) under the European Union's Horizon 2020 research and innovation programme (grant no.~646623).

\newpage

% % \bibliographystyle{apsrev4-1}
% \bibliography{refs.bib}

\begin{thebibliography}{125}%
\makeatletter
\providecommand \@ifxundefined [1]{%
 \@ifx{#1\undefined}
}%
\providecommand \@ifnum [1]{%
 \ifnum #1\expandafter \@firstoftwo
 \else \expandafter \@secondoftwo
 \fi
}%
\providecommand \@ifx [1]{%
 \ifx #1\expandafter \@firstoftwo
 \else \expandafter \@secondoftwo
 \fi
}%
\providecommand \natexlab [1]{#1}%
\providecommand \enquote  [1]{``#1''}%
\providecommand \bibnamefont  [1]{#1}%
\providecommand \bibfnamefont [1]{#1}%
\providecommand \citenamefont [1]{#1}%
\providecommand \href@noop [0]{\@secondoftwo}%
\providecommand \href [0]{\begingroup \@sanitize@url \@href}%
\providecommand \@href[1]{\@@startlink{#1}\@@href}%
\providecommand \@@href[1]{\endgroup#1\@@endlink}%
\providecommand \@sanitize@url [0]{\catcode `\\12\catcode `\$12\catcode
  `\&12\catcode `\#12\catcode `\^12\catcode `\_12\catcode `\%12\relax}%
\providecommand \@@startlink[1]{}%
\providecommand \@@endlink[0]{}%
\providecommand \url  [0]{\begingroup\@sanitize@url \@url }%
\providecommand \@url [1]{\endgroup\@href {#1}{\urlprefix }}%
\providecommand \urlprefix  [0]{URL }%
\providecommand \Eprint [0]{\href }%
\providecommand \doibase [0]{http://dx.doi.org/}%
\providecommand \selectlanguage [0]{\@gobble}%
\providecommand \bibinfo  [0]{\@secondoftwo}%
\providecommand \bibfield  [0]{\@secondoftwo}%
\providecommand \translation [1]{[#1]}%
\providecommand \BibitemOpen [0]{}%
\providecommand \bibitemStop [0]{}%
\providecommand \bibitemNoStop [0]{.\EOS\space}%
\providecommand \EOS [0]{\spacefactor3000\relax}%
\providecommand \BibitemShut  [1]{\csname bibitem#1\endcsname}%
\let\auto@bib@innerbib\@empty
%</preamble>
\bibitem [{\citenamefont {Aartsen}\ \emph
  {et~al.}(2013{\natexlab{a}})\citenamefont {Aartsen} \emph
  {et~al.}}]{Aartsen:2013bka}%
  \BibitemOpen
  \bibfield  {author} {\bibinfo {author} {\bibfnamefont {M.G.}\ \bibnamefont
  {Aartsen}} \emph {et~al.} (\bibinfo {collaboration} {IceCube
  Collaboration}),\ }\bibfield  {title} {\enquote {\bibinfo {title} {{First
  observation of PeV-energy neutrinos with IceCube}},}\ }\href {\doibase
  10.1103/PhysRevLett.111.021103} {\bibfield  {journal} {\bibinfo  {journal}
  {Phys.~Rev.~Lett.}\ }\textbf {\bibinfo {volume} {111}},\ \bibinfo {pages}
  {021103} (\bibinfo {year} {2013}{\natexlab{a}})},\ \Eprint
  {http://arxiv.org/abs/1304.5356} {arXiv:1304.5356 [astro-ph.HE]} \BibitemShut
  {NoStop}%
%%CITATION = ARXIV:1304.5356;%%
\bibitem [{\citenamefont {Aartsen}\ \emph
  {et~al.}(2013{\natexlab{b}})\citenamefont {Aartsen} \emph
  {et~al.}}]{Aartsen:2013jdh}%
  \BibitemOpen
  \bibfield  {author} {\bibinfo {author} {\bibfnamefont {M.G.}\ \bibnamefont
  {Aartsen}} \emph {et~al.} (\bibinfo {collaboration} {IceCube}),\ }\bibfield
  {title} {\enquote {\bibinfo {title} {{Evidence for High-Energy
  Extraterrestrial Neutrinos at the IceCube Detector}},}\ }\href {\doibase
  10.1126/science.1242856} {\bibfield  {journal} {\bibinfo  {journal}
  {Science}\ }\textbf {\bibinfo {volume} {342}},\ \bibinfo {pages} {1242856}
  (\bibinfo {year} {2013}{\natexlab{b}})},\ \Eprint
  {http://arxiv.org/abs/1311.5238} {arXiv:1311.5238 [astro-ph.HE]} \BibitemShut
  {NoStop}%
%%CITATION = ARXIV:1311.5238;%%
\bibitem [{\citenamefont {Aartsen}\ \emph
  {et~al.}(2014{\natexlab{a}})\citenamefont {Aartsen} \emph
  {et~al.}}]{Aartsen:2013eka}%
  \BibitemOpen
  \bibfield  {author} {\bibinfo {author} {\bibfnamefont {M.G.}\ \bibnamefont
  {Aartsen}} \emph {et~al.} (\bibinfo {collaboration} {IceCube}),\ }\bibfield
  {title} {\enquote {\bibinfo {title} {{Search for a diffuse flux of
  astrophysical muon neutrinos with the IceCube 59-string configuration}},}\
  }\href {\doibase 10.1103/PhysRevD.89.062007} {\bibfield  {journal} {\bibinfo
  {journal} {Phys.~Rev.}\ }\textbf {\bibinfo {volume} {D89}},\ \bibinfo {pages}
  {062007} (\bibinfo {year} {2014}{\natexlab{a}})},\ \Eprint
  {http://arxiv.org/abs/1311.7048} {arXiv:1311.7048 [astro-ph.HE]} \BibitemShut
  {NoStop}%
%%CITATION = ARXIV:1311.7048;%%
\bibitem [{\citenamefont {Aartsen}\ \emph
  {et~al.}(2014{\natexlab{b}})\citenamefont {Aartsen} \emph
  {et~al.}}]{Aartsen:2014gkd}%
  \BibitemOpen
  \bibfield  {author} {\bibinfo {author} {\bibfnamefont {M.G.}\ \bibnamefont
  {Aartsen}} \emph {et~al.} (\bibinfo {collaboration} {IceCube}),\ }\bibfield
  {title} {\enquote {\bibinfo {title} {{Observation of High-Energy
  Astrophysical Neutrinos in Three Years of IceCube Data}},}\ }\href {\doibase
  10.1103/PhysRevLett.113.101101} {\bibfield  {journal} {\bibinfo  {journal}
  {Phys.~Rev.~Lett.}\ }\textbf {\bibinfo {volume} {113}},\ \bibinfo {pages}
  {101101} (\bibinfo {year} {2014}{\natexlab{b}})},\ \Eprint
  {http://arxiv.org/abs/1405.5303} {arXiv:1405.5303 [astro-ph.HE]} \BibitemShut
  {NoStop}%
%%CITATION = ARXIV:1405.5303;%%
\bibitem [{\citenamefont {Lipari}\ \emph {et~al.}(2007)\citenamefont {Lipari},
  \citenamefont {Lusignoli},\ and\ \citenamefont {Meloni}}]{Lipari:2007su}%
  \BibitemOpen
  \bibfield  {author} {\bibinfo {author} {\bibfnamefont {Paolo}\ \bibnamefont
  {Lipari}}, \bibinfo {author} {\bibfnamefont {Maurizio}\ \bibnamefont
  {Lusignoli}}, \ and\ \bibinfo {author} {\bibfnamefont {Davide}\ \bibnamefont
  {Meloni}},\ }\bibfield  {title} {\enquote {\bibinfo {title} {{Flavor
  Composition and Energy Spectrum of Astrophysical Neutrinos}},}\ }\href
  {\doibase 10.1103/PhysRevD.75.123005} {\bibfield  {journal} {\bibinfo
  {journal} {Phys. Rev.}\ }\textbf {\bibinfo {volume} {D75}},\ \bibinfo {pages}
  {123005} (\bibinfo {year} {2007})},\ \Eprint {http://arxiv.org/abs/0704.0718}
  {arXiv:0704.0718 [astro-ph]} \BibitemShut {NoStop}%
%%CITATION = 0704.0718;%%
\bibitem [{\citenamefont {Murase}\ and\ \citenamefont
  {Ioka}(2013)}]{Murase:2013ffa}%
  \BibitemOpen
  \bibfield  {author} {\bibinfo {author} {\bibfnamefont {Kohta}\ \bibnamefont
  {Murase}}\ and\ \bibinfo {author} {\bibfnamefont {Kunihito}\ \bibnamefont
  {Ioka}},\ }\bibfield  {title} {\enquote {\bibinfo {title} {{TeV--PeV
  Neutrinos from Low-Power Gamma-Ray Burst Jets inside Stars}},}\ }\href
  {\doibase 10.1103/PhysRevLett.111.121102} {\bibfield  {journal} {\bibinfo
  {journal} {Phys.~Rev.~Lett.}\ }\textbf {\bibinfo {volume} {111}},\ \bibinfo
  {pages} {121102} (\bibinfo {year} {2013})},\ \Eprint
  {http://arxiv.org/abs/1306.2274} {arXiv:1306.2274 [astro-ph.HE]} \BibitemShut
  {NoStop}%
%%CITATION = ARXIV:1306.2274;%%
\bibitem [{\citenamefont {Laha}\ \emph {et~al.}(2013)\citenamefont {Laha},
  \citenamefont {Beacom}, \citenamefont {Dasgupta}, \citenamefont {Horiuchi},\
  and\ \citenamefont {Murase}}]{Laha:2013lka}%
  \BibitemOpen
  \bibfield  {author} {\bibinfo {author} {\bibfnamefont {Ranjan}\ \bibnamefont
  {Laha}}, \bibinfo {author} {\bibfnamefont {John~F.}\ \bibnamefont {Beacom}},
  \bibinfo {author} {\bibfnamefont {Basudeb}\ \bibnamefont {Dasgupta}},
  \bibinfo {author} {\bibfnamefont {Shunsaku}\ \bibnamefont {Horiuchi}}, \ and\
  \bibinfo {author} {\bibfnamefont {Kohta}\ \bibnamefont {Murase}},\ }\bibfield
   {title} {\enquote {\bibinfo {title} {{Demystifying the PeV Cascades in
  IceCube: Less (Energy) is More (Events)}},}\ }\href {\doibase
  10.1103/PhysRevD.88.043009} {\bibfield  {journal} {\bibinfo  {journal}
  {Phys.~Rev.}\ }\textbf {\bibinfo {volume} {D88}},\ \bibinfo {pages} {043009}
  (\bibinfo {year} {2013})},\ \Eprint {http://arxiv.org/abs/1306.2309}
  {arXiv:1306.2309 [astro-ph.HE]} \BibitemShut {NoStop}%
%%CITATION = ARXIV:1306.2309;%%
\bibitem [{\citenamefont {Murase}\ \emph {et~al.}(2013)\citenamefont {Murase},
  \citenamefont {Ahlers},\ and\ \citenamefont {Lacki}}]{Murase:2013rfa}%
  \BibitemOpen
  \bibfield  {author} {\bibinfo {author} {\bibfnamefont {Kohta}\ \bibnamefont
  {Murase}}, \bibinfo {author} {\bibfnamefont {Markus}\ \bibnamefont {Ahlers}},
  \ and\ \bibinfo {author} {\bibfnamefont {Brian~C.}\ \bibnamefont {Lacki}},\
  }\bibfield  {title} {\enquote {\bibinfo {title} {{Testing the Hadronuclear
  Origin of PeV Neutrinos Observed with IceCube}},}\ }\href {\doibase
  10.1103/PhysRevD.88.121301} {\bibfield  {journal} {\bibinfo  {journal}
  {Phys.~Rev.}\ }\textbf {\bibinfo {volume} {D88}},\ \bibinfo {pages} {121301}
  (\bibinfo {year} {2013})},\ \Eprint {http://arxiv.org/abs/1306.3417}
  {arXiv:1306.3417 [astro-ph.HE]} \BibitemShut {NoStop}%
%%CITATION = ARXIV:1306.3417;%%
\bibitem [{\citenamefont {Winter}(2013)}]{Winter:2013cla}%
  \BibitemOpen
  \bibfield  {author} {\bibinfo {author} {\bibfnamefont {Walter}\ \bibnamefont
  {Winter}},\ }\bibfield  {title} {\enquote {\bibinfo {title} {{Photohadronic
  Origin of the TeV-PeV Neutrinos Observed in IceCube}},}\ }\href {\doibase
  10.1103/PhysRevD.88.083007} {\bibfield  {journal} {\bibinfo  {journal}
  {Phys.~Rev.}\ }\textbf {\bibinfo {volume} {D88}},\ \bibinfo {pages} {083007}
  (\bibinfo {year} {2013})},\ \Eprint {http://arxiv.org/abs/1307.2793}
  {arXiv:1307.2793 [astro-ph.HE]} \BibitemShut {NoStop}%
%%CITATION = ARXIV:1307.2793;%%
\bibitem [{\citenamefont {Esmaili}\ and\ \citenamefont
  {Serpico}(2013)}]{Esmaili:2013gha}%
  \BibitemOpen
  \bibfield  {author} {\bibinfo {author} {\bibfnamefont {Arman}\ \bibnamefont
  {Esmaili}}\ and\ \bibinfo {author} {\bibfnamefont {Pasquale~Dario}\
  \bibnamefont {Serpico}},\ }\bibfield  {title} {\enquote {\bibinfo {title}
  {{Are IceCube neutrinos unveiling PeV-scale decaying dark matter?}}}\ }\href
  {\doibase 10.1088/1475-7516/2013/11/054} {\bibfield  {journal} {\bibinfo
  {journal} {JCAP}\ }\textbf {\bibinfo {volume} {1311}},\ \bibinfo {pages}
  {054} (\bibinfo {year} {2013})},\ \Eprint {http://arxiv.org/abs/1308.1105}
  {arXiv:1308.1105 [hep-ph]} \BibitemShut {NoStop}%
%%CITATION = ARXIV:1308.1105;%%
\bibitem [{\citenamefont {Razzaque}(2013)}]{Razzaque:2013uoa}%
  \BibitemOpen
  \bibfield  {author} {\bibinfo {author} {\bibfnamefont {Soebur}\ \bibnamefont
  {Razzaque}},\ }\bibfield  {title} {\enquote {\bibinfo {title} {{The Galactic
  Center Origin of a Subset of IceCube Neutrino Events}},}\ }\href {\doibase
  10.1103/PhysRevD.88.081302} {\bibfield  {journal} {\bibinfo  {journal}
  {Phys.~Rev.}\ }\textbf {\bibinfo {volume} {D88}},\ \bibinfo {pages} {081302}
  (\bibinfo {year} {2013})},\ \Eprint {http://arxiv.org/abs/1309.2756}
  {arXiv:1309.2756 [astro-ph.HE]} \BibitemShut {NoStop}%
%%CITATION = ARXIV:1309.2756;%%
\bibitem [{\citenamefont {Ahlers}\ and\ \citenamefont
  {Murase}(2014)}]{Ahlers:2013xia}%
  \BibitemOpen
  \bibfield  {author} {\bibinfo {author} {\bibfnamefont {Markus}\ \bibnamefont
  {Ahlers}}\ and\ \bibinfo {author} {\bibfnamefont {Kohta}\ \bibnamefont
  {Murase}},\ }\bibfield  {title} {\enquote {\bibinfo {title} {{Probing the
  Galactic Origin of the IceCube Excess with Gamma-Rays}},}\ }\href {\doibase
  10.1103/PhysRevD.90.023010} {\bibfield  {journal} {\bibinfo  {journal}
  {Phys.~Rev.}\ }\textbf {\bibinfo {volume} {D90}},\ \bibinfo {pages} {023010}
  (\bibinfo {year} {2014})},\ \Eprint {http://arxiv.org/abs/1309.4077}
  {arXiv:1309.4077 [astro-ph.HE]} \BibitemShut {NoStop}%
%%CITATION = ARXIV:1309.4077;%%
\bibitem [{\citenamefont {Liu}\ \emph {et~al.}(2014)\citenamefont {Liu},
  \citenamefont {Wang}, \citenamefont {Inoue}, \citenamefont {Crocker},\ and\
  \citenamefont {Aharonian}}]{Liu:2013wia}%
  \BibitemOpen
  \bibfield  {author} {\bibinfo {author} {\bibfnamefont {Ruo-Yu}\ \bibnamefont
  {Liu}}, \bibinfo {author} {\bibfnamefont {Xiang-Yu}\ \bibnamefont {Wang}},
  \bibinfo {author} {\bibfnamefont {Susumu}\ \bibnamefont {Inoue}}, \bibinfo
  {author} {\bibfnamefont {Roland}\ \bibnamefont {Crocker}}, \ and\ \bibinfo
  {author} {\bibfnamefont {Felix}\ \bibnamefont {Aharonian}},\ }\bibfield
  {title} {\enquote {\bibinfo {title} {{Diffuse PeV neutrinos from EeV cosmic
  ray sources: Semirelativistic hypernova remnants in star-forming
  galaxies}},}\ }\href {\doibase 10.1103/PhysRevD.89.083004} {\bibfield
  {journal} {\bibinfo  {journal} {Phys.~Rev.}\ }\textbf {\bibinfo {volume}
  {D89}},\ \bibinfo {pages} {083004} (\bibinfo {year} {2014})},\ \Eprint
  {http://arxiv.org/abs/1310.1263} {arXiv:1310.1263 [astro-ph.HE]} \BibitemShut
  {NoStop}%
%%CITATION = ARXIV:1310.1263;%%
\bibitem [{\citenamefont {Lunardini}\ \emph {et~al.}(2014)\citenamefont
  {Lunardini}, \citenamefont {Razzaque}, \citenamefont {Theodoseau},\ and\
  \citenamefont {Yang}}]{Lunardini:2013gva}%
  \BibitemOpen
  \bibfield  {author} {\bibinfo {author} {\bibfnamefont {Cecilia}\ \bibnamefont
  {Lunardini}}, \bibinfo {author} {\bibfnamefont {Soebur}\ \bibnamefont
  {Razzaque}}, \bibinfo {author} {\bibfnamefont {Kristopher~T.}\ \bibnamefont
  {Theodoseau}}, \ and\ \bibinfo {author} {\bibfnamefont {Lili}\ \bibnamefont
  {Yang}},\ }\bibfield  {title} {\enquote {\bibinfo {title} {{Neutrino Events
  at IceCube and the Fermi Bubbles}},}\ }\href {\doibase
  10.1103/PhysRevD.90.023016} {\bibfield  {journal} {\bibinfo  {journal}
  {Phys.~Rev.}\ }\textbf {\bibinfo {volume} {D90}},\ \bibinfo {pages} {023016}
  (\bibinfo {year} {2014})},\ \Eprint {http://arxiv.org/abs/1311.7188}
  {arXiv:1311.7188 [astro-ph.HE]} \BibitemShut {NoStop}%
%%CITATION = ARXIV:1311.7188;%%
\bibitem [{\citenamefont {Kashiyama}\ and\ \citenamefont
  {Meszaros}(2014)}]{Kashiyama:2014rza}%
  \BibitemOpen
  \bibfield  {author} {\bibinfo {author} {\bibfnamefont {Kazumi}\ \bibnamefont
  {Kashiyama}}\ and\ \bibinfo {author} {\bibfnamefont {Peter}\ \bibnamefont
  {Meszaros}},\ }\bibfield  {title} {\enquote {\bibinfo {title} {{Galaxy
  Mergers as a Source of Cosmic Rays, Neutrinos, and Gamma Rays}},}\ }\href
  {\doibase 10.1088/2041-8205/790/1/L14} {\bibfield  {journal} {\bibinfo
  {journal} {Astrophys.~J.}\ }\textbf {\bibinfo {volume} {790}},\ \bibinfo
  {pages} {L14} (\bibinfo {year} {2014})},\ \Eprint
  {http://arxiv.org/abs/1405.3262} {arXiv:1405.3262 [astro-ph.HE]} \BibitemShut
  {NoStop}%
%%CITATION = ARXIV:1405.3262;%%
\bibitem [{\citenamefont {Chang}\ and\ \citenamefont
  {Wang}(2014)}]{Chang:2014hua}%
  \BibitemOpen
  \bibfield  {author} {\bibinfo {author} {\bibfnamefont {Xiao-Chuan}\
  \bibnamefont {Chang}}\ and\ \bibinfo {author} {\bibfnamefont {Xiang-Yu}\
  \bibnamefont {Wang}},\ }\bibfield  {title} {\enquote {\bibinfo {title} {{The
  diffuse gamma-ray flux associated with sub-PeV/PeV neutrinos from starburst
  galaxies}},}\ }\href {\doibase 10.1088/0004-637X/793/2/131} {\bibfield
  {journal} {\bibinfo  {journal} {Astrophys.~J.}\ }\textbf {\bibinfo {volume}
  {793}},\ \bibinfo {pages} {131} (\bibinfo {year} {2014})},\ \Eprint
  {http://arxiv.org/abs/1406.1099} {arXiv:1406.1099 [astro-ph.HE]} \BibitemShut
  {NoStop}%
%%CITATION = ARXIV:1406.1099;%%
\bibitem [{\citenamefont {Aartsen}\ \emph
  {et~al.}(2014{\natexlab{c}})\citenamefont {Aartsen} \emph
  {et~al.}}]{Aartsen:2014cva}%
  \BibitemOpen
  \bibfield  {author} {\bibinfo {author} {\bibfnamefont {M.G.}\ \bibnamefont
  {Aartsen}} \emph {et~al.} (\bibinfo {collaboration} {IceCube}),\ }\bibfield
  {title} {\enquote {\bibinfo {title} {{Searches for Extended and Point-like
  Neutrino Sources with Four Years of IceCube Data}},}\ }\href {\doibase
  10.1088/0004-637X/796/2/109} {\bibfield  {journal} {\bibinfo  {journal}
  {Astrophys.~J.}\ }\textbf {\bibinfo {volume} {796}},\ \bibinfo {pages} {109}
  (\bibinfo {year} {2014}{\natexlab{c}})},\ \Eprint
  {http://arxiv.org/abs/1406.6757} {arXiv:1406.6757 [astro-ph.HE]} \BibitemShut
  {NoStop}%
%%CITATION = ARXIV:1406.6757;%%
\bibitem [{\citenamefont {Bhattacharya}\ \emph
  {et~al.}(2015{\natexlab{a}})\citenamefont {Bhattacharya}, \citenamefont
  {Enberg}, \citenamefont {Reno},\ and\ \citenamefont
  {Sarcevic}}]{Bhattacharya:2014sta}%
  \BibitemOpen
  \bibfield  {author} {\bibinfo {author} {\bibfnamefont {Atri}\ \bibnamefont
  {Bhattacharya}}, \bibinfo {author} {\bibfnamefont {Rikard}\ \bibnamefont
  {Enberg}}, \bibinfo {author} {\bibfnamefont {Mary~Hall}\ \bibnamefont
  {Reno}}, \ and\ \bibinfo {author} {\bibfnamefont {Ina}\ \bibnamefont
  {Sarcevic}},\ }\bibfield  {title} {\enquote {\bibinfo {title} {{Charm decay
  in slow-jet supernovae as the origin of the IceCube ultra-high energy
  neutrino events}},}\ }\href {\doibase 10.1088/1475-7516/2015/06/034}
  {\bibfield  {journal} {\bibinfo  {journal} {JCAP}\ }\textbf {\bibinfo
  {volume} {1506}},\ \bibinfo {pages} {034} (\bibinfo {year}
  {2015}{\natexlab{a}})},\ \Eprint {http://arxiv.org/abs/1407.2985}
  {arXiv:1407.2985 [astro-ph.HE]} \BibitemShut {NoStop}%
%%CITATION = ARXIV:1407.2985;%%
\bibitem [{\citenamefont {Bhattacharya}\ \emph
  {et~al.}(2015{\natexlab{b}})\citenamefont {Bhattacharya}, \citenamefont
  {Gandhi},\ and\ \citenamefont {Gupta}}]{Bhattacharya:2014yha}%
  \BibitemOpen
  \bibfield  {author} {\bibinfo {author} {\bibfnamefont {Atri}\ \bibnamefont
  {Bhattacharya}}, \bibinfo {author} {\bibfnamefont {Raj}\ \bibnamefont
  {Gandhi}}, \ and\ \bibinfo {author} {\bibfnamefont {Aritra}\ \bibnamefont
  {Gupta}},\ }\bibfield  {title} {\enquote {\bibinfo {title} {{The Direct
  Detection of Boosted Dark Matter at High Energies and PeV events at
  IceCube}},}\ }\href {\doibase 10.1088/1475-7516/2015/03/027} {\bibfield
  {journal} {\bibinfo  {journal} {JCAP}\ }\textbf {\bibinfo {volume} {1503}},\
  \bibinfo {pages} {027} (\bibinfo {year} {2015}{\natexlab{b}})},\ \Eprint
  {http://arxiv.org/abs/1407.3280} {arXiv:1407.3280 [hep-ph]} \BibitemShut
  {NoStop}%
%%CITATION = ARXIV:1407.3280;%%
\bibitem [{\citenamefont {Aartsen}\ \emph
  {et~al.}(2015{\natexlab{a}})\citenamefont {Aartsen} \emph
  {et~al.}}]{Aartsen:2014ivk}%
  \BibitemOpen
  \bibfield  {author} {\bibinfo {author} {\bibfnamefont {M.G.}\ \bibnamefont
  {Aartsen}} \emph {et~al.} (\bibinfo {collaboration} {IceCube}),\ }\bibfield
  {title} {\enquote {\bibinfo {title} {{Searches for small-scale anisotropies
  from neutrino point sources with three years of IceCube data}},}\ }\href
  {\doibase 10.1016/j.astropartphys.2015.01.001} {\bibfield  {journal}
  {\bibinfo  {journal} {Astropart.~Phys.}\ }\textbf {\bibinfo {volume} {66}},\
  \bibinfo {pages} {39--52} (\bibinfo {year} {2015}{\natexlab{a}})},\ \Eprint
  {http://arxiv.org/abs/1408.0634} {arXiv:1408.0634 [astro-ph.HE]} \BibitemShut
  {NoStop}%
%%CITATION = ARXIV:1408.0634;%%
\bibitem [{\citenamefont {Murase}(2015)}]{Murase:2014tsa}%
  \BibitemOpen
  \bibfield  {author} {\bibinfo {author} {\bibfnamefont {Kohta}\ \bibnamefont
  {Murase}},\ }\bibfield  {title} {\enquote {\bibinfo {title} {{On the Origin
  of High-Energy Cosmic Neutrinos}},}\ }\bibfield  {booktitle} {\emph {\bibinfo
  {booktitle} {{Proceedings of the 26th International Conference on Neutrino
  Physics and Astrophysics (Neutrino 2014)}}},\ }\href {\doibase
  10.1063/1.4915555} {\bibfield  {journal} {\bibinfo  {journal} {AIP Conf.
  Proc.}\ }\textbf {\bibinfo {volume} {1666}},\ \bibinfo {pages} {040006}
  (\bibinfo {year} {2015})},\ \Eprint {http://arxiv.org/abs/1410.3680}
  {arXiv:1410.3680 [hep-ph]} \BibitemShut {NoStop}%
%%CITATION = ARXIV:1410.3680;%%
\bibitem [{\citenamefont {Anchordoqui}(2015)}]{Anchordoqui:2014pca}%
  \BibitemOpen
  \bibfield  {author} {\bibinfo {author} {\bibfnamefont {Luis~A.}\ \bibnamefont
  {Anchordoqui}},\ }\bibfield  {title} {\enquote {\bibinfo {title} {{Neutron
  $\beta$-decay as the origin of IceCube’s PeV (anti)neutrinos}},}\ }\href
  {\doibase 10.1103/PhysRevD.91.027301} {\bibfield  {journal} {\bibinfo
  {journal} {Phys.~Rev.}\ }\textbf {\bibinfo {volume} {D91}},\ \bibinfo {pages}
  {027301} (\bibinfo {year} {2015})},\ \Eprint {http://arxiv.org/abs/1411.6457}
  {arXiv:1411.6457 [astro-ph.HE]} \BibitemShut {NoStop}%
%%CITATION = ARXIV:1411.6457;%%
\bibitem [{\citenamefont {Aartsen}\ \emph
  {et~al.}(2015{\natexlab{b}})\citenamefont {Aartsen} \emph
  {et~al.}}]{Aartsen:2014aqy}%
  \BibitemOpen
  \bibfield  {author} {\bibinfo {author} {\bibfnamefont {M.G.}\ \bibnamefont
  {Aartsen}} \emph {et~al.} (\bibinfo {collaboration} {IceCube}),\ }\bibfield
  {title} {\enquote {\bibinfo {title} {{Search for Prompt Neutrino Emission
  from Gamma-Ray Bursts with IceCube}},}\ }\href {\doibase
  10.1088/2041-8205/805/1/L5} {\bibfield  {journal} {\bibinfo  {journal}
  {Astrophys.~J.}\ }\textbf {\bibinfo {volume} {805}},\ \bibinfo {pages} {L5}
  (\bibinfo {year} {2015}{\natexlab{b}})},\ \Eprint
  {http://arxiv.org/abs/1412.6510} {arXiv:1412.6510 [astro-ph.HE]} \BibitemShut
  {NoStop}%
%%CITATION = ARXIV:1412.6510;%%
\bibitem [{\citenamefont {Ahlers}\ \emph {et~al.}(2015)\citenamefont {Ahlers},
  \citenamefont {Bai}, \citenamefont {Barger},\ and\ \citenamefont
  {Lu}}]{Ahlers:2015moa}%
  \BibitemOpen
  \bibfield  {author} {\bibinfo {author} {\bibfnamefont {Markus}\ \bibnamefont
  {Ahlers}}, \bibinfo {author} {\bibfnamefont {Yang}\ \bibnamefont {Bai}},
  \bibinfo {author} {\bibfnamefont {Vernon}\ \bibnamefont {Barger}}, \ and\
  \bibinfo {author} {\bibfnamefont {Ran}\ \bibnamefont {Lu}},\ }\bibfield
  {title} {\enquote {\bibinfo {title} {{Galactic TeV-PeV Neutrinos}},}\
  }\href@noop {} {\  (\bibinfo {year} {2015})},\ \Eprint
  {http://arxiv.org/abs/1505.03156} {arXiv:1505.03156 [hep-ph]} \BibitemShut
  {NoStop}%
%%CITATION = ARXIV:1505.03156;%%
\bibitem [{\citenamefont {Barenboim}\ and\ \citenamefont
  {Quigg}(2003)}]{Barenboim:2003jm}%
  \BibitemOpen
  \bibfield  {author} {\bibinfo {author} {\bibfnamefont {Gabriela}\
  \bibnamefont {Barenboim}}\ and\ \bibinfo {author} {\bibfnamefont {Chris}\
  \bibnamefont {Quigg}},\ }\bibfield  {title} {\enquote {\bibinfo {title}
  {Neutrino observatories can characterize cosmic sources and neutrino
  properties},}\ }\href@noop {} {\bibfield  {journal} {\bibinfo  {journal}
  {Phys. Rev.}\ }\textbf {\bibinfo {volume} {D67}},\ \bibinfo {pages} {073024}
  (\bibinfo {year} {2003})},\ \Eprint {http://arxiv.org/abs/hep-ph/0301220}
  {hep-ph/0301220} \BibitemShut {NoStop}%
%%CITATION = HEP-PH 0301220;%%
\bibitem [{\citenamefont {Xing}\ and\ \citenamefont
  {Zhou}(2006)}]{Xing:2006uk}%
  \BibitemOpen
  \bibfield  {author} {\bibinfo {author} {\bibfnamefont {Zhi-zhong}\
  \bibnamefont {Xing}}\ and\ \bibinfo {author} {\bibfnamefont {Shun}\
  \bibnamefont {Zhou}},\ }\bibfield  {title} {\enquote {\bibinfo {title}
  {Towards determination of the initial flavor composition of ultrahigh-energy
  neutrino fluxes with neutrino telescopes},}\ }\href@noop {} {\bibfield
  {journal} {\bibinfo  {journal} {Phys. Rev.}\ }\textbf {\bibinfo {volume}
  {D74}},\ \bibinfo {pages} {013010} (\bibinfo {year} {2006})},\ \Eprint
  {http://arxiv.org/abs/astro-ph/0603781} {astro-ph/0603781} \BibitemShut
  {NoStop}%
%%CITATION = ASTRO-PH/0603781;%%
\bibitem [{\citenamefont {Pakvasa}\ \emph {et~al.}(2008)\citenamefont
  {Pakvasa}, \citenamefont {Rodejohann},\ and\ \citenamefont
  {Weiler}}]{Pakvasa:2007dc}%
  \BibitemOpen
  \bibfield  {author} {\bibinfo {author} {\bibfnamefont {Sandip}\ \bibnamefont
  {Pakvasa}}, \bibinfo {author} {\bibfnamefont {Werner}\ \bibnamefont
  {Rodejohann}}, \ and\ \bibinfo {author} {\bibfnamefont {Thomas~J.}\
  \bibnamefont {Weiler}},\ }\bibfield  {title} {\enquote {\bibinfo {title}
  {{Flavor Ratios of Astrophysical Neutrinos: Implications for Precision
  Measurements}},}\ }\href {\doibase 10.1088/1126-6708/2008/02/005} {\bibfield
  {journal} {\bibinfo  {journal} {JHEP}\ }\textbf {\bibinfo {volume} {02}},\
  \bibinfo {pages} {005} (\bibinfo {year} {2008})},\ \Eprint
  {http://arxiv.org/abs/0711.4517} {arXiv:0711.4517 [hep-ph]} \BibitemShut
  {NoStop}%
%%CITATION = 0711.4517;%%
\bibitem [{\citenamefont {Esmaili}\ and\ \citenamefont
  {Farzan}(2009)}]{Esmaili:2009dz}%
  \BibitemOpen
  \bibfield  {author} {\bibinfo {author} {\bibfnamefont {Arman}\ \bibnamefont
  {Esmaili}}\ and\ \bibinfo {author} {\bibfnamefont {Yasaman}\ \bibnamefont
  {Farzan}},\ }\bibfield  {title} {\enquote {\bibinfo {title} {{An Analysis of
  Cosmic Neutrinos: Flavor Composition at Source and Neutrino Mixing
  Parameters}},}\ }\href {\doibase 10.1016/j.nuclphysb.2009.06.017} {\bibfield
  {journal} {\bibinfo  {journal} {Nucl.~Phys.}\ }\textbf {\bibinfo {volume}
  {B821}},\ \bibinfo {pages} {197--214} (\bibinfo {year} {2009})},\ \Eprint
  {http://arxiv.org/abs/0905.0259} {arXiv:0905.0259 [hep-ph]} \BibitemShut
  {NoStop}%
%%CITATION = 0905.0259;%%
\bibitem [{\citenamefont {Choubey}\ and\ \citenamefont
  {Rodejohann}(2009)}]{Choubey:2009jq}%
  \BibitemOpen
  \bibfield  {author} {\bibinfo {author} {\bibfnamefont {Sandhya}\ \bibnamefont
  {Choubey}}\ and\ \bibinfo {author} {\bibfnamefont {Werner}\ \bibnamefont
  {Rodejohann}},\ }\bibfield  {title} {\enquote {\bibinfo {title} {{Flavor
  Composition of UHE Neutrinos at Source and at Neutrino Telescopes}},}\ }\href
  {\doibase 10.1103/PhysRevD.80.113006} {\bibfield  {journal} {\bibinfo
  {journal} {Phys. Rev.}\ }\textbf {\bibinfo {volume} {D80}},\ \bibinfo {pages}
  {113006} (\bibinfo {year} {2009})},\ \Eprint {http://arxiv.org/abs/0909.1219}
  {arXiv:0909.1219 [hep-ph]} \BibitemShut {NoStop}%
\bibitem [{\citenamefont {Lai}\ \emph {et~al.}(2009)\citenamefont {Lai},
  \citenamefont {Lin},\ and\ \citenamefont {Liu}}]{Lai:2009ke}%
  \BibitemOpen
  \bibfield  {author} {\bibinfo {author} {\bibfnamefont {Kwang-Chang}\
  \bibnamefont {Lai}}, \bibinfo {author} {\bibfnamefont {Guey-Lin}\
  \bibnamefont {Lin}}, \ and\ \bibinfo {author} {\bibfnamefont {T.C.}\
  \bibnamefont {Liu}},\ }\bibfield  {title} {\enquote {\bibinfo {title}
  {{Determination of the Neutrino Flavor Ratio at the Astrophysical Source}},}\
  }\href {\doibase 10.1103/PhysRevD.80.103005} {\bibfield  {journal} {\bibinfo
  {journal} {Phys.~Rev.}\ }\textbf {\bibinfo {volume} {D80}},\ \bibinfo {pages}
  {103005} (\bibinfo {year} {2009})},\ \Eprint {http://arxiv.org/abs/0905.4003}
  {arXiv:0905.4003 [hep-ph]} \BibitemShut {NoStop}%
%%CITATION = ARXIV:0905.4003;%%
\bibitem [{\citenamefont {Mena}\ \emph {et~al.}(2014)\citenamefont {Mena},
  \citenamefont {Palomares-Ruiz},\ and\ \citenamefont
  {Vincent}}]{Mena:2014sja}%
  \BibitemOpen
  \bibfield  {author} {\bibinfo {author} {\bibfnamefont {Olga}\ \bibnamefont
  {Mena}}, \bibinfo {author} {\bibfnamefont {Sergio}\ \bibnamefont
  {Palomares-Ruiz}}, \ and\ \bibinfo {author} {\bibfnamefont {Aaron~C.}\
  \bibnamefont {Vincent}},\ }\bibfield  {title} {\enquote {\bibinfo {title}
  {{Flavor Composition of the High-Energy Neutrino Events in IceCube}},}\
  }\href {\doibase 10.1103/PhysRevLett.113.091103} {\bibfield  {journal}
  {\bibinfo  {journal} {Phys.~Rev.~Lett.}\ }\textbf {\bibinfo {volume} {113}},\
  \bibinfo {pages} {091103} (\bibinfo {year} {2014})},\ \Eprint
  {http://arxiv.org/abs/1404.0017} {arXiv:1404.0017 [astro-ph.HE]} \BibitemShut
  {NoStop}%
%%CITATION = ARXIV:1404.0017;%%
\bibitem [{\citenamefont {Xu}\ \emph {et~al.}(2014)\citenamefont {Xu},
  \citenamefont {He},\ and\ \citenamefont {Rodejohann}}]{Xu:2014via}%
  \BibitemOpen
  \bibfield  {author} {\bibinfo {author} {\bibfnamefont {Xun-Jie}\ \bibnamefont
  {Xu}}, \bibinfo {author} {\bibfnamefont {Hong-Jian}\ \bibnamefont {He}}, \
  and\ \bibinfo {author} {\bibfnamefont {Werner}\ \bibnamefont {Rodejohann}},\
  }\bibfield  {title} {\enquote {\bibinfo {title} {{Constraining Astrophysical
  Neutrino Flavor Composition from Leptonic Unitarity}},}\ }\href {\doibase
  10.1088/1475-7516/2014/12/039} {\bibfield  {journal} {\bibinfo  {journal}
  {JCAP}\ }\textbf {\bibinfo {volume} {1412}},\ \bibinfo {pages} {039}
  (\bibinfo {year} {2014})},\ \Eprint {http://arxiv.org/abs/1407.3736}
  {arXiv:1407.3736 [hep-ph]} \BibitemShut {NoStop}%
%%CITATION = ARXIV:1407.3736;%%
\bibitem [{\citenamefont {Fu}\ \emph {et~al.}(2015)\citenamefont {Fu},
  \citenamefont {Ho},\ and\ \citenamefont {Weiler}}]{Fu:2014isa}%
  \BibitemOpen
  \bibfield  {author} {\bibinfo {author} {\bibfnamefont {Lingjun}\ \bibnamefont
  {Fu}}, \bibinfo {author} {\bibfnamefont {Chiu~Man}\ \bibnamefont {Ho}}, \
  and\ \bibinfo {author} {\bibfnamefont {Thomas~J.}\ \bibnamefont {Weiler}},\
  }\bibfield  {title} {\enquote {\bibinfo {title} {{Aspects of the Flavor
  Triangle for Cosmic Neutrino Propagation}},}\ }\href {\doibase
  10.1103/PhysRevD.91.053001} {\bibfield  {journal} {\bibinfo  {journal}
  {Phys.~Rev.}\ }\textbf {\bibinfo {volume} {D91}},\ \bibinfo {pages} {053001}
  (\bibinfo {year} {2015})},\ \Eprint {http://arxiv.org/abs/1411.1174}
  {arXiv:1411.1174 [hep-ph]} \BibitemShut {NoStop}%
%%CITATION = ARXIV:1411.1174;%%
\bibitem [{\citenamefont {Palomares-Ruiz}\ \emph {et~al.}(2015)\citenamefont
  {Palomares-Ruiz}, \citenamefont {Vincent},\ and\ \citenamefont
  {Mena}}]{Palomares-Ruiz:2015mka}%
  \BibitemOpen
  \bibfield  {author} {\bibinfo {author} {\bibfnamefont {Sergio}\ \bibnamefont
  {Palomares-Ruiz}}, \bibinfo {author} {\bibfnamefont {Aaron~C.}\ \bibnamefont
  {Vincent}}, \ and\ \bibinfo {author} {\bibfnamefont {Olga}\ \bibnamefont
  {Mena}},\ }\bibfield  {title} {\enquote {\bibinfo {title} {{Spectral analysis
  of the high-energy IceCube neutrinos}},}\ }\href {\doibase
  10.1103/PhysRevD.91.103008} {\bibfield  {journal} {\bibinfo  {journal}
  {Phys.~Rev.}\ }\textbf {\bibinfo {volume} {D91}},\ \bibinfo {pages} {103008}
  (\bibinfo {year} {2015})},\ \Eprint {http://arxiv.org/abs/1502.02649}
  {arXiv:1502.02649 [astro-ph.HE]} \BibitemShut {NoStop}%
%%CITATION = ARXIV:1502.02649;%%
\bibitem [{\citenamefont {Aartsen}\ \emph
  {et~al.}(2015{\natexlab{c}})\citenamefont {Aartsen} \emph
  {et~al.}}]{Aartsen:2015ivb}%
  \BibitemOpen
  \bibfield  {author} {\bibinfo {author} {\bibfnamefont {M.G.}\ \bibnamefont
  {Aartsen}} \emph {et~al.} (\bibinfo {collaboration} {IceCube}),\ }\bibfield
  {title} {\enquote {\bibinfo {title} {{Flavor Ratio of Astrophysical Neutrinos
  above 35 TeV in IceCube}},}\ }\href {\doibase 10.1103/PhysRevLett.114.171102}
  {\bibfield  {journal} {\bibinfo  {journal} {Phys.~Rev.~Lett.}\ }\textbf
  {\bibinfo {volume} {114}},\ \bibinfo {pages} {171102} (\bibinfo {year}
  {2015}{\natexlab{c}})},\ \Eprint {http://arxiv.org/abs/1502.03376}
  {arXiv:1502.03376 [astro-ph.HE]} \BibitemShut {NoStop}%
%%CITATION = ARXIV:1502.03376;%%
\bibitem [{\citenamefont {Palladino}\ and\ \citenamefont
  {Vissani}(2015)}]{Palladino:2015vna}%
  \BibitemOpen
  \bibfield  {author} {\bibinfo {author} {\bibfnamefont {Andrea}\ \bibnamefont
  {Palladino}}\ and\ \bibinfo {author} {\bibfnamefont {Francesco}\ \bibnamefont
  {Vissani}},\ }\bibfield  {title} {\enquote {\bibinfo {title} {{New
  parametrization of cosmic neutrino oscillations}},}\ }\href@noop {} {\
  (\bibinfo {year} {2015})},\ \Eprint {http://arxiv.org/abs/1504.05238}
  {arXiv:1504.05238 [hep-ph]} \BibitemShut {NoStop}%
%%CITATION = ARXIV:1504.05238;%%
\bibitem [{\citenamefont {Aartsen}\ \emph
  {et~al.}(2015{\natexlab{d}})\citenamefont {Aartsen} \emph
  {et~al.}}]{Aartsen:2015ita}%
  \BibitemOpen
  \bibfield  {author} {\bibinfo {author} {\bibfnamefont {M.~G.}\ \bibnamefont
  {Aartsen}} \emph {et~al.} (\bibinfo {collaboration} {IceCube}),\ }\bibfield
  {title} {\enquote {\bibinfo {title} {{A combined maximum-likelihood analysis
  of the high-energy astrophysical neutrino flux measured with IceCube}},}\
  }\href {\doibase 10.1088/0004-637X/809/1/98} {\bibfield  {journal} {\bibinfo
  {journal} {Astrophys.~J.}\ }\textbf {\bibinfo {volume} {809}},\ \bibinfo
  {pages} {98} (\bibinfo {year} {2015}{\natexlab{d}})},\ \Eprint
  {http://arxiv.org/abs/1507.03991} {arXiv:1507.03991 [astro-ph.HE]}
  \BibitemShut {NoStop}%
%%CITATION = ARXIV:1507.03991;%%
\bibitem [{\citenamefont {Winter}(2014)}]{Winter:2014pya}%
  \BibitemOpen
  \bibfield  {author} {\bibinfo {author} {\bibfnamefont {Walter}\ \bibnamefont
  {Winter}},\ }\bibfield  {title} {\enquote {\bibinfo {title} {{Describing the
  Observed Cosmic Neutrinos by Interactions of Nuclei with Matter}},}\ }\href
  {\doibase 10.1103/PhysRevD.90.103003} {\bibfield  {journal} {\bibinfo
  {journal} {Phys.~Rev.}\ }\textbf {\bibinfo {volume} {D90}},\ \bibinfo {pages}
  {103003} (\bibinfo {year} {2014})},\ \Eprint {http://arxiv.org/abs/1407.7536}
  {arXiv:1407.7536 [astro-ph.HE]} \BibitemShut {NoStop}%
%%CITATION = ARXIV:1407.7536;%%
\bibitem [{\citenamefont {Palladino}\ \emph {et~al.}(2015)\citenamefont
  {Palladino}, \citenamefont {Pagliaroli}, \citenamefont {Villante},\ and\
  \citenamefont {Vissani}}]{Palladino:2015zua}%
  \BibitemOpen
  \bibfield  {author} {\bibinfo {author} {\bibfnamefont {A.}~\bibnamefont
  {Palladino}}, \bibinfo {author} {\bibfnamefont {G.}~\bibnamefont
  {Pagliaroli}}, \bibinfo {author} {\bibfnamefont {F. L.}\ \bibnamefont
  {Villante}}, \ and\ \bibinfo {author} {\bibfnamefont {F.}~\bibnamefont
  {Vissani}},\ }\bibfield  {title} {\enquote {\bibinfo {title} {{What is the
  Flavor of the Cosmic Neutrinos Seen by IceCube?}}}\ }\href {\doibase
  10.1103/PhysRevLett.114.171101} {\bibfield  {journal} {\bibinfo  {journal}
  {Phys.~Rev.~Lett.}\ }\textbf {\bibinfo {volume} {114}},\ \bibinfo {pages}
  {171101} (\bibinfo {year} {2015})},\ \Eprint
  {http://arxiv.org/abs/1502.02923} {arXiv:1502.02923 [astro-ph.HE]}
  \BibitemShut {NoStop}%
%%CITATION = ARXIV:1502.02923;%%
\bibitem [{\citenamefont {Aartsen}\ \emph
  {et~al.}(2014{\natexlab{d}})\citenamefont {Aartsen} \emph
  {et~al.}}]{Aartsen:2014njl}%
  \BibitemOpen
  \bibfield  {author} {\bibinfo {author} {\bibfnamefont {M.G.}\ \bibnamefont
  {Aartsen}} \emph {et~al.} (\bibinfo {collaboration} {IceCube}),\ }\bibfield
  {title} {\enquote {\bibinfo {title} {{IceCube-Gen2: A Vision for the Future
  of Neutrino Astronomy in Antarctica}},}\ }\href@noop {} {\  (\bibinfo {year}
  {2014}{\natexlab{d}})},\ \Eprint {http://arxiv.org/abs/1412.5106}
  {arXiv:1412.5106 [astro-ph.HE]} \BibitemShut {NoStop}%
%%CITATION = ARXIV:1412.5106;%%
\bibitem [{\citenamefont {Gandhi}\ \emph {et~al.}(1996)\citenamefont {Gandhi},
  \citenamefont {Quigg}, \citenamefont {Reno},\ and\ \citenamefont
  {Sarcevic}}]{Gandhi:1995tf}%
  \BibitemOpen
  \bibfield  {author} {\bibinfo {author} {\bibfnamefont {Raj}\ \bibnamefont
  {Gandhi}}, \bibinfo {author} {\bibfnamefont {Chris}\ \bibnamefont {Quigg}},
  \bibinfo {author} {\bibfnamefont {Mary~Hall}\ \bibnamefont {Reno}}, \ and\
  \bibinfo {author} {\bibfnamefont {Ina}\ \bibnamefont {Sarcevic}},\ }\bibfield
   {title} {\enquote {\bibinfo {title} {{Ultrahigh-energy neutrino
  interactions}},}\ }\href@noop {} {\bibfield  {journal} {\bibinfo  {journal}
  {Astropart.~Phys.}\ }\textbf {\bibinfo {volume} {5}},\ \bibinfo {pages}
  {81--110} (\bibinfo {year} {1996})},\ \Eprint
  {http://arxiv.org/abs/hep-ph/9512364} {hep-ph/9512364} \BibitemShut {NoStop}%
%%CITATION = HEP-PH/9512364;%%
\bibitem [{\citenamefont {Gandhi}\ \emph {et~al.}(1998)\citenamefont {Gandhi},
  \citenamefont {Quigg}, \citenamefont {Reno},\ and\ \citenamefont
  {Sarcevic}}]{Gandhi:1998ri}%
  \BibitemOpen
  \bibfield  {author} {\bibinfo {author} {\bibfnamefont {Raj}\ \bibnamefont
  {Gandhi}}, \bibinfo {author} {\bibfnamefont {Chris}\ \bibnamefont {Quigg}},
  \bibinfo {author} {\bibfnamefont {Mary~Hall}\ \bibnamefont {Reno}}, \ and\
  \bibinfo {author} {\bibfnamefont {Ina}\ \bibnamefont {Sarcevic}},\ }\bibfield
   {title} {\enquote {\bibinfo {title} {{Neutrino interactions at ultrahigh
  energies}},}\ }\href@noop {} {\bibfield  {journal} {\bibinfo  {journal}
  {Phys. Rev.}\ }\textbf {\bibinfo {volume} {D58}},\ \bibinfo {pages} {093009}
  (\bibinfo {year} {1998})},\ \Eprint {http://arxiv.org/abs/hep-ph/9807264}
  {hep-ph/9807264} \BibitemShut {NoStop}%
%%CITATION = HEP-PH/9807264;%%
\bibitem [{\citenamefont {Anchordoqui}\ \emph
  {et~al.}(2005{\natexlab{a}})\citenamefont {Anchordoqui}, \citenamefont
  {Goldberg}, \citenamefont {Halzen},\ and\ \citenamefont
  {Weiler}}]{Anchordoqui:2004eb}%
  \BibitemOpen
  \bibfield  {author} {\bibinfo {author} {\bibfnamefont {Luis~A.}\ \bibnamefont
  {Anchordoqui}}, \bibinfo {author} {\bibfnamefont {Haim}\ \bibnamefont
  {Goldberg}}, \bibinfo {author} {\bibfnamefont {Francis}\ \bibnamefont
  {Halzen}}, \ and\ \bibinfo {author} {\bibfnamefont {Thomas~J.}\ \bibnamefont
  {Weiler}},\ }\bibfield  {title} {\enquote {\bibinfo {title} {{Neutrinos as a
  diagnostic of high energy astrophysical processes}},}\ }\href {\doibase
  10.1016/j.physletb.2005.06.056} {\bibfield  {journal} {\bibinfo  {journal}
  {Phys.~Lett.}\ }\textbf {\bibinfo {volume} {B621}},\ \bibinfo {pages}
  {18--21} (\bibinfo {year} {2005}{\natexlab{a}})},\ \Eprint
  {http://arxiv.org/abs/hep-ph/0410003} {arXiv:hep-ph/0410003 [hep-ph]}
  \BibitemShut {NoStop}%
%%CITATION = HEP-PH/0410003;%%
\bibitem [{\citenamefont {Bhattacharya}\ \emph {et~al.}(2011)\citenamefont
  {Bhattacharya}, \citenamefont {Gandhi}, \citenamefont {Rodejohann},\ and\
  \citenamefont {Watanabe}}]{Bhattacharya:2011qu}%
  \BibitemOpen
  \bibfield  {author} {\bibinfo {author} {\bibfnamefont {Atri}\ \bibnamefont
  {Bhattacharya}}, \bibinfo {author} {\bibfnamefont {Raj}\ \bibnamefont
  {Gandhi}}, \bibinfo {author} {\bibfnamefont {Werner}\ \bibnamefont
  {Rodejohann}}, \ and\ \bibinfo {author} {\bibfnamefont {Atsushi}\
  \bibnamefont {Watanabe}},\ }\bibfield  {title} {\enquote {\bibinfo {title}
  {{The Glashow resonance at IceCube: signatures, event rates and $pp$ vs.
  $p\gamma$ interactions}},}\ }\href {\doibase 10.1088/1475-7516/2011/10/017}
  {\bibfield  {journal} {\bibinfo  {journal} {JCAP}\ }\textbf {\bibinfo
  {volume} {1110}},\ \bibinfo {pages} {017} (\bibinfo {year} {2011})},\ \Eprint
  {http://arxiv.org/abs/1108.3163} {arXiv:1108.3163 [astro-ph.HE]} \BibitemShut
  {NoStop}%
%%CITATION = ARXIV:1108.3163;%%
\bibitem [{\citenamefont {Barger}\ \emph {et~al.}(2014)\citenamefont {Barger},
  \citenamefont {Fu}, \citenamefont {Learned}, \citenamefont {Marfatia},
  \citenamefont {Pakvasa} \emph {et~al.}}]{Barger:2014iua}%
  \BibitemOpen
  \bibfield  {author} {\bibinfo {author} {\bibfnamefont {V.}~\bibnamefont
  {Barger}}, \bibinfo {author} {\bibfnamefont {Lingjun}\ \bibnamefont {Fu}},
  \bibinfo {author} {\bibfnamefont {J.G.}\ \bibnamefont {Learned}}, \bibinfo
  {author} {\bibfnamefont {D.}~\bibnamefont {Marfatia}}, \bibinfo {author}
  {\bibfnamefont {S.}~\bibnamefont {Pakvasa}},  \emph {et~al.},\ }\bibfield
  {title} {\enquote {\bibinfo {title} {{Glashow resonance as a window into
  cosmic neutrino sources}},}\ }\href {\doibase 10.1103/PhysRevD.90.121301}
  {\bibfield  {journal} {\bibinfo  {journal} {Phys.~Rev.}\ }\textbf {\bibinfo
  {volume} {D90}},\ \bibinfo {pages} {121301} (\bibinfo {year} {2014})},\
  \Eprint {http://arxiv.org/abs/1407.3255} {arXiv:1407.3255 [astro-ph.HE]}
  \BibitemShut {NoStop}%
%%CITATION = ARXIV:1407.3255;%%
\bibitem [{\citenamefont {Learned}\ and\ \citenamefont
  {Pakvasa}(1995)}]{Learned:1994wg}%
  \BibitemOpen
  \bibfield  {author} {\bibinfo {author} {\bibfnamefont {John~G.}\ \bibnamefont
  {Learned}}\ and\ \bibinfo {author} {\bibfnamefont {Sandip}\ \bibnamefont
  {Pakvasa}},\ }\bibfield  {title} {\enquote {\bibinfo {title} {{Detecting
  tau-neutrino oscillations at PeV energies}},}\ }\href {\doibase
  10.1016/0927-6505(94)00043-3} {\bibfield  {journal} {\bibinfo  {journal}
  {Astropart.~Phys.}\ }\textbf {\bibinfo {volume} {3}},\ \bibinfo {pages}
  {267--274} (\bibinfo {year} {1995})},\ \Eprint
  {http://arxiv.org/abs/hep-ph/9405296} {arXiv:hep-ph/9405296} \BibitemShut
  {NoStop}%
%%CITATION = HEP-PH/9405296;%%
\bibitem [{\citenamefont {Beacom}\ \emph
  {et~al.}(2003{\natexlab{a}})\citenamefont {Beacom}, \citenamefont {Bell},
  \citenamefont {Hooper}, \citenamefont {Pakvasa},\ and\ \citenamefont
  {Weiler}}]{Beacom:2003nh}%
  \BibitemOpen
  \bibfield  {author} {\bibinfo {author} {\bibfnamefont {John~F.}\ \bibnamefont
  {Beacom}}, \bibinfo {author} {\bibfnamefont {Nicole~F.}\ \bibnamefont
  {Bell}}, \bibinfo {author} {\bibfnamefont {Dan}\ \bibnamefont {Hooper}},
  \bibinfo {author} {\bibfnamefont {Sandip}\ \bibnamefont {Pakvasa}}, \ and\
  \bibinfo {author} {\bibfnamefont {Thomas~J.}\ \bibnamefont {Weiler}},\
  }\bibfield  {title} {\enquote {\bibinfo {title} {Measuring flavor ratios of
  high-energy astrophysical neutrinos},}\ }\href@noop {} {\bibfield  {journal}
  {\bibinfo  {journal} {Phys. Rev.}\ }\textbf {\bibinfo {volume} {D68}},\
  \bibinfo {pages} {093005} (\bibinfo {year} {2003}{\natexlab{a}})},\ \bibinfo
  {note} {erratum-ibid.D72, 019901 (2005)},\ \Eprint
  {http://arxiv.org/abs/hep-ph/0307025} {hep-ph/0307025} \BibitemShut {NoStop}%
%%CITATION = HEP-PH 0307025;%%
\bibitem [{\citenamefont {Bugaev}\ \emph {et~al.}(2004)\citenamefont {Bugaev},
  \citenamefont {Montaruli}, \citenamefont {Shlepin},\ and\ \citenamefont
  {Sokalski}}]{Bugaev:2003sw}%
  \BibitemOpen
  \bibfield  {author} {\bibinfo {author} {\bibfnamefont {Edgar}\ \bibnamefont
  {Bugaev}}, \bibinfo {author} {\bibfnamefont {Teresa}\ \bibnamefont
  {Montaruli}}, \bibinfo {author} {\bibfnamefont {Yuri}\ \bibnamefont
  {Shlepin}}, \ and\ \bibinfo {author} {\bibfnamefont {Igor~A.}\ \bibnamefont
  {Sokalski}},\ }\bibfield  {title} {\enquote {\bibinfo {title} {{Propagation
  of tau neutrinos and tau leptons through the earth and their detection in
  underwater / ice neutrino telescopes}},}\ }\href {\doibase
  10.1016/j.astropartphys.2004.03.002} {\bibfield  {journal} {\bibinfo
  {journal} {Astropart.~Phys.}\ }\textbf {\bibinfo {volume} {21}},\ \bibinfo
  {pages} {491--509} (\bibinfo {year} {2004})},\ \Eprint
  {http://arxiv.org/abs/hep-ph/0312295} {arXiv:hep-ph/0312295 [hep-ph]}
  \BibitemShut {NoStop}%
%%CITATION = HEP-PH/0312295;%%
\bibitem [{\citenamefont {Kashti}\ and\ \citenamefont
  {Waxman}(2005)}]{Kashti:2005qa}%
  \BibitemOpen
  \bibfield  {author} {\bibinfo {author} {\bibfnamefont {Tamar}\ \bibnamefont
  {Kashti}}\ and\ \bibinfo {author} {\bibfnamefont {Eli}\ \bibnamefont
  {Waxman}},\ }\bibfield  {title} {\enquote {\bibinfo {title} {{Flavoring
  astrophysical neutrinos: Flavor ratios depend on energy}},}\ }\href {\doibase
  10.1103/PhysRevLett.95.181101} {\bibfield  {journal} {\bibinfo  {journal}
  {Phys. Rev. Lett.}\ }\textbf {\bibinfo {volume} {95}},\ \bibinfo {pages}
  {181101} (\bibinfo {year} {2005})},\ \Eprint
  {http://arxiv.org/abs/astro-ph/0507599} {arXiv:astro-ph/0507599} \BibitemShut
  {NoStop}%
%%CITATION = ASTRO-PH/0507599;%%
\bibitem [{\citenamefont {Kachelriess}\ \emph {et~al.}(2008)\citenamefont
  {Kachelriess}, \citenamefont {Ostapchenko},\ and\ \citenamefont
  {Tomas}}]{Kachelriess:2007tr}%
  \BibitemOpen
  \bibfield  {author} {\bibinfo {author} {\bibfnamefont {M.}~\bibnamefont
  {Kachelriess}}, \bibinfo {author} {\bibfnamefont {S.}~\bibnamefont
  {Ostapchenko}}, \ and\ \bibinfo {author} {\bibfnamefont {R.}~\bibnamefont
  {Tomas}},\ }\bibfield  {title} {\enquote {\bibinfo {title} {{High energy
  neutrino yields from astrophysical sources. 2. Magnetized sources}},}\ }\href
  {\doibase 10.1103/PhysRevD.77.023007} {\bibfield  {journal} {\bibinfo
  {journal} {Phys. Rev.}\ }\textbf {\bibinfo {volume} {D77}},\ \bibinfo {pages}
  {023007} (\bibinfo {year} {2008})},\ \Eprint {http://arxiv.org/abs/0708.3047}
  {arXiv:0708.3047 [astro-ph]} \BibitemShut {NoStop}%
%%CITATION = ARXIV:0708.3047;%%
\bibitem [{\citenamefont {H{\"u}mmer}\ \emph
  {et~al.}(2010{\natexlab{a}})\citenamefont {H{\"u}mmer}, \citenamefont
  {Maltoni}, \citenamefont {Winter},\ and\ \citenamefont
  {Yaguna}}]{Hummer:2010ai}%
  \BibitemOpen
  \bibfield  {author} {\bibinfo {author} {\bibfnamefont {S.}~\bibnamefont
  {H{\"u}mmer}}, \bibinfo {author} {\bibfnamefont {M.}~\bibnamefont {Maltoni}},
  \bibinfo {author} {\bibfnamefont {W.}~\bibnamefont {Winter}}, \ and\ \bibinfo
  {author} {\bibfnamefont {C.}~\bibnamefont {Yaguna}},\ }\bibfield  {title}
  {\enquote {\bibinfo {title} {{Energy dependent neutrino flavor ratios from
  cosmic accelerators on the Hillas plot}},}\ }\href {\doibase
  10.1016/j.astropartphys.2010.07.003} {\bibfield  {journal} {\bibinfo
  {journal} {Astropart.~Phys.}\ }\textbf {\bibinfo {volume} {34}},\ \bibinfo
  {pages} {205--224} (\bibinfo {year} {2010}{\natexlab{a}})},\ \Eprint
  {http://arxiv.org/abs/1007.0006} {arXiv:1007.0006 [astro-ph.HE]} \BibitemShut
  {NoStop}%
\bibitem [{\citenamefont {Kachelriess}\ and\ \citenamefont
  {Tomas}(2006)}]{Kachelriess:2006fi}%
  \BibitemOpen
  \bibfield  {author} {\bibinfo {author} {\bibfnamefont {M.}~\bibnamefont
  {Kachelriess}}\ and\ \bibinfo {author} {\bibfnamefont {R.}~\bibnamefont
  {Tomas}},\ }\bibfield  {title} {\enquote {\bibinfo {title} {{High energy
  neutrino yields from astrophysical sources. 1. Weakly magnetized sources}},}\
  }\href {\doibase 10.1103/PhysRevD.74.063009} {\bibfield  {journal} {\bibinfo
  {journal} {Phys. Rev.}\ }\textbf {\bibinfo {volume} {D74}},\ \bibinfo {pages}
  {063009} (\bibinfo {year} {2006})},\ \Eprint
  {http://arxiv.org/abs/astro-ph/0606406} {arXiv:astro-ph/0606406} \BibitemShut
  {NoStop}%
%%CITATION = ASTRO-PH/0606406;%%
\bibitem [{\citenamefont {Enberg}\ \emph {et~al.}(2009)\citenamefont {Enberg},
  \citenamefont {Reno},\ and\ \citenamefont {Sarcevic}}]{Enberg:2008jm}%
  \BibitemOpen
  \bibfield  {author} {\bibinfo {author} {\bibfnamefont {Rikard}\ \bibnamefont
  {Enberg}}, \bibinfo {author} {\bibfnamefont {Mary~Hall}\ \bibnamefont
  {Reno}}, \ and\ \bibinfo {author} {\bibfnamefont {Ina}\ \bibnamefont
  {Sarcevic}},\ }\bibfield  {title} {\enquote {\bibinfo {title} {{High energy
  neutrinos from charm in astrophysical sources}},}\ }\href {\doibase
  10.1103/PhysRevD.79.053006} {\bibfield  {journal} {\bibinfo  {journal}
  {Phys.~Rev.}\ }\textbf {\bibinfo {volume} {D79}},\ \bibinfo {pages} {053006}
  (\bibinfo {year} {2009})},\ \Eprint {http://arxiv.org/abs/0808.2807}
  {arXiv:0808.2807 [astro-ph]} \BibitemShut {NoStop}%
%%CITATION = ARXIV:0808.2807;%%
\bibitem [{\citenamefont {H{\"u}mmer}\ \emph
  {et~al.}(2010{\natexlab{b}})\citenamefont {H{\"u}mmer}, \citenamefont
  {R{\"u}ger}, \citenamefont {Spanier},\ and\ \citenamefont
  {Winter}}]{Hummer:2010vx}%
  \BibitemOpen
  \bibfield  {author} {\bibinfo {author} {\bibfnamefont {S.}~\bibnamefont
  {H{\"u}mmer}}, \bibinfo {author} {\bibfnamefont {M.}~\bibnamefont
  {R{\"u}ger}}, \bibinfo {author} {\bibfnamefont {F.}~\bibnamefont {Spanier}},
  \ and\ \bibinfo {author} {\bibfnamefont {W.}~\bibnamefont {Winter}},\
  }\bibfield  {title} {\enquote {\bibinfo {title} {{Simplified models for
  photohadronic interactions in cosmic accelerators}},}\ }\href {\doibase
  10.1088/0004-637X/721/1/630} {\bibfield  {journal} {\bibinfo  {journal}
  {Astrophys.~J.}\ }\textbf {\bibinfo {volume} {721}},\ \bibinfo {pages}
  {630--652} (\bibinfo {year} {2010}{\natexlab{b}})},\ \Eprint
  {http://arxiv.org/abs/1002.1310} {arXiv:1002.1310 [astro-ph.HE]} \BibitemShut
  {NoStop}%
\bibitem [{\citenamefont {Lunardini}\ and\ \citenamefont
  {Smirnov}(2001)}]{Lunardini:2000fy}%
  \BibitemOpen
  \bibfield  {author} {\bibinfo {author} {\bibfnamefont {C.}~\bibnamefont
  {Lunardini}}\ and\ \bibinfo {author} {\bibfnamefont {A.~Yu.}\ \bibnamefont
  {Smirnov}},\ }\bibfield  {title} {\enquote {\bibinfo {title} {{High-energy
  neutrino conversion and the lepton asymmetry in the universe}},}\ }\href
  {\doibase 10.1103/PhysRevD.64.073006} {\bibfield  {journal} {\bibinfo
  {journal} {Phys.~Rev.}\ }\textbf {\bibinfo {volume} {D64}},\ \bibinfo {pages}
  {073006} (\bibinfo {year} {2001})},\ \Eprint
  {http://arxiv.org/abs/hep-ph/0012056} {arXiv:hep-ph/0012056 [hep-ph]}
  \BibitemShut {NoStop}%
%%CITATION = HEP-PH/0012056;%%
\bibitem [{\citenamefont {Razzaque}\ and\ \citenamefont
  {Smirnov}(2010)}]{Razzaque:2009kq}%
  \BibitemOpen
  \bibfield  {author} {\bibinfo {author} {\bibfnamefont {Soebur}\ \bibnamefont
  {Razzaque}}\ and\ \bibinfo {author} {\bibfnamefont {A.~Yu.}\ \bibnamefont
  {Smirnov}},\ }\bibfield  {title} {\enquote {\bibinfo {title} {{Flavor
  conversion of cosmic neutrinos from hidden jets}},}\ }\href {\doibase
  10.1007/JHEP03(2010)031} {\bibfield  {journal} {\bibinfo  {journal} {JHEP}\
  }\textbf {\bibinfo {volume} {1003}},\ \bibinfo {pages} {031} (\bibinfo {year}
  {2010})},\ \Eprint {http://arxiv.org/abs/0912.4028} {arXiv:0912.4028
  [hep-ph]} \BibitemShut {NoStop}%
%%CITATION = ARXIV:0912.4028;%%
\bibitem [{\citenamefont {Sahu}\ and\ \citenamefont
  {Zhang}(2010)}]{Sahu:2010ap}%
  \BibitemOpen
  \bibfield  {author} {\bibinfo {author} {\bibfnamefont {Sarira}\ \bibnamefont
  {Sahu}}\ and\ \bibinfo {author} {\bibfnamefont {Bing}\ \bibnamefont
  {Zhang}},\ }\bibfield  {title} {\enquote {\bibinfo {title} {{Effect of
  Resonant Neutrino Oscillation on TeV Neutrino Flavor Ratio from Choked
  GRBs}},}\ }\href {\doibase 10.1088/1674-4527/10/10/001} {\bibfield  {journal}
  {\bibinfo  {journal} {Res.~Astron.~Astrophys.}\ }\textbf {\bibinfo {volume}
  {10}},\ \bibinfo {pages} {943--949} (\bibinfo {year} {2010})},\ \Eprint
  {http://arxiv.org/abs/1007.4582} {arXiv:1007.4582 [hep-ph]} \BibitemShut
  {NoStop}%
%%CITATION = ARXIV:1007.4582;%%
\bibitem [{\citenamefont {Farzan}\ and\ \citenamefont
  {Smirnov}(2008)}]{Farzan:2008eg}%
  \BibitemOpen
  \bibfield  {author} {\bibinfo {author} {\bibfnamefont {Yasaman}\ \bibnamefont
  {Farzan}}\ and\ \bibinfo {author} {\bibfnamefont {Alexei~Yu.}\ \bibnamefont
  {Smirnov}},\ }\bibfield  {title} {\enquote {\bibinfo {title} {{Coherence and
  oscillations of cosmic neutrinos}},}\ }\href {\doibase
  10.1016/j.nuclphysb.2008.07.028} {\bibfield  {journal} {\bibinfo  {journal}
  {Nucl.~Phys.}\ }\textbf {\bibinfo {volume} {B805}},\ \bibinfo {pages}
  {356--376} (\bibinfo {year} {2008})},\ \Eprint
  {http://arxiv.org/abs/0803.0495} {arXiv:0803.0495 [hep-ph]} \BibitemShut
  {NoStop}%
%%CITATION = ARXIV:0803.0495;%%
\bibitem [{\citenamefont {Akhmedov}\ \emph {et~al.}(2012)\citenamefont
  {Akhmedov}, \citenamefont {Hernandez},\ and\ \citenamefont
  {Smirnov}}]{Akhmedov:2012uu}%
  \BibitemOpen
  \bibfield  {author} {\bibinfo {author} {\bibfnamefont {Evgeny}\ \bibnamefont
  {Akhmedov}}, \bibinfo {author} {\bibfnamefont {Daniel}\ \bibnamefont
  {Hernandez}}, \ and\ \bibinfo {author} {\bibfnamefont {Alexei}\ \bibnamefont
  {Smirnov}},\ }\bibfield  {title} {\enquote {\bibinfo {title} {{Neutrino
  production coherence and oscillation experiments}},}\ }\href {\doibase
  10.1007/JHEP04(2012)052} {\bibfield  {journal} {\bibinfo  {journal} {JHEP}\
  }\textbf {\bibinfo {volume} {1204}},\ \bibinfo {pages} {052} (\bibinfo {year}
  {2012})},\ \Eprint {http://arxiv.org/abs/1201.4128} {arXiv:1201.4128
  [hep-ph]} \BibitemShut {NoStop}%
%%CITATION = ARXIV:1201.4128;%%
\bibitem [{\citenamefont {Jones}(2015)}]{Jones:2014sfa}%
  \BibitemOpen
  \bibfield  {author} {\bibinfo {author} {\bibfnamefont {B.J.P.}\ \bibnamefont
  {Jones}},\ }\bibfield  {title} {\enquote {\bibinfo {title} {{Dynamical pion
  collapse and the coherence of conventional neutrino beams}},}\ }\href
  {\doibase 10.1103/PhysRevD.91.053002} {\bibfield  {journal} {\bibinfo
  {journal} {Phys.~Rev.}\ }\textbf {\bibinfo {volume} {D91}},\ \bibinfo {pages}
  {053002} (\bibinfo {year} {2015})},\ \Eprint {http://arxiv.org/abs/1412.2264}
  {arXiv:1412.2264 [hep-ph]} \BibitemShut {NoStop}%
%%CITATION = ARXIV:1412.2264;%%
\bibitem [{\citenamefont {Olive}\ \emph {et~al.}(2014)\citenamefont {Olive}
  \emph {et~al.}}]{Agashe:2014kda}%
  \BibitemOpen
  \bibfield  {author} {\bibinfo {author} {\bibfnamefont {K.A.}\ \bibnamefont
  {Olive}} \emph {et~al.} (\bibinfo {collaboration} {Particle Data Group}),\
  }\bibfield  {title} {\enquote {\bibinfo {title} {{Review of Particle
  Physics}},}\ }\href {\doibase 10.1088/1674-1137/38/9/090001} {\bibfield
  {journal} {\bibinfo  {journal} {Chin.Phys.}\ }\textbf {\bibinfo {volume}
  {C38}},\ \bibinfo {pages} {090001} (\bibinfo {year} {2014})}\BibitemShut
  {NoStop}%
%%CITATION = CHPHD,C38,090001;%%
\bibitem [{\citenamefont {Gonzalez-Garcia}\ \emph {et~al.}(2014)\citenamefont
  {Gonzalez-Garcia}, \citenamefont {Maltoni},\ and\ \citenamefont
  {Schwetz}}]{Gonzalez-Garcia:2014bfa}%
  \BibitemOpen
  \bibfield  {author} {\bibinfo {author} {\bibfnamefont {M.C.}\ \bibnamefont
  {Gonzalez-Garcia}}, \bibinfo {author} {\bibfnamefont {Michele}\ \bibnamefont
  {Maltoni}}, \ and\ \bibinfo {author} {\bibfnamefont {Thomas}\ \bibnamefont
  {Schwetz}},\ }\bibfield  {title} {\enquote {\bibinfo {title} {{Updated fit to
  three neutrino mixing: status of leptonic CP violation}},}\ }\href {\doibase
  10.1007/JHEP11(2014)052} {\bibfield  {journal} {\bibinfo  {journal} {JHEP}\
  }\textbf {\bibinfo {volume} {1411}},\ \bibinfo {pages} {052} (\bibinfo {year}
  {2014})},\ \Eprint {http://arxiv.org/abs/1409.5439} {arXiv:1409.5439
  [hep-ph]} \BibitemShut {NoStop}%
%%CITATION = ARXIV:1409.5439;%%
\bibitem [{\citenamefont {Dalitz}(1953)}]{Dalitz:1953cp}%
  \BibitemOpen
  \bibfield  {author} {\bibinfo {author} {\bibfnamefont {R.H.}\ \bibnamefont
  {Dalitz}},\ }\bibfield  {title} {\enquote {\bibinfo {title} {{On the analysis
  of tau-meson data and the nature of the tau-meson}},}\ }\href {\doibase
  10.1080/14786441008520365} {\bibfield  {journal} {\bibinfo  {journal}
  {Phil.~Mag.}\ }\textbf {\bibinfo {volume} {44}},\ \bibinfo {pages}
  {1068--1080} (\bibinfo {year} {1953})}\BibitemShut {NoStop}%
%%CITATION = PHMAA,44,1068;%%
\bibitem [{\citenamefont {Kopp}\ \emph {et~al.}(2013)\citenamefont {Kopp},
  \citenamefont {Machado}, \citenamefont {Maltoni},\ and\ \citenamefont
  {Schwetz}}]{Kopp:2013vaa}%
  \BibitemOpen
  \bibfield  {author} {\bibinfo {author} {\bibfnamefont {Joachim}\ \bibnamefont
  {Kopp}}, \bibinfo {author} {\bibfnamefont {Pedro A.~N.}\ \bibnamefont
  {Machado}}, \bibinfo {author} {\bibfnamefont {Michele}\ \bibnamefont
  {Maltoni}}, \ and\ \bibinfo {author} {\bibfnamefont {Thomas}\ \bibnamefont
  {Schwetz}},\ }\bibfield  {title} {\enquote {\bibinfo {title} {{Sterile
  Neutrino Oscillations: The Global Picture}},}\ }\href {\doibase
  10.1007/JHEP05(2013)050} {\bibfield  {journal} {\bibinfo  {journal} {JHEP}\
  }\textbf {\bibinfo {volume} {1305}},\ \bibinfo {pages} {050} (\bibinfo {year}
  {2013})},\ \Eprint {http://arxiv.org/abs/1303.3011} {arXiv:1303.3011
  [hep-ph]} \BibitemShut {NoStop}%
%%CITATION = ARXIV:1303.3011;%%
\bibitem [{\citenamefont {Giunti}\ \emph {et~al.}(2013)\citenamefont {Giunti},
  \citenamefont {Laveder}, \citenamefont {Li},\ and\ \citenamefont
  {Long}}]{Giunti:2013aea}%
  \BibitemOpen
  \bibfield  {author} {\bibinfo {author} {\bibfnamefont {C.}~\bibnamefont
  {Giunti}}, \bibinfo {author} {\bibfnamefont {M.}~\bibnamefont {Laveder}},
  \bibinfo {author} {\bibfnamefont {Y.F.}\ \bibnamefont {Li}}, \ and\ \bibinfo
  {author} {\bibfnamefont {H.W.}\ \bibnamefont {Long}},\ }\bibfield  {title}
  {\enquote {\bibinfo {title} {{Pragmatic View of Short-Baseline Neutrino
  Oscillations}},}\ }\href {\doibase 10.1103/PhysRevD.88.073008} {\bibfield
  {journal} {\bibinfo  {journal} {Phys.~Rev.}\ }\textbf {\bibinfo {volume}
  {D88}},\ \bibinfo {pages} {073008} (\bibinfo {year} {2013})},\ \Eprint
  {http://arxiv.org/abs/1308.5288} {arXiv:1308.5288 [hep-ph]} \BibitemShut
  {NoStop}%
%%CITATION = ARXIV:1308.5288;%%
\bibitem [{\citenamefont {Glashow}(1960)}]{Glashow:1960zz}%
  \BibitemOpen
  \bibfield  {author} {\bibinfo {author} {\bibfnamefont {Sheldon~L.}\
  \bibnamefont {Glashow}},\ }\bibfield  {title} {\enquote {\bibinfo {title}
  {{Resonant Scattering of Antineutrinos}},}\ }\href {\doibase
  10.1103/PhysRev.118.316} {\bibfield  {journal} {\bibinfo  {journal}
  {Phys.~Rev.}\ }\textbf {\bibinfo {volume} {118}},\ \bibinfo {pages}
  {316--317} (\bibinfo {year} {1960})}\BibitemShut {NoStop}%
%%CITATION = PHRVA,118,316;%%
\bibitem [{\citenamefont {Berezinsky}\ and\ \citenamefont
  {Gazizov}(1977)}]{Berezinsky:1977sf}%
  \BibitemOpen
  \bibfield  {author} {\bibinfo {author} {\bibfnamefont {V.S.}\ \bibnamefont
  {Berezinsky}}\ and\ \bibinfo {author} {\bibfnamefont {A.Z.}\ \bibnamefont
  {Gazizov}},\ }\bibfield  {title} {\enquote {\bibinfo {title} {{Cosmic
  Neutrinos and Possibility to Search for W Bosons Having 30-GeV-100-GeV Masses
  in Underwater Experiments}},}\ }\href@noop {} {\bibfield  {journal} {\bibinfo
   {journal} {JETP~Lett.}\ }\textbf {\bibinfo {volume} {25}},\ \bibinfo {pages}
  {254--256} (\bibinfo {year} {1977})}\BibitemShut {NoStop}%
%%CITATION = JTPLA,25,254;%%
\bibitem [{\citenamefont {Berezinsky}\ and\ \citenamefont
  {Gazizov}(1981)}]{Berezinsky:1981bt}%
  \BibitemOpen
  \bibfield  {author} {\bibinfo {author} {\bibfnamefont {V.S.}\ \bibnamefont
  {Berezinsky}}\ and\ \bibinfo {author} {\bibfnamefont {A.Z.}\ \bibnamefont
  {Gazizov}},\ }\bibfield  {title} {\enquote {\bibinfo {title} {{Neutrino -
  electron scattering at energies above the W boson production threshold}},}\
  }\href@noop {} {\bibfield  {journal} {\bibinfo  {journal}
  {Sov.~J.~Nucl.~Phys.}\ }\textbf {\bibinfo {volume} {33}},\ \bibinfo {pages}
  {120--125} (\bibinfo {year} {1981})}\BibitemShut {NoStop}%
%%CITATION = SJNCA,33,120;%%
\bibitem [{\citenamefont {Athar}\ \emph {et~al.}(2000)\citenamefont {Athar},
  \citenamefont {Parente},\ and\ \citenamefont {Zas}}]{Athar:2000rx}%
  \BibitemOpen
  \bibfield  {author} {\bibinfo {author} {\bibfnamefont {H.}~\bibnamefont
  {Athar}}, \bibinfo {author} {\bibfnamefont {G.}~\bibnamefont {Parente}}, \
  and\ \bibinfo {author} {\bibfnamefont {E.}~\bibnamefont {Zas}},\ }\bibfield
  {title} {\enquote {\bibinfo {title} {{Prospects for observations of
  high-energy cosmic tau neutrinos}},}\ }\href {\doibase
  10.1103/PhysRevD.62.093010} {\bibfield  {journal} {\bibinfo  {journal}
  {Phys.~Rev.}\ }\textbf {\bibinfo {volume} {D62}},\ \bibinfo {pages} {093010}
  (\bibinfo {year} {2000})},\ \Eprint {http://arxiv.org/abs/hep-ph/0006123}
  {arXiv:hep-ph/0006123 [hep-ph]} \BibitemShut {NoStop}%
%%CITATION = HEP-PH/0006123;%%
\bibitem [{\citenamefont {DeYoung}\ \emph {et~al.}(2007)\citenamefont
  {DeYoung}, \citenamefont {Razzaque},\ and\ \citenamefont
  {Cowen}}]{DeYoung:2006fg}%
  \BibitemOpen
  \bibfield  {author} {\bibinfo {author} {\bibfnamefont {Tyce}\ \bibnamefont
  {DeYoung}}, \bibinfo {author} {\bibfnamefont {S.}~\bibnamefont {Razzaque}}, \
  and\ \bibinfo {author} {\bibfnamefont {D.F.}\ \bibnamefont {Cowen}},\
  }\bibfield  {title} {\enquote {\bibinfo {title} {{Astrophysical tau neutrino
  detection in kilometer-scale Cherenkov detectors via muonic tau decay}},}\
  }\href {\doibase 10.1016/j.astropartphys.2006.11.003} {\bibfield  {journal}
  {\bibinfo  {journal} {Astropart.~Phys.}\ }\textbf {\bibinfo {volume} {27}},\
  \bibinfo {pages} {238--243} (\bibinfo {year} {2007})},\ \Eprint
  {http://arxiv.org/abs/astro-ph/0608486} {arXiv:astro-ph/0608486 [astro-ph]}
  \BibitemShut {NoStop}%
%%CITATION = ASTRO-PH/0608486;%%
\bibitem [{\citenamefont {Xing}\ and\ \citenamefont
  {Zhou}(2011)}]{Xing:2011zm}%
  \BibitemOpen
  \bibfield  {author} {\bibinfo {author} {\bibfnamefont {Zhi-zhong}\
  \bibnamefont {Xing}}\ and\ \bibinfo {author} {\bibfnamefont {Shun}\
  \bibnamefont {Zhou}},\ }\bibfield  {title} {\enquote {\bibinfo {title} {{The
  Glashow resonance as a discriminator of UHE cosmic neutrinos originating from
  p-gamma and p-p collisions}},}\ }\href {\doibase 10.1103/PhysRevD.84.033006}
  {\bibfield  {journal} {\bibinfo  {journal} {Phys.~Rev.}\ }\textbf {\bibinfo
  {volume} {D84}},\ \bibinfo {pages} {033006} (\bibinfo {year} {2011})},\
  \Eprint {http://arxiv.org/abs/1105.4114} {arXiv:1105.4114 [hep-ph]}
  \BibitemShut {NoStop}%
%%CITATION = ARXIV:1105.4114;%%
\bibitem [{\citenamefont {Bhattacharya}\ \emph {et~al.}(2012)\citenamefont
  {Bhattacharya}, \citenamefont {Gandhi}, \citenamefont {Rodejohann},\ and\
  \citenamefont {Watanabe}}]{Bhattacharya:2012fh}%
  \BibitemOpen
  \bibfield  {author} {\bibinfo {author} {\bibfnamefont {Atri}\ \bibnamefont
  {Bhattacharya}}, \bibinfo {author} {\bibfnamefont {Raj}\ \bibnamefont
  {Gandhi}}, \bibinfo {author} {\bibfnamefont {Werner}\ \bibnamefont
  {Rodejohann}}, \ and\ \bibinfo {author} {\bibfnamefont {Atsushi}\
  \bibnamefont {Watanabe}},\ }\bibfield  {title} {\enquote {\bibinfo {title}
  {{On the interpretation of IceCube cascade events in terms of the Glashow
  resonance}},}\ }\href@noop {} {\  (\bibinfo {year} {2012})},\ \Eprint
  {http://arxiv.org/abs/1209.2422} {arXiv:1209.2422 [hep-ph]} \BibitemShut
  {NoStop}%
%%CITATION = ARXIV:1209.2422;%%
\bibitem [{\citenamefont {Abbasi}\ \emph {et~al.}(2012)\citenamefont {Abbasi}
  \emph {et~al.}}]{Abbasi:2012cu}%
  \BibitemOpen
  \bibfield  {author} {\bibinfo {author} {\bibfnamefont {R.}~\bibnamefont
  {Abbasi}} \emph {et~al.} (\bibinfo {collaboration} {IceCube}),\ }\bibfield
  {title} {\enquote {\bibinfo {title} {{A Search for UHE Tau Neutrinos with
  IceCube}},}\ }\href {\doibase 10.1103/PhysRevD.86.022005} {\bibfield
  {journal} {\bibinfo  {journal} {Phys.~Rev.}\ }\textbf {\bibinfo {volume}
  {D86}},\ \bibinfo {pages} {022005} (\bibinfo {year} {2012})},\ \Eprint
  {http://arxiv.org/abs/1202.4564} {arXiv:1202.4564 [astro-ph.HE]} \BibitemShut
  {NoStop}%
%%CITATION = ARXIV:1202.4564;%%
\bibitem [{\citenamefont {{Barger}}\ \emph {et~al.}(2013)\citenamefont
  {{Barger}}, \citenamefont {{Learned}},\ and\ \citenamefont
  {{Pakvasa}}}]{2013PhRvD..87c7302B}%
  \BibitemOpen
  \bibfield  {author} {\bibinfo {author} {\bibfnamefont {V.}~\bibnamefont
  {{Barger}}}, \bibinfo {author} {\bibfnamefont {J.}~\bibnamefont {{Learned}}},
  \ and\ \bibinfo {author} {\bibfnamefont {S.}~\bibnamefont {{Pakvasa}}},\
  }\bibfield  {title} {\enquote {\bibinfo {title} {{IceCube PeV cascade events
  initiated by electron-antineutrinos at Glashow resonance}},}\ }\href
  {\doibase 10.1103/PhysRevD.87.037302} {\bibfield  {journal} {\bibinfo
  {journal} {Phys.~Rev.}\ }\textbf {\bibinfo {volume} {D87}},\ \bibinfo {eid}
  {037302} (\bibinfo {year} {2013})},\ \Eprint {http://arxiv.org/abs/1207.4571}
  {arXiv:1207.4571 [astro-ph.HE]} \BibitemShut {NoStop}%
\bibitem [{\citenamefont {Blennow}\ and\ \citenamefont
  {Meloni}(2009)}]{Blennow:2009rp}%
  \BibitemOpen
  \bibfield  {author} {\bibinfo {author} {\bibfnamefont {Mattias}\ \bibnamefont
  {Blennow}}\ and\ \bibinfo {author} {\bibfnamefont {Davide}\ \bibnamefont
  {Meloni}},\ }\bibfield  {title} {\enquote {\bibinfo {title} {{Non-standard
  interaction effects on astrophysical neutrino fluxes}},}\ }\href {\doibase
  10.1103/PhysRevD.80.065009} {\bibfield  {journal} {\bibinfo  {journal} {Phys.
  Rev.}\ }\textbf {\bibinfo {volume} {D80}},\ \bibinfo {pages} {065009}
  (\bibinfo {year} {2009})},\ \Eprint {http://arxiv.org/abs/0901.2110}
  {arXiv:0901.2110 [hep-ph]} \BibitemShut {NoStop}%
%%CITATION = 0901.2110;%%
\bibitem [{\citenamefont {Bustamante}\ \emph {et~al.}(2011)\citenamefont
  {Bustamante}, \citenamefont {Gago},\ and\ \citenamefont
  {Jones~Perez}}]{Bustamante:2010bf}%
  \BibitemOpen
  \bibfield  {author} {\bibinfo {author} {\bibfnamefont {M.}~\bibnamefont
  {Bustamante}}, \bibinfo {author} {\bibfnamefont {A.M.}\ \bibnamefont {Gago}},
  \ and\ \bibinfo {author} {\bibfnamefont {Joel}\ \bibnamefont {Jones~Perez}},\
  }\bibfield  {title} {\enquote {\bibinfo {title} {{SUSY Renormalization Group
  Effects in Ultra High Energy Neutrinos}},}\ }\href {\doibase
  10.1007/JHEP05(2011)133} {\bibfield  {journal} {\bibinfo  {journal} {JHEP}\
  }\textbf {\bibinfo {volume} {1105}},\ \bibinfo {pages} {133} (\bibinfo {year}
  {2011})},\ \Eprint {http://arxiv.org/abs/1012.2728} {arXiv:1012.2728
  [hep-ph]} \BibitemShut {NoStop}%
%%CITATION = ARXIV:1012.2728;%%
\bibitem [{\citenamefont {Jain}\ \emph {et~al.}(2002)\citenamefont {Jain},
  \citenamefont {Kar}, \citenamefont {McKay}, \citenamefont {Panda},\ and\
  \citenamefont {Ralston}}]{Jain:2002kz}%
  \BibitemOpen
  \bibfield  {author} {\bibinfo {author} {\bibfnamefont {Pankaj}\ \bibnamefont
  {Jain}}, \bibinfo {author} {\bibfnamefont {Supriya}\ \bibnamefont {Kar}},
  \bibinfo {author} {\bibfnamefont {Douglas~W.}\ \bibnamefont {McKay}},
  \bibinfo {author} {\bibfnamefont {Sukanta}\ \bibnamefont {Panda}}, \ and\
  \bibinfo {author} {\bibfnamefont {John~P.}\ \bibnamefont {Ralston}},\
  }\bibfield  {title} {\enquote {\bibinfo {title} {{Angular dependence of
  neutrino flux in KM$^3$ detectors in low scale gravity models}},}\ }\href
  {\doibase 10.1103/PhysRevD.66.065018} {\bibfield  {journal} {\bibinfo
  {journal} {Phys.~Rev.}\ }\textbf {\bibinfo {volume} {D66}},\ \bibinfo {pages}
  {065018} (\bibinfo {year} {2002})},\ \Eprint
  {http://arxiv.org/abs/hep-ph/0205052} {arXiv:hep-ph/0205052 [hep-ph]}
  \BibitemShut {NoStop}%
%%CITATION = HEP-PH/0205052;%%
\bibitem [{\citenamefont {Hussain}\ and\ \citenamefont
  {McKay}(2004)}]{Hussain:2003vi}%
  \BibitemOpen
  \bibfield  {author} {\bibinfo {author} {\bibfnamefont {Shahid}\ \bibnamefont
  {Hussain}}\ and\ \bibinfo {author} {\bibfnamefont {Douglas~W.}\ \bibnamefont
  {McKay}},\ }\bibfield  {title} {\enquote {\bibinfo {title} {{Energy and
  angular distribution of upward UHE neutrinos and signals of low scale
  gravity: Role of tau decay}},}\ }\href {\doibase 10.1103/PhysRevD.69.085004}
  {\bibfield  {journal} {\bibinfo  {journal} {Phys.~Rev.}\ }\textbf {\bibinfo
  {volume} {D69}},\ \bibinfo {pages} {085004} (\bibinfo {year} {2004})},\
  \Eprint {http://arxiv.org/abs/hep-ph/0310091} {arXiv:hep-ph/0310091 [hep-ph]}
  \BibitemShut {NoStop}%
%%CITATION = HEP-PH/0310091;%%
\bibitem [{\citenamefont {Borriello}\ \emph {et~al.}(2008)\citenamefont
  {Borriello}, \citenamefont {Cuoco}, \citenamefont {Mangano}, \citenamefont
  {Miele}, \citenamefont {Pastor} \emph {et~al.}}]{Borriello:2007cs}%
  \BibitemOpen
  \bibfield  {author} {\bibinfo {author} {\bibfnamefont {E.}~\bibnamefont
  {Borriello}}, \bibinfo {author} {\bibfnamefont {A.}~\bibnamefont {Cuoco}},
  \bibinfo {author} {\bibfnamefont {G.}~\bibnamefont {Mangano}}, \bibinfo
  {author} {\bibfnamefont {G.}~\bibnamefont {Miele}}, \bibinfo {author}
  {\bibfnamefont {S.}~\bibnamefont {Pastor}},  \emph {et~al.},\ }\bibfield
  {title} {\enquote {\bibinfo {title} {{Disentangling neutrino-nucleon cross
  section and high energy neutrino flux with a km$^3$ neutrino telescope}},}\
  }\href {\doibase 10.1103/PhysRevD.77.045019} {\bibfield  {journal} {\bibinfo
  {journal} {Phys.~Rev.}\ }\textbf {\bibinfo {volume} {D77}},\ \bibinfo {pages}
  {045019} (\bibinfo {year} {2008})},\ \Eprint {http://arxiv.org/abs/0711.0152}
  {arXiv:0711.0152 [astro-ph]} \BibitemShut {NoStop}%
%%CITATION = ARXIV:0711.0152;%%
\bibitem [{\citenamefont {Marfatia}\ \emph {et~al.}(2015)\citenamefont
  {Marfatia}, \citenamefont {McKay},\ and\ \citenamefont
  {Weiler}}]{Marfatia:2015hva}%
  \BibitemOpen
  \bibfield  {author} {\bibinfo {author} {\bibfnamefont {D.}~\bibnamefont
  {Marfatia}}, \bibinfo {author} {\bibfnamefont {D.~W.}\ \bibnamefont {McKay}},
  \ and\ \bibinfo {author} {\bibfnamefont {T.~J.}\ \bibnamefont {Weiler}},\
  }\bibfield  {title} {\enquote {\bibinfo {title} {{New physics with
  ultra-high-energy neutrinos}},}\ }\href {\doibase
  10.1016/j.physletb.2015.07.002} {\bibfield  {journal} {\bibinfo  {journal}
  {Phys.~Lett.}\ }\textbf {\bibinfo {volume} {B748}},\ \bibinfo {pages}
  {113--116} (\bibinfo {year} {2015})},\ \Eprint
  {http://arxiv.org/abs/1502.06337} {arXiv:1502.06337 [hep-ph]} \BibitemShut
  {NoStop}%
%%CITATION = ARXIV:1502.06337;%%
\bibitem [{\citenamefont {Beacom}\ \emph
  {et~al.}(2003{\natexlab{b}})\citenamefont {Beacom}, \citenamefont {Bell},
  \citenamefont {Hooper}, \citenamefont {Pakvasa},\ and\ \citenamefont
  {Weiler}}]{Beacom:2002vi}%
  \BibitemOpen
  \bibfield  {author} {\bibinfo {author} {\bibfnamefont {John~F.}\ \bibnamefont
  {Beacom}}, \bibinfo {author} {\bibfnamefont {Nicole~F.}\ \bibnamefont
  {Bell}}, \bibinfo {author} {\bibfnamefont {Dan}\ \bibnamefont {Hooper}},
  \bibinfo {author} {\bibfnamefont {Sandip}\ \bibnamefont {Pakvasa}}, \ and\
  \bibinfo {author} {\bibfnamefont {Thomas~J.}\ \bibnamefont {Weiler}},\
  }\bibfield  {title} {\enquote {\bibinfo {title} {{Decay of high-energy
  astrophysical neutrinos}},}\ }\href {\doibase 10.1103/PhysRevLett.90.181301}
  {\bibfield  {journal} {\bibinfo  {journal} {Phys.~Rev.~Lett.}\ }\textbf
  {\bibinfo {volume} {90}},\ \bibinfo {pages} {181301} (\bibinfo {year}
  {2003}{\natexlab{b}})},\ \Eprint {http://arxiv.org/abs/hep-ph/0211305}
  {arXiv:hep-ph/0211305 [hep-ph]} \BibitemShut {NoStop}%
%%CITATION = HEP-PH/0211305;%%
\bibitem [{\citenamefont {Maltoni}\ and\ \citenamefont
  {Winter}(2008)}]{Maltoni:2008jr}%
  \BibitemOpen
  \bibfield  {author} {\bibinfo {author} {\bibfnamefont {Michele}\ \bibnamefont
  {Maltoni}}\ and\ \bibinfo {author} {\bibfnamefont {Walter}\ \bibnamefont
  {Winter}},\ }\bibfield  {title} {\enquote {\bibinfo {title} {{Testing
  neutrino oscillations plus decay with neutrino telescopes}},}\ }\href
  {\doibase 10.1088/1126-6708/2008/07/064} {\bibfield  {journal} {\bibinfo
  {journal} {JHEP}\ }\textbf {\bibinfo {volume} {07}},\ \bibinfo {pages} {064}
  (\bibinfo {year} {2008})},\ \Eprint {http://arxiv.org/abs/0803.2050}
  {arXiv:0803.2050 [hep-ph]} \BibitemShut {NoStop}%
%%CITATION = 0803.2050;%%
\bibitem [{\citenamefont {Baerwald}\ \emph {et~al.}(2012)\citenamefont
  {Baerwald}, \citenamefont {Bustamante},\ and\ \citenamefont
  {Winter}}]{Baerwald:2012kc}%
  \BibitemOpen
  \bibfield  {author} {\bibinfo {author} {\bibfnamefont {Philipp}\ \bibnamefont
  {Baerwald}}, \bibinfo {author} {\bibfnamefont {Mauricio}\ \bibnamefont
  {Bustamante}}, \ and\ \bibinfo {author} {\bibfnamefont {Walter}\ \bibnamefont
  {Winter}},\ }\bibfield  {title} {\enquote {\bibinfo {title} {{Neutrino Decays
  over Cosmological Distances and the Implications for Neutrino Telescopes}},}\
  }\href {\doibase 10.1088/1475-7516/2012/10/020} {\bibfield  {journal}
  {\bibinfo  {journal} {JCAP}\ }\textbf {\bibinfo {volume} {1210}},\ \bibinfo
  {pages} {020} (\bibinfo {year} {2012})},\ \Eprint
  {http://arxiv.org/abs/1208.4600} {arXiv:1208.4600 [astro-ph.CO]} \BibitemShut
  {NoStop}%
%%CITATION = ARXIV:1208.4600;%%
\bibitem [{\citenamefont {Pagliaroli}\ \emph {et~al.}(2015)\citenamefont
  {Pagliaroli}, \citenamefont {Palladino}, \citenamefont {Vissani},\ and\
  \citenamefont {Villante}}]{Pagliaroli:2015rca}%
  \BibitemOpen
  \bibfield  {author} {\bibinfo {author} {\bibfnamefont {G.}~\bibnamefont
  {Pagliaroli}}, \bibinfo {author} {\bibfnamefont {A.}~\bibnamefont
  {Palladino}}, \bibinfo {author} {\bibfnamefont {F.}~\bibnamefont {Vissani}},
  \ and\ \bibinfo {author} {\bibfnamefont {F.L.}\ \bibnamefont {Villante}},\
  }\bibfield  {title} {\enquote {\bibinfo {title} {{Testing neutrino decay
  scenarios with IceCube data}},}\ }\href@noop {} {\  (\bibinfo {year}
  {2015})},\ \Eprint {http://arxiv.org/abs/1506.02624} {arXiv:1506.02624
  [hep-ph]} \BibitemShut {NoStop}%
%%CITATION = ARXIV:1506.02624;%%
\bibitem [{\citenamefont {Kolb}\ and\ \citenamefont
  {Turner}(1987)}]{Kolb:1987qy}%
  \BibitemOpen
  \bibfield  {author} {\bibinfo {author} {\bibfnamefont {Edward~W.}\
  \bibnamefont {Kolb}}\ and\ \bibinfo {author} {\bibfnamefont {Michael~S.}\
  \bibnamefont {Turner}},\ }\bibfield  {title} {\enquote {\bibinfo {title}
  {{Supernova SN 1987a and the Secret Interactions of Neutrinos}},}\ }\href
  {\doibase 10.1103/PhysRevD.36.2895} {\bibfield  {journal} {\bibinfo
  {journal} {Phys.~Rev.}\ }\textbf {\bibinfo {volume} {D36}},\ \bibinfo {pages}
  {2895} (\bibinfo {year} {1987})}\BibitemShut {NoStop}%
%%CITATION = PHRVA,D36,2895;%%
\bibitem [{\citenamefont {Ioka}\ and\ \citenamefont
  {Murase}(2014)}]{Ioka:2014kca}%
  \BibitemOpen
  \bibfield  {author} {\bibinfo {author} {\bibfnamefont {Kunihto}\ \bibnamefont
  {Ioka}}\ and\ \bibinfo {author} {\bibfnamefont {Kohta}\ \bibnamefont
  {Murase}},\ }\bibfield  {title} {\enquote {\bibinfo {title} {{IceCube
  PeV--–EeV neutrinos and secret interactions of neutrinos}},}\ }\href
  {\doibase 10.1093/ptep/ptu090} {\bibfield  {journal} {\bibinfo  {journal}
  {PTEP}\ }\textbf {\bibinfo {volume} {2014}},\ \bibinfo {pages} {061E01}
  (\bibinfo {year} {2014})},\ \Eprint {http://arxiv.org/abs/1404.2279}
  {arXiv:1404.2279 [astro-ph.HE]} \BibitemShut {NoStop}%
%%CITATION = ARXIV:1404.2279;%%
\bibitem [{\citenamefont {Ng}\ and\ \citenamefont {Beacom}(2014)}]{Ng:2014pca}%
  \BibitemOpen
  \bibfield  {author} {\bibinfo {author} {\bibfnamefont {Kenny C.~Y.}\
  \bibnamefont {Ng}}\ and\ \bibinfo {author} {\bibfnamefont {John~F.}\
  \bibnamefont {Beacom}},\ }\bibfield  {title} {\enquote {\bibinfo {title}
  {{Cosmic neutrino cascades from secret neutrino interactions}},}\ }\href
  {\doibase 10.1103/PhysRevD.90.065035, 10.1103/PhysRevD.90.089904} {\bibfield
  {journal} {\bibinfo  {journal} {Phys.~Rev.}\ }\textbf {\bibinfo {volume}
  {D90}},\ \bibinfo {pages} {065035} (\bibinfo {year} {2014})},\ \Eprint
  {http://arxiv.org/abs/1404.2288} {arXiv:1404.2288 [astro-ph.HE]} \BibitemShut
  {NoStop}%
%%CITATION = ARXIV:1404.2288;%%
\bibitem [{\citenamefont {Blum}\ \emph {et~al.}(2014)\citenamefont {Blum},
  \citenamefont {Hook},\ and\ \citenamefont {Murase}}]{Blum:2014ewa}%
  \BibitemOpen
  \bibfield  {author} {\bibinfo {author} {\bibfnamefont {Kfir}\ \bibnamefont
  {Blum}}, \bibinfo {author} {\bibfnamefont {Anson}\ \bibnamefont {Hook}}, \
  and\ \bibinfo {author} {\bibfnamefont {Kohta}\ \bibnamefont {Murase}},\
  }\bibfield  {title} {\enquote {\bibinfo {title} {{High energy neutrino
  telescopes as a probe of the neutrino mass mechanism}},}\ }\href@noop {} {\
  (\bibinfo {year} {2014})},\ \Eprint {http://arxiv.org/abs/1408.3799}
  {arXiv:1408.3799 [hep-ph]} \BibitemShut {NoStop}%
%%CITATION = ARXIV:1408.3799;%%
\bibitem [{\citenamefont {Cherry}\ \emph {et~al.}(2014)\citenamefont {Cherry},
  \citenamefont {Friedland},\ and\ \citenamefont {Shoemaker}}]{Cherry:2014xra}%
  \BibitemOpen
  \bibfield  {author} {\bibinfo {author} {\bibfnamefont {John~F.}\ \bibnamefont
  {Cherry}}, \bibinfo {author} {\bibfnamefont {Alexander}\ \bibnamefont
  {Friedland}}, \ and\ \bibinfo {author} {\bibfnamefont {Ian~M.}\ \bibnamefont
  {Shoemaker}},\ }\bibfield  {title} {\enquote {\bibinfo {title} {{Neutrino
  Portal Dark Matter: From Dwarf Galaxies to IceCube}},}\ }\href@noop {} {\
  (\bibinfo {year} {2014})},\ \Eprint {http://arxiv.org/abs/1411.1071}
  {arXiv:1411.1071 [hep-ph]} \BibitemShut {NoStop}%
%%CITATION = ARXIV:1411.1071;%%
\bibitem [{\citenamefont {Kamada}\ and\ \citenamefont
  {Yu}(2015)}]{Kamada:2015era}%
  \BibitemOpen
  \bibfield  {author} {\bibinfo {author} {\bibfnamefont {Ayuki}\ \bibnamefont
  {Kamada}}\ and\ \bibinfo {author} {\bibfnamefont {Hai-Bo}\ \bibnamefont
  {Yu}},\ }\bibfield  {title} {\enquote {\bibinfo {title} {{Coherent
  Propagation of PeV Neutrinos and the Dip in the Neutrino Spectrum at
  IceCube}},}\ }\href@noop {} {\  (\bibinfo {year} {2015})},\ \Eprint
  {http://arxiv.org/abs/1504.00711} {arXiv:1504.00711 [hep-ph]} \BibitemShut
  {NoStop}%
%%CITATION = ARXIV:1504.00711;%%
\bibitem [{\citenamefont {DiFranzo}\ and\ \citenamefont
  {Hooper}(2015)}]{DiFranzo:2015qea}%
  \BibitemOpen
  \bibfield  {author} {\bibinfo {author} {\bibfnamefont {Anthony}\ \bibnamefont
  {DiFranzo}}\ and\ \bibinfo {author} {\bibfnamefont {Dan}\ \bibnamefont
  {Hooper}},\ }\bibfield  {title} {\enquote {\bibinfo {title} {{Searching for
  MeV-Scale Gauge Bosons with IceCube}},}\ }\href@noop {} {\  (\bibinfo {year}
  {2015})},\ \Eprint {http://arxiv.org/abs/1507.03015} {arXiv:1507.03015
  [hep-ph]} \BibitemShut {NoStop}%
%%CITATION = ARXIV:1507.03015;%%
\bibitem [{\citenamefont {Beacom}\ \emph
  {et~al.}(2004{\natexlab{a}})\citenamefont {Beacom} \emph
  {et~al.}}]{Beacom:2003eu}%
  \BibitemOpen
  \bibfield  {author} {\bibinfo {author} {\bibfnamefont {John~F.}\ \bibnamefont
  {Beacom}} \emph {et~al.},\ }\bibfield  {title} {\enquote {\bibinfo {title}
  {{Pseudo-Dirac neutrinos, a challenge for neutrino telescopes}},}\ }\href
  {\doibase 10.1103/PhysRevLett.92.011101} {\bibfield  {journal} {\bibinfo
  {journal} {Phys. Rev. Lett.}\ }\textbf {\bibinfo {volume} {92}},\ \bibinfo
  {pages} {011101} (\bibinfo {year} {2004}{\natexlab{a}})},\ \Eprint
  {http://arxiv.org/abs/hep-ph/0307151} {arXiv:hep-ph/0307151} \BibitemShut
  {NoStop}%
%%CITATION = HEP-PH/0307151;%%
\bibitem [{\citenamefont {Esmaili}\ and\ \citenamefont
  {Farzan}(2012)}]{Esmaili:2012ac}%
  \BibitemOpen
  \bibfield  {author} {\bibinfo {author} {\bibfnamefont {Arman}\ \bibnamefont
  {Esmaili}}\ and\ \bibinfo {author} {\bibfnamefont {Yasaman}\ \bibnamefont
  {Farzan}},\ }\bibfield  {title} {\enquote {\bibinfo {title} {{Implications of
  the Pseudo-Dirac Scenario for Ultra High Energy Neutrinos from GRBs}},}\
  }\href {\doibase 10.1088/1475-7516/2012/12/014} {\bibfield  {journal}
  {\bibinfo  {journal} {JCAP}\ }\textbf {\bibinfo {volume} {1212}},\ \bibinfo
  {pages} {014} (\bibinfo {year} {2012})},\ \Eprint
  {http://arxiv.org/abs/1208.6012} {arXiv:1208.6012 [hep-ph]} \BibitemShut
  {NoStop}%
%%CITATION = ARXIV:1208.6012;%%
\bibitem [{\citenamefont {Joshipura}\ \emph {et~al.}(2014)\citenamefont
  {Joshipura}, \citenamefont {Mohanty},\ and\ \citenamefont
  {Pakvasa}}]{Joshipura:2013yba}%
  \BibitemOpen
  \bibfield  {author} {\bibinfo {author} {\bibfnamefont {Anjan~S.}\
  \bibnamefont {Joshipura}}, \bibinfo {author} {\bibfnamefont {Subhendra}\
  \bibnamefont {Mohanty}}, \ and\ \bibinfo {author} {\bibfnamefont {Sandip}\
  \bibnamefont {Pakvasa}},\ }\bibfield  {title} {\enquote {\bibinfo {title}
  {{Pseudo-Dirac neutrinos via a mirror world and depletion of ultrahigh energy
  neutrinos}},}\ }\href {\doibase 10.1103/PhysRevD.89.033003} {\bibfield
  {journal} {\bibinfo  {journal} {Phys.~Rev.}\ }\textbf {\bibinfo {volume}
  {D89}},\ \bibinfo {pages} {033003} (\bibinfo {year} {2014})},\ \Eprint
  {http://arxiv.org/abs/1307.5712} {arXiv:1307.5712 [hep-ph]} \BibitemShut
  {NoStop}%
%%CITATION = ARXIV:1307.5712;%%
\bibitem [{\citenamefont {Ellis}\ \emph {et~al.}(1984)\citenamefont {Ellis},
  \citenamefont {Hagelin}, \citenamefont {Nanopoulos},\ and\ \citenamefont
  {Srednicki}}]{Ellis:1983jz}%
  \BibitemOpen
  \bibfield  {author} {\bibinfo {author} {\bibfnamefont {John~R.}\ \bibnamefont
  {Ellis}}, \bibinfo {author} {\bibfnamefont {J.S.}\ \bibnamefont {Hagelin}},
  \bibinfo {author} {\bibfnamefont {Dimitri~V.}\ \bibnamefont {Nanopoulos}}, \
  and\ \bibinfo {author} {\bibfnamefont {M.}~\bibnamefont {Srednicki}},\
  }\bibfield  {title} {\enquote {\bibinfo {title} {{Search for Violations of
  Quantum Mechanics}},}\ }\href {\doibase 10.1016/0550-3213(84)90053-1}
  {\bibfield  {journal} {\bibinfo  {journal} {Nucl.~Phys.}\ }\textbf {\bibinfo
  {volume} {B241}},\ \bibinfo {pages} {381} (\bibinfo {year}
  {1984})}\BibitemShut {NoStop}%
%%CITATION = NUPHA,B241,381;%%
\bibitem [{\citenamefont {Banks}\ \emph {et~al.}(1984)\citenamefont {Banks},
  \citenamefont {Susskind},\ and\ \citenamefont {Peskin}}]{Banks:1983by}%
  \BibitemOpen
  \bibfield  {author} {\bibinfo {author} {\bibfnamefont {Tom}\ \bibnamefont
  {Banks}}, \bibinfo {author} {\bibfnamefont {Leonard}\ \bibnamefont
  {Susskind}}, \ and\ \bibinfo {author} {\bibfnamefont {Michael~E.}\
  \bibnamefont {Peskin}},\ }\bibfield  {title} {\enquote {\bibinfo {title}
  {{Difficulties for the Evolution of Pure States Into Mixed States}},}\ }\href
  {\doibase 10.1016/0550-3213(84)90184-6} {\bibfield  {journal} {\bibinfo
  {journal} {Nucl.~Phys.}\ }\textbf {\bibinfo {volume} {B244}},\ \bibinfo
  {pages} {125} (\bibinfo {year} {1984})}\BibitemShut {NoStop}%
%%CITATION = NUPHA,B244,125;%%
\bibitem [{\citenamefont {Benatti}\ and\ \citenamefont
  {Floreanini}(2000)}]{Benatti:2000ph}%
  \BibitemOpen
  \bibfield  {author} {\bibinfo {author} {\bibfnamefont {F.}~\bibnamefont
  {Benatti}}\ and\ \bibinfo {author} {\bibfnamefont {R.}~\bibnamefont
  {Floreanini}},\ }\bibfield  {title} {\enquote {\bibinfo {title} {{Open system
  approach to neutrino oscillations}},}\ }\href@noop {} {\bibfield  {journal}
  {\bibinfo  {journal} {JHEP}\ }\textbf {\bibinfo {volume} {02}},\ \bibinfo
  {pages} {032} (\bibinfo {year} {2000})},\ \Eprint
  {http://arxiv.org/abs/hep-ph/0002221} {arXiv:hep-ph/0002221} \BibitemShut
  {NoStop}%
%%CITATION = HEP-PH/0002221;%%
\bibitem [{\citenamefont {Gago}\ \emph {et~al.}(2002)\citenamefont {Gago},
  \citenamefont {Santos}, \citenamefont {Teves},\ and\ \citenamefont
  {Zukanovich~Funchal}}]{Gago:2002na}%
  \BibitemOpen
  \bibfield  {author} {\bibinfo {author} {\bibfnamefont {A.M.}\ \bibnamefont
  {Gago}}, \bibinfo {author} {\bibfnamefont {E.M.}\ \bibnamefont {Santos}},
  \bibinfo {author} {\bibfnamefont {W.J.C.}\ \bibnamefont {Teves}}, \ and\
  \bibinfo {author} {\bibfnamefont {R.}~\bibnamefont {Zukanovich~Funchal}},\
  }\bibfield  {title} {\enquote {\bibinfo {title} {{A Study on quantum
  decoherence phenomena with three generations of neutrinos}},}\ }\href@noop {}
  {\  (\bibinfo {year} {2002})},\ \Eprint {http://arxiv.org/abs/hep-ph/0208166}
  {arXiv:hep-ph/0208166 [hep-ph]} \BibitemShut {NoStop}%
%%CITATION = HEP-PH/0208166;%%
\bibitem [{\citenamefont {Morgan}\ \emph {et~al.}(2006)\citenamefont {Morgan},
  \citenamefont {Winstanley}, \citenamefont {Brunner},\ and\ \citenamefont
  {Thompson}}]{Morgan:2004vv}%
  \BibitemOpen
  \bibfield  {author} {\bibinfo {author} {\bibfnamefont {Dean}\ \bibnamefont
  {Morgan}}, \bibinfo {author} {\bibfnamefont {Elizabeth}\ \bibnamefont
  {Winstanley}}, \bibinfo {author} {\bibfnamefont {Jurgen}\ \bibnamefont
  {Brunner}}, \ and\ \bibinfo {author} {\bibfnamefont {Lee~F.}\ \bibnamefont
  {Thompson}},\ }\bibfield  {title} {\enquote {\bibinfo {title} {{Probing
  quantum decoherence in atmospheric neutrino oscillations with a neutrino
  telescope}},}\ }\href {\doibase 10.1016/j.astropartphys.2006.03.001}
  {\bibfield  {journal} {\bibinfo  {journal} {Astropart.~Phys.}\ }\textbf
  {\bibinfo {volume} {25}},\ \bibinfo {pages} {311--327} (\bibinfo {year}
  {2006})},\ \Eprint {http://arxiv.org/abs/astro-ph/0412618}
  {arXiv:astro-ph/0412618} \BibitemShut {NoStop}%
%%CITATION = ASTRO-PH/0412618;%%
\bibitem [{\citenamefont {Hooper}\ \emph {et~al.}(2005)\citenamefont {Hooper},
  \citenamefont {Morgan},\ and\ \citenamefont {Winstanley}}]{Hooper:2005jp}%
  \BibitemOpen
  \bibfield  {author} {\bibinfo {author} {\bibfnamefont {Dan}\ \bibnamefont
  {Hooper}}, \bibinfo {author} {\bibfnamefont {Dean}\ \bibnamefont {Morgan}}, \
  and\ \bibinfo {author} {\bibfnamefont {Elizabeth}\ \bibnamefont
  {Winstanley}},\ }\bibfield  {title} {\enquote {\bibinfo {title} {{Lorentz and
  CPT invariance violation in high-energy neutrinos}},}\ }\href {\doibase
  10.1103/PhysRevD.72.065009} {\bibfield  {journal} {\bibinfo  {journal} {Phys.
  Rev.}\ }\textbf {\bibinfo {volume} {D72}},\ \bibinfo {pages} {065009}
  (\bibinfo {year} {2005})},\ \Eprint {http://arxiv.org/abs/hep-ph/0506091}
  {arXiv:hep-ph/0506091} \BibitemShut {NoStop}%
%%CITATION = HEP-PH/0506091;%%
\bibitem [{\citenamefont {Anchordoqui}\ \emph
  {et~al.}(2005{\natexlab{b}})\citenamefont {Anchordoqui}, \citenamefont
  {Goldberg}, \citenamefont {Gonzalez-Garcia}, \citenamefont {Halzen},
  \citenamefont {Hooper} \emph {et~al.}}]{Anchordoqui:2005gj}%
  \BibitemOpen
  \bibfield  {author} {\bibinfo {author} {\bibfnamefont {Luis~A.}\ \bibnamefont
  {Anchordoqui}}, \bibinfo {author} {\bibfnamefont {Haim}\ \bibnamefont
  {Goldberg}}, \bibinfo {author} {\bibfnamefont {M.C.}\ \bibnamefont
  {Gonzalez-Garcia}}, \bibinfo {author} {\bibfnamefont {Francis}\ \bibnamefont
  {Halzen}}, \bibinfo {author} {\bibfnamefont {Dan}\ \bibnamefont {Hooper}},
  \emph {et~al.},\ }\bibfield  {title} {\enquote {\bibinfo {title} {{Probing
  Planck scale physics with IceCube}},}\ }\href {\doibase
  10.1103/PhysRevD.72.065019} {\bibfield  {journal} {\bibinfo  {journal}
  {Phys.~Rev.}\ }\textbf {\bibinfo {volume} {D72}},\ \bibinfo {pages} {065019}
  (\bibinfo {year} {2005}{\natexlab{b}})},\ \Eprint
  {http://arxiv.org/abs/hep-ph/0506168} {arXiv:hep-ph/0506168 [hep-ph]}
  \BibitemShut {NoStop}%
%%CITATION = HEP-PH/0506168;%%
\bibitem [{\citenamefont {Bustamante}\ \emph {et~al.}()\citenamefont
  {Bustamante}, \citenamefont {Beacom},\ and\ \citenamefont
  {Murase}}]{BBWprep}%
  \BibitemOpen
  \bibfield  {author} {\bibinfo {author} {\bibfnamefont {Mauricio}\
  \bibnamefont {Bustamante}}, \bibinfo {author} {\bibfnamefont {John~F.}\
  \bibnamefont {Beacom}}, \ and\ \bibinfo {author} {\bibfnamefont {Kohta}\
  \bibnamefont {Murase}},\ }\href@noop {} {\bibinfo  {journal} {{In
  preparation}}\ }\BibitemShut {NoStop}%
\bibitem [{\citenamefont {Colladay}\ and\ \citenamefont
  {Kostelecky}(1998)}]{Colladay:1998fq}%
  \BibitemOpen
\bibfield  {journal} {  }\bibfield  {author} {\bibinfo {author} {\bibfnamefont
  {Don}\ \bibnamefont {Colladay}}\ and\ \bibinfo {author} {\bibfnamefont
  {V.~Alan}\ \bibnamefont {Kostelecky}},\ }\bibfield  {title} {\enquote
  {\bibinfo {title} {{Lorentz violating extension of the standard model}},}\
  }\href {\doibase 10.1103/PhysRevD.58.116002} {\bibfield  {journal} {\bibinfo
  {journal} {Phys.~Rev.}\ }\textbf {\bibinfo {volume} {D58}},\ \bibinfo {pages}
  {116002} (\bibinfo {year} {1998})},\ \Eprint
  {http://arxiv.org/abs/hep-ph/9809521} {arXiv:hep-ph/9809521 [hep-ph]}
  \BibitemShut {NoStop}%
%%CITATION = HEP-PH/9809521;%%
\bibitem [{\citenamefont {Kostelecky}\ and\ \citenamefont
  {Mewes}(2004)}]{Kostelecky:2003xn}%
  \BibitemOpen
  \bibfield  {author} {\bibinfo {author} {\bibfnamefont {V.~Alan}\ \bibnamefont
  {Kostelecky}}\ and\ \bibinfo {author} {\bibfnamefont {Matthew}\ \bibnamefont
  {Mewes}},\ }\bibfield  {title} {\enquote {\bibinfo {title} {{Lorentz and CPT
  violation in the neutrino sector}},}\ }\href {\doibase
  10.1103/PhysRevD.70.031902} {\bibfield  {journal} {\bibinfo  {journal}
  {Phys.~Rev.}\ }\textbf {\bibinfo {volume} {D70}},\ \bibinfo {pages} {031902}
  (\bibinfo {year} {2004})},\ \Eprint {http://arxiv.org/abs/hep-ph/0308300}
  {arXiv:hep-ph/0308300 [hep-ph]} \BibitemShut {NoStop}%
%%CITATION = HEP-PH/0308300;%%
\bibitem [{\citenamefont {Ellis}\ and\ \citenamefont
  {Mavromatos}(2013)}]{Ellis:2011ek}%
  \BibitemOpen
  \bibfield  {author} {\bibinfo {author} {\bibfnamefont {John}\ \bibnamefont
  {Ellis}}\ and\ \bibinfo {author} {\bibfnamefont {Nick~E.}\ \bibnamefont
  {Mavromatos}},\ }\bibfield  {title} {\enquote {\bibinfo {title} {{Probes of
  Lorentz Violation}},}\ }\href {\doibase 10.1016/j.astropartphys.2012.05.004}
  {\bibfield  {journal} {\bibinfo  {journal} {Astropart.~Phys.}\ }\textbf
  {\bibinfo {volume} {43}},\ \bibinfo {pages} {50--55} (\bibinfo {year}
  {2013})},\ \Eprint {http://arxiv.org/abs/1111.1178} {arXiv:1111.1178
  [astro-ph.HE]} \BibitemShut {NoStop}%
%%CITATION = ARXIV:1111.1178;%%
\bibitem [{\citenamefont {Bustamante}\ \emph {et~al.}(2010)\citenamefont
  {Bustamante}, \citenamefont {Gago},\ and\ \citenamefont
  {Pena-Garay}}]{Bustamante:2010nq}%
  \BibitemOpen
  \bibfield  {author} {\bibinfo {author} {\bibfnamefont {M.}~\bibnamefont
  {Bustamante}}, \bibinfo {author} {\bibfnamefont {A.M.}\ \bibnamefont {Gago}},
  \ and\ \bibinfo {author} {\bibfnamefont {C.}~\bibnamefont {Pena-Garay}},\
  }\bibfield  {title} {\enquote {\bibinfo {title} {{Energy-independent new
  physics in the flavour ratios of high-energy astrophysical neutrinos}},}\
  }\href {\doibase 10.1007/JHEP04(2010)066} {\bibfield  {journal} {\bibinfo
  {journal} {JHEP}\ }\textbf {\bibinfo {volume} {1004}},\ \bibinfo {pages}
  {066} (\bibinfo {year} {2010})},\ \Eprint {http://arxiv.org/abs/1001.4878}
  {arXiv:1001.4878 [hep-ph]} \BibitemShut {NoStop}%
\bibitem [{\citenamefont {Gasperini}(1989)}]{Gasperini:1989rt}%
  \BibitemOpen
  \bibfield  {author} {\bibinfo {author} {\bibfnamefont {M.}~\bibnamefont
  {Gasperini}},\ }\bibfield  {title} {\enquote {\bibinfo {title} {{Experimental
  Constraints on a Minimal and Nonminimal Violation of the Equivalence
  Principle in the Oscillations of Massive Neutrinos}},}\ }\href {\doibase
  10.1103/PhysRevD.39.3606} {\bibfield  {journal} {\bibinfo  {journal}
  {Phys.~Rev.}\ }\textbf {\bibinfo {volume} {D39}},\ \bibinfo {pages}
  {3606--3611} (\bibinfo {year} {1989})}\BibitemShut {NoStop}%
%%CITATION = PHRVA,D39,3606;%%
\bibitem [{\citenamefont {Butler}\ \emph {et~al.}(1993)\citenamefont {Butler},
  \citenamefont {Nozawa}, \citenamefont {Malaney},\ and\ \citenamefont
  {Boothroyd}}]{Butler:1993wi}%
  \BibitemOpen
  \bibfield  {author} {\bibinfo {author} {\bibfnamefont {M.N.}\ \bibnamefont
  {Butler}}, \bibinfo {author} {\bibfnamefont {S.}~\bibnamefont {Nozawa}},
  \bibinfo {author} {\bibfnamefont {R.A.}\ \bibnamefont {Malaney}}, \ and\
  \bibinfo {author} {\bibfnamefont {A.I.}\ \bibnamefont {Boothroyd}},\
  }\bibfield  {title} {\enquote {\bibinfo {title} {{Gravitationally induced
  neutrino oscillations}},}\ }\href {\doibase 10.1103/PhysRevD.47.2615}
  {\bibfield  {journal} {\bibinfo  {journal} {Phys.~Rev.}\ }\textbf {\bibinfo
  {volume} {D47}},\ \bibinfo {pages} {2615--2618} (\bibinfo {year}
  {1993})}\BibitemShut {NoStop}%
%%CITATION = PHRVA,D47,2615;%%
\bibitem [{\citenamefont {Glashow}\ \emph {et~al.}(1997)\citenamefont
  {Glashow}, \citenamefont {Halprin}, \citenamefont {Krastev}, \citenamefont
  {Leung},\ and\ \citenamefont {Pantaleone}}]{Glashow:1997gx}%
  \BibitemOpen
  \bibfield  {author} {\bibinfo {author} {\bibfnamefont {S.L.}\ \bibnamefont
  {Glashow}}, \bibinfo {author} {\bibfnamefont {A.}~\bibnamefont {Halprin}},
  \bibinfo {author} {\bibfnamefont {P.I.}\ \bibnamefont {Krastev}}, \bibinfo
  {author} {\bibfnamefont {Chung~Ngoc}\ \bibnamefont {Leung}}, \ and\ \bibinfo
  {author} {\bibfnamefont {James~T.}\ \bibnamefont {Pantaleone}},\ }\bibfield
  {title} {\enquote {\bibinfo {title} {{Comments on neutrino tests of special
  relativity}},}\ }\href {\doibase 10.1103/PhysRevD.56.2433} {\bibfield
  {journal} {\bibinfo  {journal} {Phys.~Rev.}\ }\textbf {\bibinfo {volume}
  {D56}},\ \bibinfo {pages} {2433--2434} (\bibinfo {year} {1997})},\ \Eprint
  {http://arxiv.org/abs/hep-ph/9703454} {arXiv:hep-ph/9703454 [hep-ph]}
  \BibitemShut {NoStop}%
%%CITATION = HEP-PH/9703454;%%
\bibitem [{\citenamefont {De~Sabbata}\ and\ \citenamefont
  {Gasperini}(1981)}]{DeSabbata:1981ek}%
  \BibitemOpen
  \bibfield  {author} {\bibinfo {author} {\bibfnamefont {V.}~\bibnamefont
  {De~Sabbata}}\ and\ \bibinfo {author} {\bibfnamefont {M.}~\bibnamefont
  {Gasperini}},\ }\bibfield  {title} {\enquote {\bibinfo {title} {{Neutrino
  Oscillations in the Presence of Torsion}},}\ }\href {\doibase
  10.1007/BF02902051} {\bibfield  {journal} {\bibinfo  {journal} {Nuovo Cim.}\
  }\textbf {\bibinfo {volume} {A65}},\ \bibinfo {pages} {479--500} (\bibinfo
  {year} {1981})}\BibitemShut {NoStop}%
%%CITATION = NUCIA,A65,479;%%
\bibitem [{\citenamefont {Argüelles}\ \emph {et~al.}(2015)\citenamefont
  {Argüelles}, \citenamefont {Katori},\ and\ \citenamefont
  {Salvado}}]{Arguelles:2015dca}%
  \BibitemOpen
  \bibfield  {author} {\bibinfo {author} {\bibfnamefont {Carlos~A.}\
  \bibnamefont {Argüelles}}, \bibinfo {author} {\bibfnamefont {Teppei}\
  \bibnamefont {Katori}}, \ and\ \bibinfo {author} {\bibfnamefont {Jordi}\
  \bibnamefont {Salvado}},\ }\bibfield  {title} {\enquote {\bibinfo {title}
  {{New Physics in Astrophysical Neutrino Flavor}},}\ }\href@noop {} {\
  (\bibinfo {year} {2015})},\ \Eprint {http://arxiv.org/abs/1506.02043}
  {arXiv:1506.02043 [hep-ph]} \BibitemShut {NoStop}%
%%CITATION = ARXIV:1506.02043;%%
\bibitem [{\citenamefont {Aeikens}\ \emph {et~al.}(2014)\citenamefont
  {Aeikens}, \citenamefont {Pas}, \citenamefont {Pakvasa},\ and\ \citenamefont
  {Sicking}}]{Aeikens:2014yga}%
  \BibitemOpen
  \bibfield  {author} {\bibinfo {author} {\bibfnamefont {Elke}\ \bibnamefont
  {Aeikens}}, \bibinfo {author} {\bibfnamefont {Heinrich}\ \bibnamefont {Pas}},
  \bibinfo {author} {\bibfnamefont {Sandip}\ \bibnamefont {Pakvasa}}, \ and\
  \bibinfo {author} {\bibfnamefont {Philipp}\ \bibnamefont {Sicking}},\
  }\bibfield  {title} {\enquote {\bibinfo {title} {{Flavor ratios of
  extragalactical neutrinos and neutrino shortcuts in extra dimensions}},}\
  }\href@noop {} {\  (\bibinfo {year} {2014})},\ \Eprint
  {http://arxiv.org/abs/1410.0408} {arXiv:1410.0408 [hep-ph]} \BibitemShut
  {NoStop}%
%%CITATION = ARXIV:1410.0408;%%
\bibitem [{\citenamefont {Agashe}\ \emph {et~al.}(2014)\citenamefont {Agashe},
  \citenamefont {Cui}, \citenamefont {Necib},\ and\ \citenamefont
  {Thaler}}]{Agashe:2014yua}%
  \BibitemOpen
  \bibfield  {author} {\bibinfo {author} {\bibfnamefont {Kaustubh}\
  \bibnamefont {Agashe}}, \bibinfo {author} {\bibfnamefont {Yanou}\
  \bibnamefont {Cui}}, \bibinfo {author} {\bibfnamefont {Lina}\ \bibnamefont
  {Necib}}, \ and\ \bibinfo {author} {\bibfnamefont {Jesse}\ \bibnamefont
  {Thaler}},\ }\bibfield  {title} {\enquote {\bibinfo {title} {{(In)direct
  Detection of Boosted Dark Matter}},}\ }\href {\doibase
  10.1088/1475-7516/2014/10/062} {\bibfield  {journal} {\bibinfo  {journal}
  {JCAP}\ }\textbf {\bibinfo {volume} {1410}},\ \bibinfo {pages} {062}
  (\bibinfo {year} {2014})},\ \Eprint {http://arxiv.org/abs/1405.7370}
  {arXiv:1405.7370 [hep-ph]} \BibitemShut {NoStop}%
%%CITATION = ARXIV:1405.7370;%%
\bibitem [{\citenamefont {Kopp}\ \emph {et~al.}(2015)\citenamefont {Kopp},
  \citenamefont {Liu},\ and\ \citenamefont {Wang}}]{Kopp:2015bfa}%
  \BibitemOpen
  \bibfield  {author} {\bibinfo {author} {\bibfnamefont {Joachim}\ \bibnamefont
  {Kopp}}, \bibinfo {author} {\bibfnamefont {Jia}\ \bibnamefont {Liu}}, \ and\
  \bibinfo {author} {\bibfnamefont {Xiao-Ping}\ \bibnamefont {Wang}},\
  }\bibfield  {title} {\enquote {\bibinfo {title} {{Boosted Dark Matter in
  IceCube and at the Galactic Center}},}\ }\href {\doibase
  10.1007/JHEP04(2015)105} {\bibfield  {journal} {\bibinfo  {journal} {JHEP}\
  }\textbf {\bibinfo {volume} {1504}},\ \bibinfo {pages} {105} (\bibinfo {year}
  {2015})},\ \Eprint {http://arxiv.org/abs/1503.02669} {arXiv:1503.02669
  [hep-ph]} \BibitemShut {NoStop}%
%%CITATION = ARXIV:1503.02669;%%
\bibitem [{\citenamefont {Mehta}\ and\ \citenamefont
  {Winter}(2011)}]{Mehta:2011qb}%
  \BibitemOpen
  \bibfield  {author} {\bibinfo {author} {\bibfnamefont {Poonam}\ \bibnamefont
  {Mehta}}\ and\ \bibinfo {author} {\bibfnamefont {Walter}\ \bibnamefont
  {Winter}},\ }\bibfield  {title} {\enquote {\bibinfo {title} {{Interplay of
  energy dependent astrophysical neutrino flavor ratios and new physics
  effects}},}\ }\href@noop {} {\bibfield  {journal} {\bibinfo  {journal}
  {JCAP}\ }\textbf {\bibinfo {volume} {1103}},\ \bibinfo {pages} {041}
  (\bibinfo {year} {2011})},\ \Eprint {http://arxiv.org/abs/1101.2673}
  {arXiv:1101.2673 [hep-ph]} \BibitemShut {NoStop}%
\bibitem [{\citenamefont {Kappes}(2007)}]{Kappes:2007ci}%
  \BibitemOpen
  \bibfield  {author} {\bibinfo {author} {\bibfnamefont {A.}~\bibnamefont
  {Kappes}} (\bibinfo {collaboration} {KM3NeT}),\ }\bibfield  {title} {\enquote
  {\bibinfo {title} {{KM3NeT: A Next Generation Neutrino Telescope in the
  Mediterranean Sea}},}\ }\bibfield  {booktitle} {\emph {\bibinfo {booktitle}
  {{Proceedings of the {Sixth International Workshop on New Worlds in
  Astroparticle Physics}}}},\ }\href@noop {} {\  (\bibinfo {year} {2007})},\
  \Eprint {http://arxiv.org/abs/0711.0563} {arXiv:0711.0563 [astro-ph]}
  \BibitemShut {NoStop}%
%%CITATION = ARXIV:0711.0563;%%
\bibitem [{\citenamefont {H{\"u}mmer}\ \emph {et~al.}(2012)\citenamefont
  {H{\"u}mmer}, \citenamefont {Baerwald},\ and\ \citenamefont
  {Winter}}]{Hummer:2011ms}%
  \BibitemOpen
  \bibfield  {author} {\bibinfo {author} {\bibfnamefont {Svenja}\ \bibnamefont
  {H{\"u}mmer}}, \bibinfo {author} {\bibfnamefont {Philipp}\ \bibnamefont
  {Baerwald}}, \ and\ \bibinfo {author} {\bibfnamefont {Walter}\ \bibnamefont
  {Winter}},\ }\bibfield  {title} {\enquote {\bibinfo {title} {{Neutrino
  Emission from Gamma-Ray Burst Fireballs, Revised}},}\ }\href {\doibase
  10.1103/PhysRevLett.108.231101} {\bibfield  {journal} {\bibinfo  {journal}
  {Phys. Rev. Lett.}\ }\textbf {\bibinfo {volume} {108}},\ \bibinfo {pages}
  {231101} (\bibinfo {year} {2012})},\ \Eprint {http://arxiv.org/abs/1112.1076}
  {arXiv:1112.1076 [astro-ph.HE]} \BibitemShut {NoStop}%
%%CITATION = ARXIV:1112.1076;%%
\bibitem [{\citenamefont {Serpico}\ and\ \citenamefont
  {Kachelriess}(2005)}]{Serpico:2005sz}%
  \BibitemOpen
  \bibfield  {author} {\bibinfo {author} {\bibfnamefont {P.~D.}\ \bibnamefont
  {Serpico}}\ and\ \bibinfo {author} {\bibfnamefont {M.}~\bibnamefont
  {Kachelriess}},\ }\bibfield  {title} {\enquote {\bibinfo {title} {{Measuring
  the 13-mixing angle and the CP phase with neutrino telescopes}},}\
  }\href@noop {} {\bibfield  {journal} {\bibinfo  {journal} {Phys. Rev. Lett.}\
  }\textbf {\bibinfo {volume} {94}},\ \bibinfo {pages} {211102} (\bibinfo
  {year} {2005})},\ \Eprint {http://arxiv.org/abs/hep-ph/0502088}
  {hep-ph/0502088} \BibitemShut {NoStop}%
%%CITATION = HEP-PH 0502088;%%
\bibitem [{\citenamefont {Winter}(2006)}]{Winter:2006ce}%
  \BibitemOpen
  \bibfield  {author} {\bibinfo {author} {\bibfnamefont {Walter}\ \bibnamefont
  {Winter}},\ }\bibfield  {title} {\enquote {\bibinfo {title} {How
  astrophysical neutrino sources could be used for early measurements of
  neutrino mass hierarchy and leptonic {CP} phase},}\ }\href@noop {} {\bibfield
   {journal} {\bibinfo  {journal} {Phys. Rev.}\ }\textbf {\bibinfo {volume}
  {D74}},\ \bibinfo {pages} {033015} (\bibinfo {year} {2006})},\ \Eprint
  {http://arxiv.org/abs/hep-ph/0604191} {hep-ph/0604191} \BibitemShut {NoStop}%
%%CITATION = HEP-PH/0604191;%%
\bibitem [{\citenamefont {Blum}\ \emph {et~al.}(2007)\citenamefont {Blum},
  \citenamefont {Nir},\ and\ \citenamefont {Waxman}}]{Blum:2007ie}%
  \BibitemOpen
  \bibfield  {author} {\bibinfo {author} {\bibfnamefont {Kfir}\ \bibnamefont
  {Blum}}, \bibinfo {author} {\bibfnamefont {Yosef}\ \bibnamefont {Nir}}, \
  and\ \bibinfo {author} {\bibfnamefont {Eli}\ \bibnamefont {Waxman}},\
  }\bibfield  {title} {\enquote {\bibinfo {title} {{Probing CP violation in
  neutrino oscillations with neutrino telescopes}},}\ }\href@noop {} {\
  (\bibinfo {year} {2007})},\ \Eprint {http://arxiv.org/abs/arXiv:0706.2070
  [hep-ph]} {arXiv:0706.2070 [hep-ph]} \BibitemShut {NoStop}%
%%CITATION = ARXIV:0706.2070;%%
\bibitem [{\citenamefont {Meloni}\ and\ \citenamefont
  {Ohlsson}(2012)}]{Meloni:2012nk}%
  \BibitemOpen
  \bibfield  {author} {\bibinfo {author} {\bibfnamefont {Davide}\ \bibnamefont
  {Meloni}}\ and\ \bibinfo {author} {\bibfnamefont {Tommy}\ \bibnamefont
  {Ohlsson}},\ }\bibfield  {title} {\enquote {\bibinfo {title} {{Leptonic CP
  violation and mixing patterns at neutrino telescopes}},}\ }\href {\doibase
  10.1103/PhysRevD.86.067701} {\bibfield  {journal} {\bibinfo  {journal}
  {Phys.~Rev.}\ }\textbf {\bibinfo {volume} {D86}},\ \bibinfo {pages} {067701}
  (\bibinfo {year} {2012})},\ \Eprint {http://arxiv.org/abs/1206.6886}
  {arXiv:1206.6886 [hep-ph]} \BibitemShut {NoStop}%
%%CITATION = ARXIV:1206.6886;%%
\bibitem [{\citenamefont {Chatterjee}\ \emph {et~al.}(2014)\citenamefont
  {Chatterjee}, \citenamefont {Devi}, \citenamefont {Ghosh}, \citenamefont
  {Moharana},\ and\ \citenamefont {Raut}}]{Chatterjee:2013tza}%
  \BibitemOpen
  \bibfield  {author} {\bibinfo {author} {\bibfnamefont {Animesh}\ \bibnamefont
  {Chatterjee}}, \bibinfo {author} {\bibfnamefont {Moon~Moon}\ \bibnamefont
  {Devi}}, \bibinfo {author} {\bibfnamefont {Monojit}\ \bibnamefont {Ghosh}},
  \bibinfo {author} {\bibfnamefont {Reetanjali}\ \bibnamefont {Moharana}}, \
  and\ \bibinfo {author} {\bibfnamefont {Sushant~K.}\ \bibnamefont {Raut}},\
  }\bibfield  {title} {\enquote {\bibinfo {title} {{Probing CP violation with
  the first three years of ultrahigh energy neutrinos from IceCube}},}\ }\href
  {\doibase 10.1103/PhysRevD.90.073003} {\bibfield  {journal} {\bibinfo
  {journal} {Phys.~Rev.}\ }\textbf {\bibinfo {volume} {D90}},\ \bibinfo {pages}
  {073003} (\bibinfo {year} {2014})},\ \Eprint {http://arxiv.org/abs/1312.6593}
  {arXiv:1312.6593 [hep-ph]} \BibitemShut {NoStop}%
%%CITATION = ARXIV:1312.6593;%%
\bibitem [{\citenamefont {Fu}\ and\ \citenamefont {Ho}(2014)}]{Fu:2014gja}%
  \BibitemOpen
  \bibfield  {author} {\bibinfo {author} {\bibfnamefont {Lingjun}\ \bibnamefont
  {Fu}}\ and\ \bibinfo {author} {\bibfnamefont {Chiu~Man}\ \bibnamefont {Ho}},\
  }\bibfield  {title} {\enquote {\bibinfo {title} {{Flavor Ratios and Mass
  Hierarchy at Neutrino Telescopes}},}\ }\href@noop {} {\  (\bibinfo {year}
  {2014})},\ \Eprint {http://arxiv.org/abs/1407.1090} {arXiv:1407.1090
  [hep-ph]} \BibitemShut {NoStop}%
%%CITATION = ARXIV:1407.1090;%%
\bibitem [{\citenamefont {Adams}\ \emph {et~al.}(2013)\citenamefont {Adams}
  \emph {et~al.}}]{Adams:2013qkq}%
  \BibitemOpen
  \bibfield  {author} {\bibinfo {author} {\bibfnamefont {C.}~\bibnamefont
  {Adams}} \emph {et~al.} (\bibinfo {collaboration} {LBNE}),\ }\bibfield
  {title} {\enquote {\bibinfo {title} {{The Long-Baseline Neutrino Experiment:
  Exploring Fundamental Symmetries of the Universe}},}\ }\href@noop {} {\
  (\bibinfo {year} {2013})},\ \Eprint {http://arxiv.org/abs/1307.7335}
  {arXiv:1307.7335 [hep-ex]} \BibitemShut {NoStop}%
%%CITATION = ARXIV:1307.7335;%%
\bibitem [{\citenamefont {Beacom}\ \emph
  {et~al.}(2004{\natexlab{b}})\citenamefont {Beacom}, \citenamefont {Bell},
  \citenamefont {Hooper}, \citenamefont {Pakvasa},\ and\ \citenamefont
  {Weiler}}]{Beacom:2003zg}%
  \BibitemOpen
  \bibfield  {author} {\bibinfo {author} {\bibfnamefont {John~F.}\ \bibnamefont
  {Beacom}}, \bibinfo {author} {\bibfnamefont {Nicole~F.}\ \bibnamefont
  {Bell}}, \bibinfo {author} {\bibfnamefont {Dan}\ \bibnamefont {Hooper}},
  \bibinfo {author} {\bibfnamefont {Sandip}\ \bibnamefont {Pakvasa}}, \ and\
  \bibinfo {author} {\bibfnamefont {Thomas~J.}\ \bibnamefont {Weiler}},\
  }\bibfield  {title} {\enquote {\bibinfo {title} {{Sensitivity to
  $\theta_{13}$ and $\delta$ in the decaying astrophysical neutrino
  scenario}},}\ }\href {\doibase 10.1103/PhysRevD.69.017303} {\bibfield
  {journal} {\bibinfo  {journal} {Phys.~Rev.}\ }\textbf {\bibinfo {volume}
  {D69}},\ \bibinfo {pages} {017303} (\bibinfo {year} {2004}{\natexlab{b}})},\
  \Eprint {http://arxiv.org/abs/hep-ph/0309267} {arXiv:hep-ph/0309267 [hep-ph]}
  \BibitemShut {NoStop}%
%%CITATION = HEP-PH/0309267;%%
\end{thebibliography}


\begin{appendix}

\newpage
\clearpage

\onecolumngrid
\begin{center}
 \bf Supplemental Material
\end{center}
\vspace*{0.2cm}
\twocolumngrid

\section{Plots for inverted mass hierarchy}\label{app:PlotsIH}

We repeat the figures shown in the main text of the paper, but now for the inverted hierarchy (IH) instead of the normal hierarchy. The differences are modest and are vanishing for the $3\sigma$ regions.

\begin{figure}[htb!]
 \centering
 \includegraphics[width=0.45\textwidth]{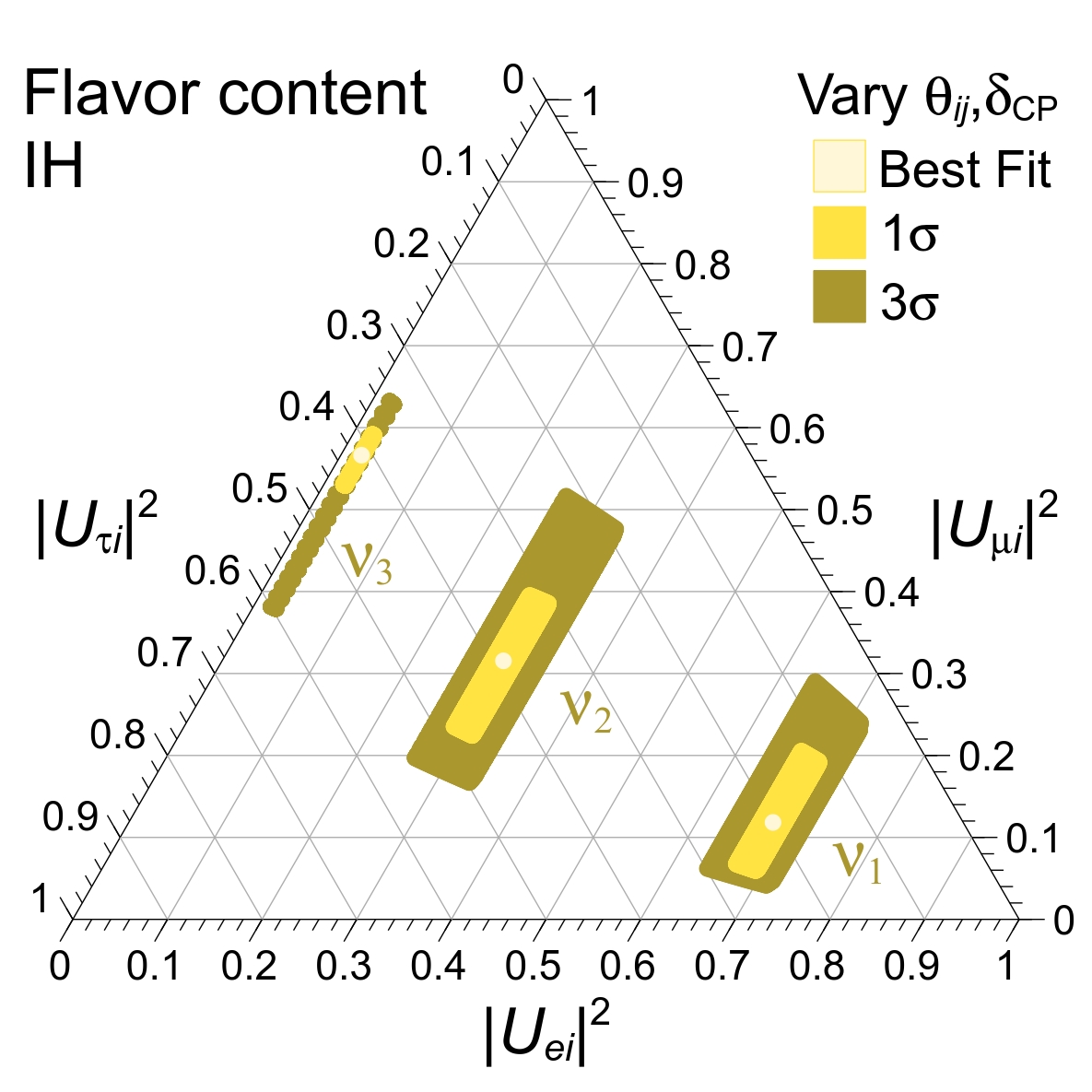}
 \caption{\label{fig:FlavorContentIH}Same as \figu{FlavorContentNH}, but for an inverted mass hierarchy.}
\end{figure}

\begin{figure}[htb!]
 \centering
 \includegraphics[width=0.45\textwidth]{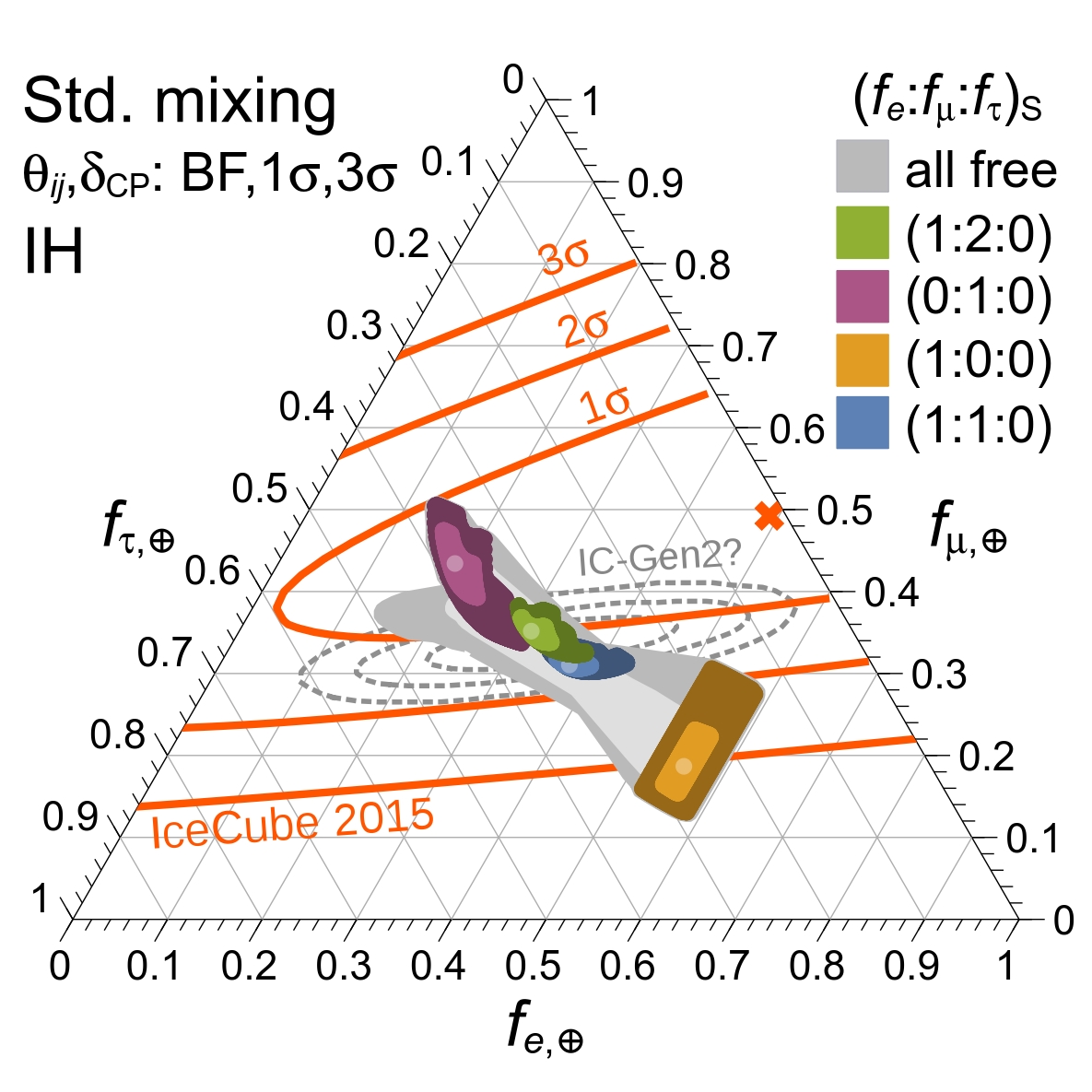}
 \caption{\label{fig:StdMixingSelectedSourceRatiosIH}Same as \figu{StdMixingSelectedSourceRatiosNH}, but for an inverted mass hierarchy.}
\end{figure}

\vspace*{2.7cm}

\begin{figure}[htb!]
 \centering
 \includegraphics[width=0.45\textwidth]{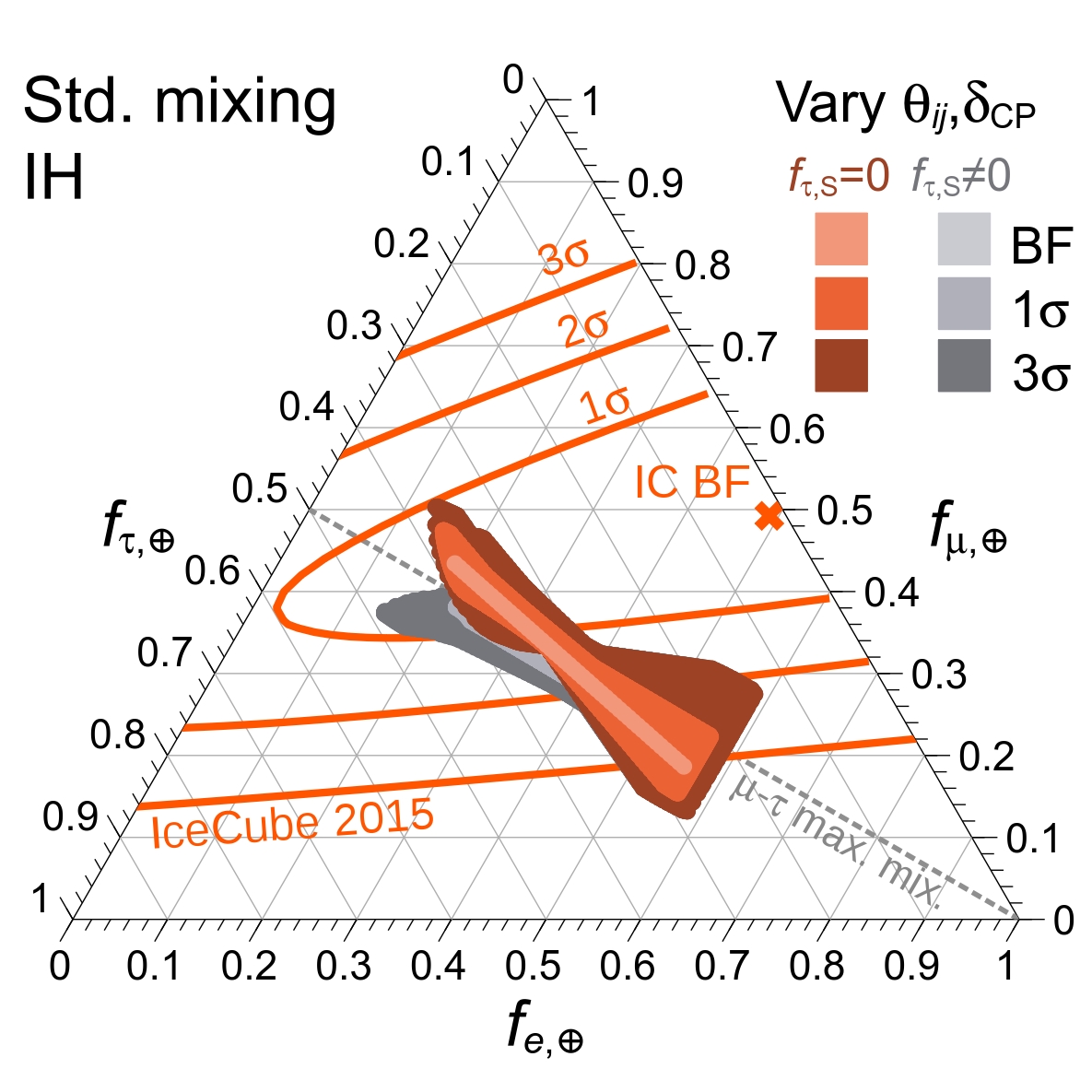}
 \caption{\label{fig:StdMixingFullVarVsNoTauIH}Same as \figu{StdMixingFullVarVsNoTauNH}, but for an inverted mass hierarchy.}
\end{figure}

\vspace*{0.45cm}

\begin{figure}[htb!]
 \centering
 \includegraphics[width=0.45\textwidth]{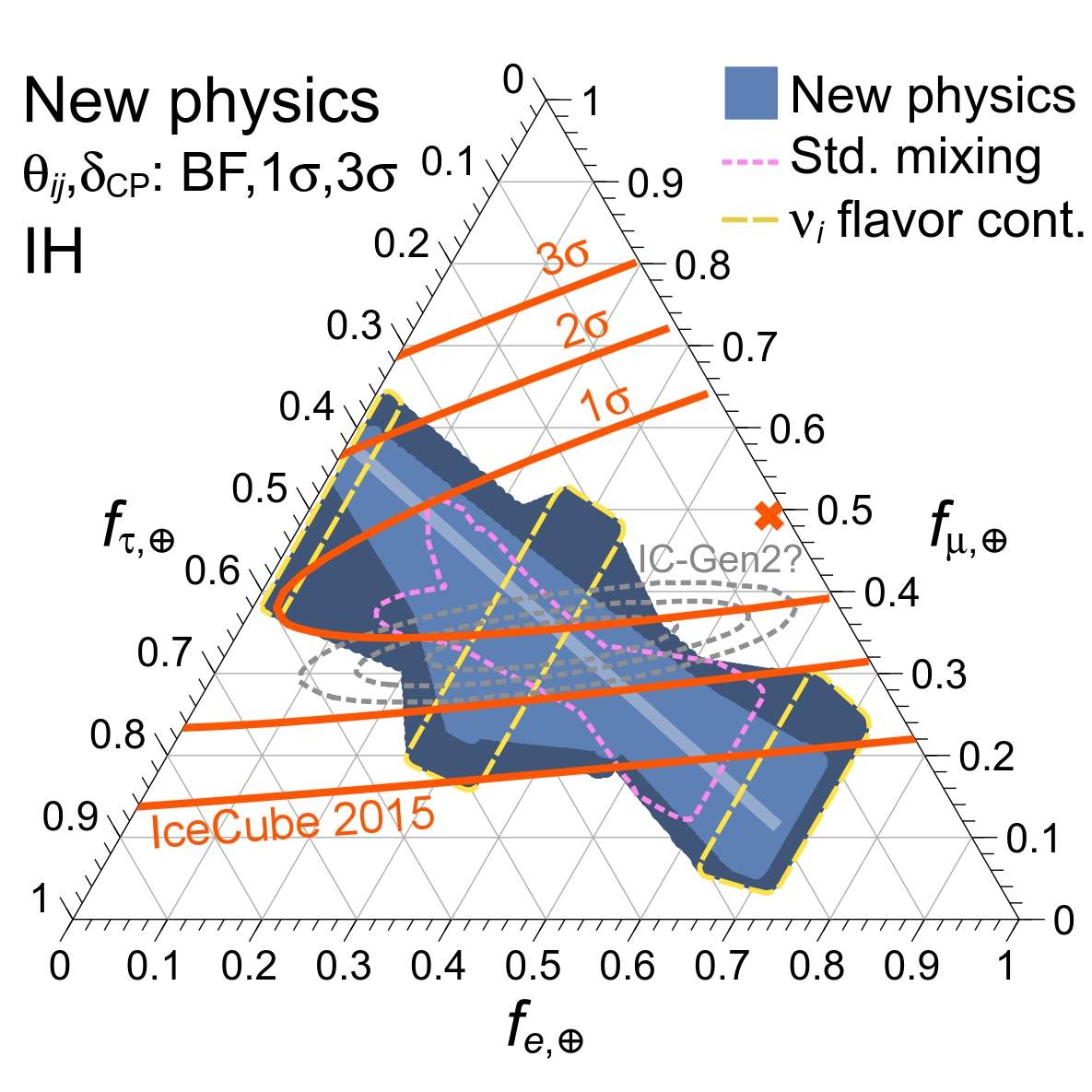}
 \caption{\label{fig:RegionNewPhysicsIH}Same as \figu{RegionNewPhysicsNH}, but for an inverted mass hierarchy.}
\end{figure}

\newpage
\clearpage

%%%%%%%%%%%%%%%%%%%%%%%%%%%%%%%%%%%%%%%%%%%%%%%%%%%%%%%%%%%%%%%%%%%%%%%%%%%%%%%
%%%%%%%%%%%%%%%%%%%%%%%%%%%%%%%%%%%%%%%%%%%%%%%%%%%%%%%%%%%%%%%%%%%%%%%%%%%%%%%

\section{Energy-dependent effects on the flavor composition}\label{app:energy}

The energy dependence of the flavor composition at the source has been considered in, \eg, \Refs~\cite{Kashti:2005qa,Kachelriess:2006fi,Kachelriess:2007tr,Hummer:2010ai}. At lower energies, a $\left(\frac{1}{3}:\frac{2}{3}:0\right)_\text{S}$ composition is expected, coming from the full pion decay chain. At higher energies, where the synchrotron losses of the muons created by the pion decays become larger and where additional neutrino production channels become accessible, the source flavor composition changes. Non-trivial flavor compositions are obtained if several processes compete or the cooled muons pile up at lower energies~\cite{Hummer:2010ai}.

If neutrinos in a broad energy interval are considered, then the inferred flavor composition at Earth will be a superposition of those at different energies. It will be challenging for IceCube to pinpoint the exact flavor ratios unless a large volume of data is available for fine energy binning to be feasible.

Let us assume that such a binning is in fact feasible. In that case, by assuming that the energy dependence of the source flavor composition is the same for all of the sources contributing to the neutrino diffuse flux, we can predict the ratios at Earth at different energies.

\begin{figure*}[htb!]
 \centering
 \includegraphics[width=0.495\textwidth]{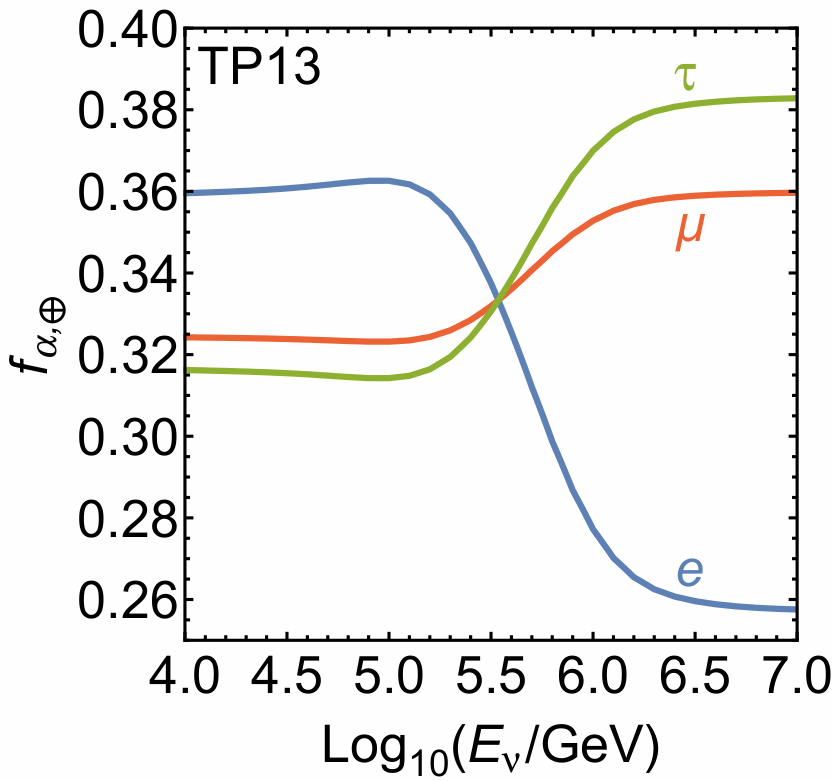} 
 \includegraphics[width=0.495\textwidth]{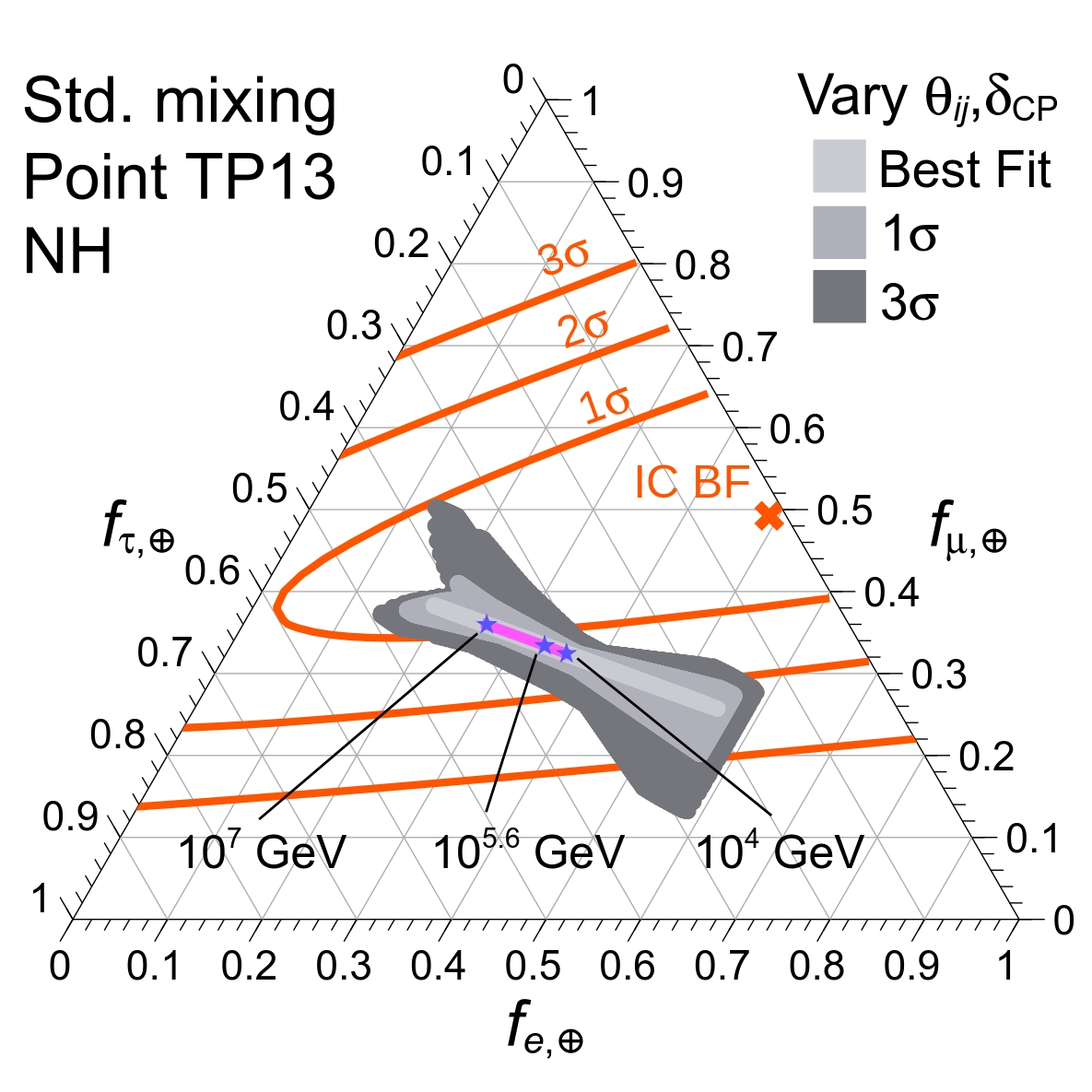}
 \caption{\label{fig:EnergyDependenceTP13}{\it Left:} Variation with energy of the flavor ratios at Earth, for neutrinos coming from a source with emission parameters given by test point TP13 from \Ref~\cite{Hummer:2010ai}. The neutrinos were produced in $p\gamma$ interactions, calculated using NeuCosmA~\cite{Hummer:2010vx,Hummer:2011ms}. {\it Right:} Ternary plots showing the trajectory of the flavor ratios of the same source, marking the values at different energies.}
\end{figure*}

Figure \ref{fig:EnergyDependenceTP13} shows the variation of the flavor ratios at Earth from a source with emission parameters given by a ``classical'' pion beam source evolving into a muon-damped source at higher energies (test point TP13 from \Ref~\cite{Hummer:2010ai}, a photohadronic model where the target photons come from synchrotron emission of co-accelerated electrons). As expected, the trajectory in flavor space lies within the Standard Model region shown in \figu{StdMixingFullVarVsNoTauNH}.

\begin{figure*}[htb!]
 \centering
 \includegraphics[width=0.495\textwidth]{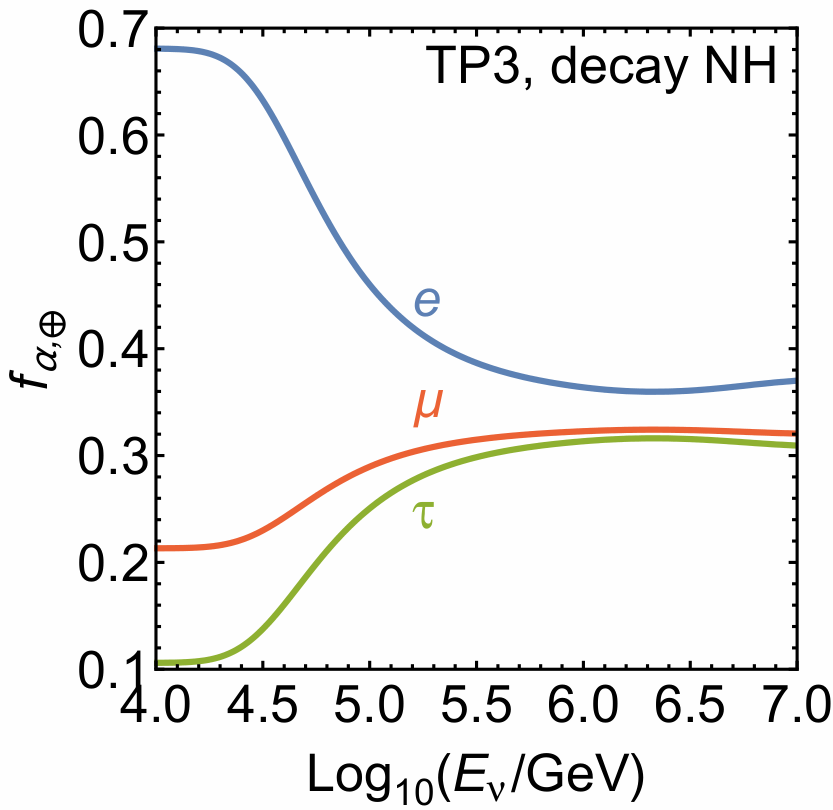} 
 \includegraphics[width=0.495\textwidth]{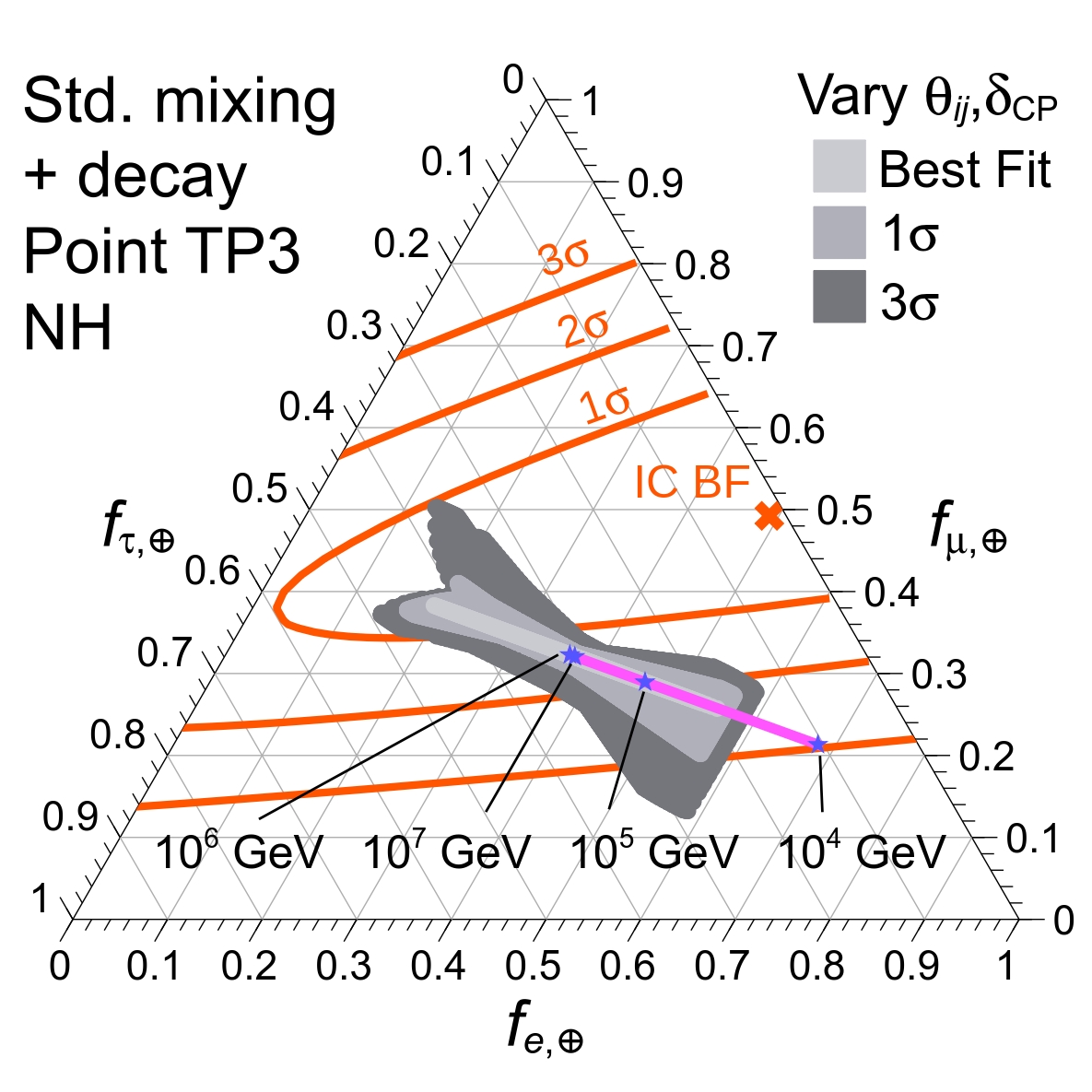}
 \caption{\label{fig:EnergyDependenceTP3DecayNH}{\it Left:} Variation with energy of the flavor ratios at Earth, for neutrinos coming from a source with emission parameters given by test point TP3 from \Ref~\cite{Hummer:2010ai}, including decay of $\nu_2$ and $\nu_3$. The neutrinos were produced in $p\gamma$ interactions, calculated using NeuCosmA~\cite{Hummer:2010vx,Hummer:2011ms}. {\it Right:} Ternary plots showing the trajectory of the flavor ratios of the same source, marking the values at different energies.}
\end{figure*}

Figure \ref{fig:EnergyDependenceTP3DecayNH} shows the results for a different parameter set (test point TP3 from \Ref~\cite{Hummer:2010ai}, corresponding to AGN cores, a mixed source at low energies, where muon pile-ups add to the pion decay chain, evolving into muon damped source at high energies), this time taking into account in addition invisible decays of $\nu_2$ and $\nu_3$, and a stable $\nu_1$, \ie, in the normal hierarchy. The fact that decays are invisible means that the decay products of $\nu_2$ and $\nu_3$ do not contribute to the $\nu_1$ flux.

In this example, the energy dependence of the flavor composition at the source competes with the energy dependence of the new physics effect; see discussion in \Ref~\cite{Mehta:2011qb}.
We have fixed the decay damping parameter to $\hat \alpha L = 10^5 \, \mathrm{GeV}$, following the notation in Eq.~(18) of \Ref~\cite{Mehta:2011qb}.
As expected, the trajectory in flavor space now leaves the Standard Model region at the lowest energies, where the suppression from decay is stronger, and reaches the region corresponding to the flavor content of $\nu_1$ (see \figu{FlavorContentNH}). The full trajectory is still completely contained within the new physics region from \figu{RegionNewPhysicsNH}.

\begin{figure*}[htb!]
 \centering
 \includegraphics[width=0.495\textwidth]{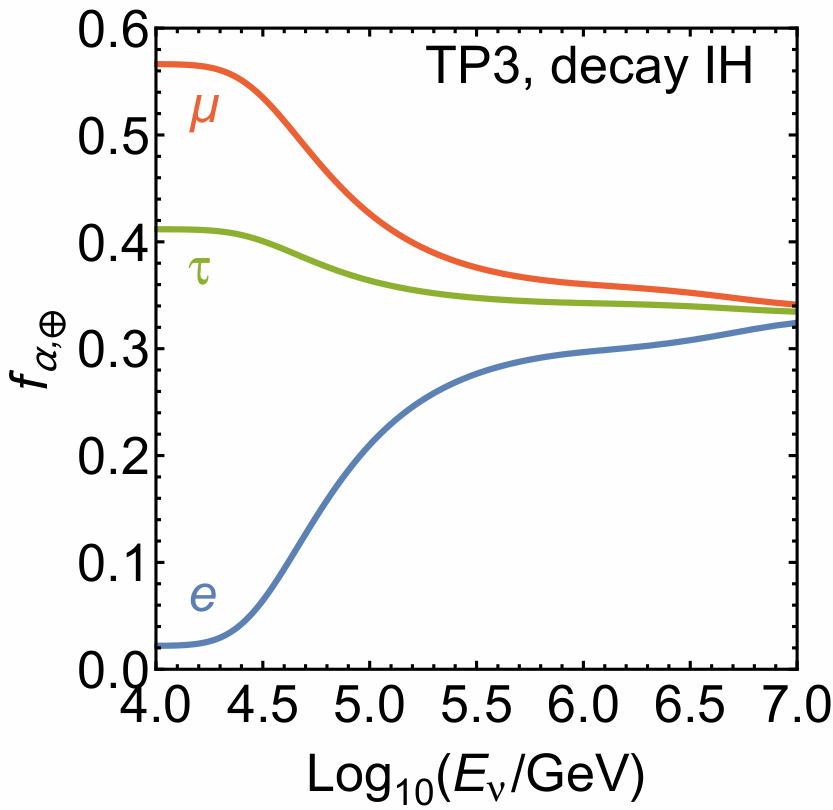} 
 \includegraphics[width=0.495\textwidth]{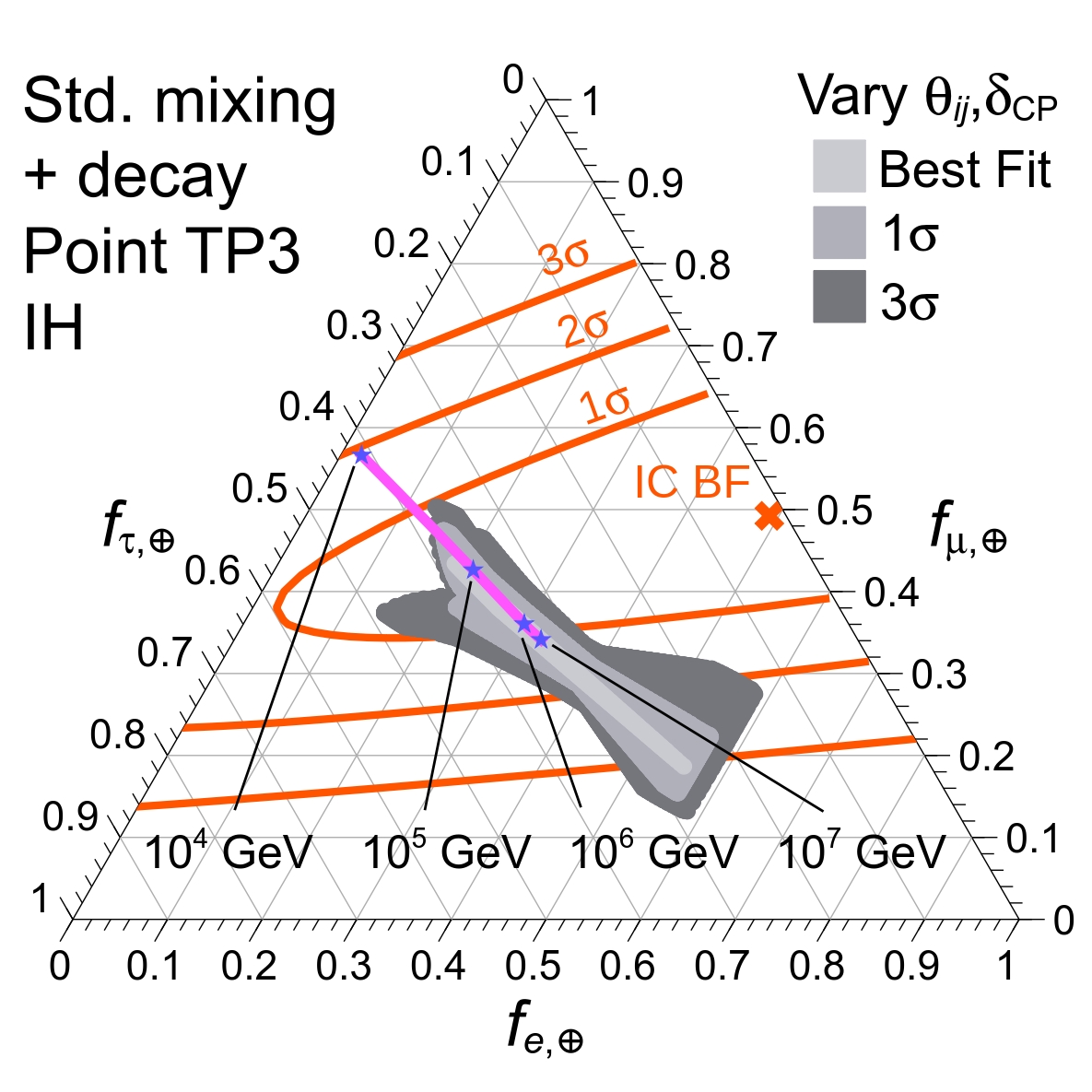}
 \caption{\label{fig:EnergyDependenceTP3DecayIH}{\it Left:} Variation with energy of the flavor ratios at Earth, for neutrinos coming from a source with emission parameters given by test point TP3 from \Ref~\cite{Hummer:2010ai}, including decay of $\nu_1$ and $\nu_2$. The neutrinos were produced in $p\gamma$ interactions, calculated using NeuCosmA~\cite{Hummer:2010vx,Hummer:2011ms}. {\it Right:} Ternary plots showing the trajectory of the flavor ratios of the same source, marking the values at different energies.}
\end{figure*}

Figure \ref{fig:EnergyDependenceTP3DecayIH} shows complementary results for test point TP3 in the inverted hierarchy case, when $\nu_3$ is stable while $\nu_1$ and $\nu_2$ decay. In this case, the trajectory leaves the Standard Model region and reaches the region corresponding to the flavor content of $\nu_3$. Note that this case may be disfavored by the observation of neutrinos from SN 1987A.

As long as the flavor ratios at Earth are confined within the gray region of Figs.~\ref{fig:EnergyDependenceTP13}--\ref{fig:EnergyDependenceTP3DecayIH}, no distinction between the effects of energy-dependent new physics and those of energy-dependent flavor ratios at the source is possible. Given sufficient energy resolution and statistics, however, it might become possible to disentangle them: if, for some energy range, the flavor ratios leave the gray region and enter the region that is reachable exclusively with new physics, shown in Figs.~\ref{fig:RegionNewPhysicsNH} and \ref{fig:RegionNewPhysicsIH}, then this would represent evidence of energy-dependent new physics. As neutrino mixing parameters become better measured, the gray region will shrink, improving the ability to determine the cause of any observed energy variation.

If the experimental determination of the flavor ratios finds no energy dependence, then we could interpret this as a hint that neutrino production occurs in a narrow energy window, inside of which the flavor composition at the sources is approximately constant.

\onecolumngrid

\twocolumngrid

\newpage
\clearpage

\onecolumngrid

\begin{figure*}[htb!]
 \centering
 \includegraphics[width=0.45\textwidth]{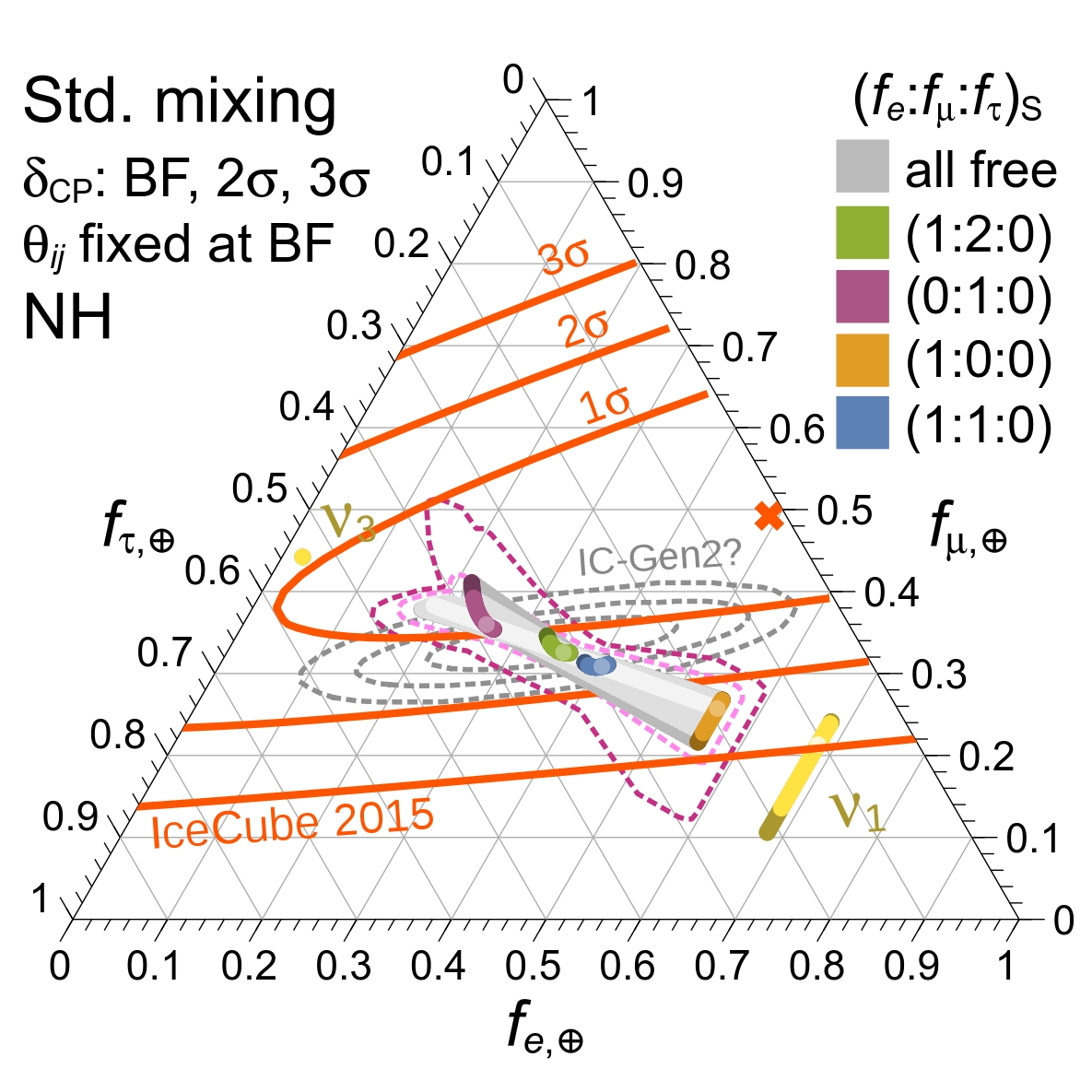}
 \includegraphics[width=0.45\textwidth]{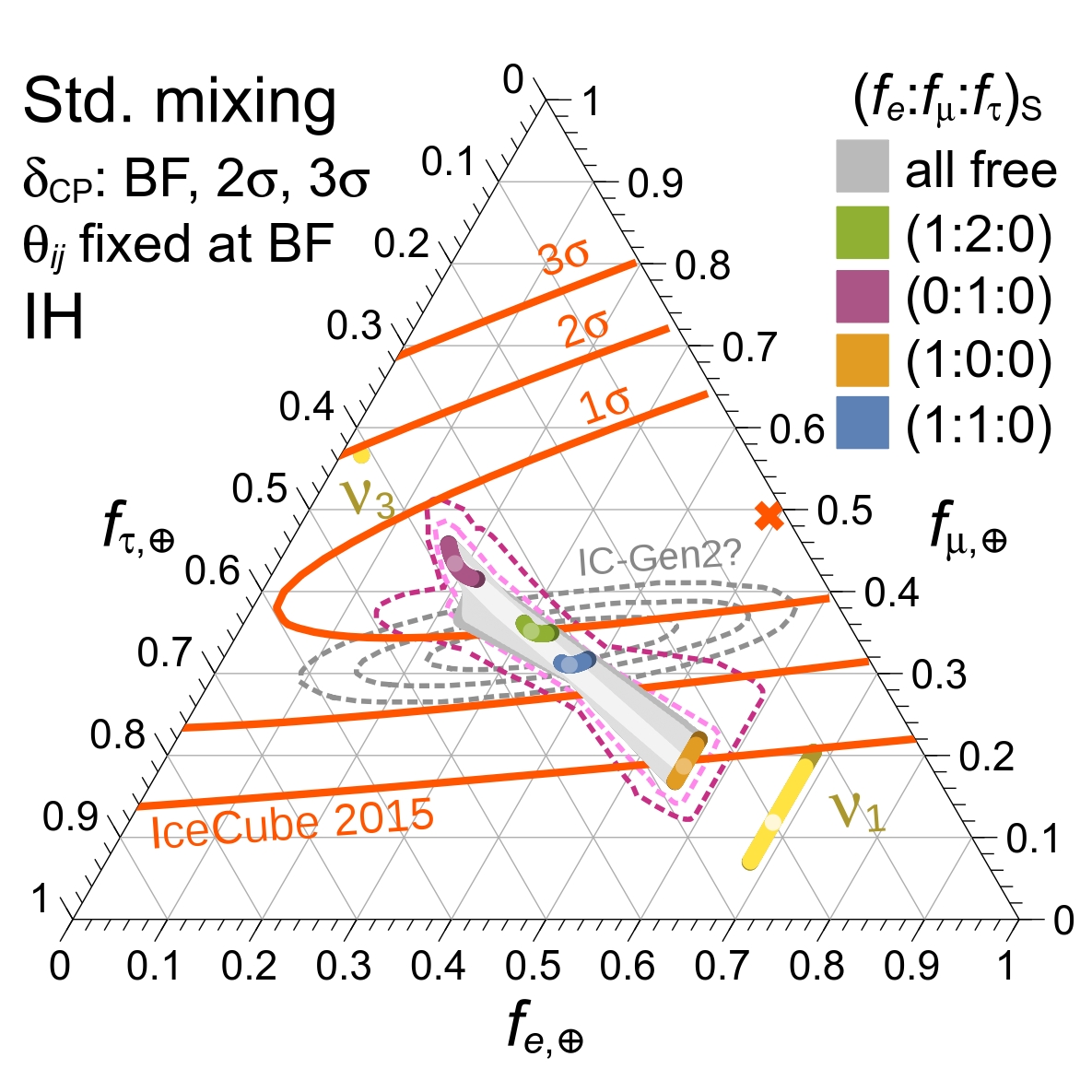}
 \caption{\label{fig:StdMixingFullVarOnlyDelta}Flavor ratios at Earth with unconstrained source flavor ratios and four particular choices of source flavor ratios. The mixing angles are kept fixed at their current best-fit values; the CP-violation phase $\deltacp$ is the only mixing parameter that has been varied to create the colored regions. The regions corresponding to the flavor content of $\nu_1$ and $\nu_3$ at the $1\sigma$ and $3\sigma$ regions are also shown; note that the flavor content of $\nu_3$ has no dependence on $\deltacp$.}
\end{figure*}

\twocolumngrid

%%%%%%%%%%%%%%%%%%%%%%%%%%%%%%%%%%%%%%%%%%%%%%%%%%%%%%%%%%%%%%%%%%%%%%%%%%%%%%%
%%%%%%%%%%%%%%%%%%%%%%%%%%%%%%%%%%%%%%%%%%%%%%%%%%%%%%%%%%%%%%%%%%%%%%%%%%%%%%%

\section{Measurement of $\boldsymbol{\deltacp}$?}\label{appendix:DeltaCP}

There have been proposals to determine the leptonic CP phase, $\deltacp$, using high-energy astrophysical neutrinos; see, \eg, \Refs~\cite{Serpico:2005sz,Winter:2006ce,Blum:2007ie,Meloni:2012nk,Chatterjee:2013tza,Fu:2014gja}. These studies considered specific flavor ratios at the source, typically, $\left(\frac{1}{3}:\frac{2}{3}:0\right)_\text{S}$, $\left(0:1:0\right)_\text{S}$, or $\left(1:0:0\right)_\text{S}$, and studied the effect of $\deltacp$ on the ratio at Earth of the $\nu_\mu+\bar{\nu}_\mu$ flux to the $\nu_e+\bar{\nu}_e+\nu_\tau+\bar{\nu}_\tau$ flux. For a value of $\theta_{13} \approx 10^\circ$, this ratio was found to vary as a function of $\deltacp$ in up to $\sim 20\%$, with the muon-damped and neutron decay scenarios being the most sensitive. At higher energies, if tau-neutrinos are identified, the ratio of $\nu_e+\bar{\nu}_e$ to $\nu_\tau+\bar{\nu}_\tau$ can be used to improve the measurement of $\deltacp$. While optimistic studies~\cite{Blum:2007ie} find that astrophysical sources by themselves might be able to claim a measurement of $\deltacp$, most studies (\eg, \cite{Serpico:2005sz,Winter:2006ce}) focus on a combined analysis of oscillation and astrophysical neutrino data in order to ascertain discovery.

We therefore show in \figu{StdMixingFullVarOnlyDelta} the allowed regions of flavor composition if all of the mixing parameters except $\deltacp$ are fixed to their current best-fit values. This assumption corresponds to future improved measurements of the mixing angles coming from other experiments, whereas $\deltacp$ is to be determined from neutrino telescope data. We include the region corresponding to unconstrained flavor ratios at the source, not only to particular ratios. For comparison, we include the outlines of the regions generated when the mixing angles are also allowed to vary within their $1\sigma$ and $3\sigma$ regions.

The comparison of Figs.~\ref{fig:StdMixingSelectedSourceRatiosNH} \& \ref{fig:StdMixingSelectedSourceRatiosIH} and \figu{StdMixingFullVarOnlyDelta} shows that, in the absence of knowledge of the flavor composition at the source, $\deltacp$ cannot be extracted. However, if the source flavor composition were known to be muon-damped ($(0:1:0)_\text{S}$) or neutron source ($(1:0:0)_\text{S}$), then, given the statistics of IceCube-Gen2, one would expect a marginal statistically significant contribution to the mass hierarchy or CP-violation sensitivities of T2K and NO$\nu$A~\cite{Winter:2006ce}. A new long-baseline experiment, such as DUNE \cite{Adams:2013qkq}, would be more precise.

Notably, if neutrinos fully decay in the NH, the measurement of $\deltacp$ may even be more attractive~\cite{Beacom:2003zg}. Figure \ref{fig:StdMixingSelectedSourceRatiosNH} shows that, in this case, IceCube-Gen2 could actually measure the phase, since the flavor ratios would be determined solely by the flavor content of the one stable mass eigenstate, $\nu_1$ (lower right yellow region). This decay scenario is, however, disfavored by current IceCube data at $\gtrsim 2\sigma$; see \figu{RegionNewPhysicsNH}.

\end{appendix}

\end{document}